\documentclass{ws-rv975x65}

\def\ltwid{\mathrel{\raise.3ex\hbox{$<$\kern-.75em\lower1ex\hbox{$\sim$}}}}
\def\gtwid{\mathrel{\raise.3ex\hbox{$>$\kern-.75em\lower1ex\hbox{$\sim$}}}}

\def\Fint{\rlap{$\Biggl\rfloor$}\Biggl\lceil}
\def\square{\kern1pt\vbox{\hrule height 1.2pt\hbox{\vrule width 1.2pt\hskip 3pt
   \vbox{\vskip 6pt}\hskip 3pt\vrule width 0.6pt}\hrule height 0.6pt}\kern1pt}
\def\overleftrightarrow#1{\vbox{\ialign{##\crcr
     $\leftrightarrow$\crcr\noalign{\kern-1pt\nointerlineskip}
     $\hfil\displaystyle{#1}\hfil$\crcr}}}

\usepackage{graphicx}

\usepackage{ws-rv-van}            
\usepackage{ws-rv-thm}  
\makeindex
\begin{document}
\chapter[Perturbative Quantum Gravity Comes of Age]{Perturbative Quantum Gravity Comes of Age}
\author[R. P. Woodard]{R. P. Woodard}
\address{Department of Physics, University of Florida \\
Gainesville, FL 32611, UNITED STATES \\
woodard@phys.ufl.edu}

\begin{abstract}
I argue that cosmological data from the epoch of primordial
inflation is catalyzing the maturation of quantum gravity 
from speculation into a hard science. I explain why quantum 
gravitational effects from primordial inflation are
observable. I then review what has been done, both 
theoretically and observationally, and what the future holds.
I also discuss what this tells us about quantum gravity. 
\end{abstract}
\body

\section{Introduction}\label{intro}

Gravity was the first of the fundamental forces to impress its
existence upon our ancestors because it is universally attractive
and long range. These same features ensure its precedence in 
cosmology. Gravity also couples to stress-energy, which is why 
quantum general relativity is not perturbatively renormalizable
\cite{DeWitt:1967yk,DeWitt:1967ub,DeWitt:1967uc,'tHooft:1974bx,
Deser:1974zzd,Deser:1974cz,Deser:1974cy,Deser:1974nb,Deser:1974xq,
Stelle:1976gc,Goroff:1985sz,Goroff:1985th,vandeVen:1991gw}, and 
why identifiable effects are unobservably weak at low energies 
\cite{Woodard:2009ns}. These problems have hindered the study of 
quantum gravity until recently. This article is about how 
interlocking developments in the theory and observation of 
inflationary cosmology have changed that situation, and what the 
future holds.

The experiences of two Harvard graduate students serve to 
illustrate the situation before inflation. The first is Leonard
Parker who took his degree in 1967, based on his justly famous 
work quantifying particle production in an expanding universe
\cite{Parker:1968mv,Parker:1969au,Parker:1971pt}. Back then 
people believed that the expansion of the universe had been 
constantly slowing down or ``decelerating''. Parker's work
was greeted with indifference on account of the small particle 
production associated with the current expansion, and on the 
inability of a decelerating universe to preserve memories of 
early times when the expansion rate was much higher. The ruling 
dogma of the 1960's was S-matrix theory, whose more extreme 
proponents believed they could guess the strong interaction 
S-matrix based on a very few properties such as analyticity and 
unitarity. Through a curious process this later morphed into 
string theory. Quantum field theory was regarded as a failed 
formalism whose success for quantum electrodynamics was an 
accident.

Confirmation of the Standard Model had changed opinions about 
quantum field theory by my own time at Harvard (1977-1983).
However, the perturbative nonrenormalizability of quantum 
general relativity led to dismissive statements such as, ``only 
old men should work on quantum gravity.'' The formalism of 
quantum field theory had also become completely tied to 
asymptotic scattering experiments. For example, no one worried 
about correcting free vacuum because infinite time evolution 
from ``in'' states to ``out'' states was supposed to do this 
automatically. Little attention was paid to making observations 
at finite times because the S-matrix was deemed the only valid 
observable, the knowledge of which completely defined a quantum
field theory. My thesis on developing an invariant extension of 
local Green's functions for quantum gravity was only accepted 
because Brandeis Professor Stanley Deser vouched for it. I left
it unpublished for eight years \cite{Tsamis:1989yu}.

The situation was no better during the early stages of my 
career. As a postdoc I worked with a very bright graduate 
student who dismissed the quantum gravity community as ``la-la 
land'' and made no secret of his plan to change fields. 
And there is no denying that any number of crank ideas were
treated with perfect seriousness in those days, which validated 
our critics. I recall knowledgeable people questioning why anyone 
bothered trying to quantize gravity in view of the classical 
theory's success. That opinion was never viable in view of the 
fact that the lowest divergences of quantum gravity 
\cite{'tHooft:1974bx,Deser:1974zzd,Deser:1974cz,
Deser:1974cy,Deser:1974nb,Deser:1974xq} derive from the 
gravitational response to matter theories which are certainly
quantum, whether or not gravitons exist \cite{Woodard:2009ns}. 
The difference between then and now is that I can point to data 
--- and quite a lot of it --- from the same gravitational 
response to quantum matter.

Today cosmological particle production is recognized as the
source of the primordial perturbations which seeded structure 
formation. There is a growing realization that these perturbations 
are quantum gravitational phenomena \cite{Woodard:2009ns,
Krauss:2013pha}, and that they cannot be described by any sort 
of S-matrix or by the use of in-out quantum field theory 
\cite{Weinberg:2005vy,Weinberg:2006ac}. This poses a challenge 
for fundamental theory and an opportunity for its practitioners, 
which dismays some physicists and delights others. All of the 
problems that had to be solved for flat space scattering theory 
in the mid 20th century are being re-examined, in particular, 
defining observables which are infrared finite, renormalizable 
(at least in the sense of low energy effective field theory) and 
in rough agreement with the way things are measured 
\cite{Miao:2012xc,Miao:2013oko}. People are also thinking 
seriously about how to perturbatively correct the initial state 
\cite{Kahya:2009sz}.

This revolutionary change of attitude did not result from any
outbreak of sobriety within the quantum gravity community, or
of toleration from our colleagues. The transformation was forced 
upon us by the overwhelming data in support of inflationary 
cosmology. In the coming sections of this article I review the 
theory behind that data, in particular:
\begin{itemize}
\item{Why quantum gravitational effects from inflation are 
observable;}
\item{Why the tree order power spectra are quantum gravitational 
effects;}
\item{Loop corrections to the primordial power spectra;}
\item{Other potentially observable effects; and}
\item{What the future holds.}
\end{itemize}

\section{Why Quantum Gravitational Effects from Primordial Inflation 
Are Observable}\label{why}

Three things are responsible for the remarkable fact that quantum
gravitational effects from the epoch of primordial inflation can be
observed today:
\begin{itemize}
\item{The inflationary Hubble parameter is large enough that quantum
gravitational effects are small, but not negligible;}
\item{Long wave length gravitons and massless, minimally coupled 
scalars experience explosive particle production during inflation; 
and}
\item{The process of first horizon crossing results in long wave
length gravitons and massless, minimally coupled scalars becoming 
fossilized so that they can survive down to the current epoch.}
\end{itemize}
I will make the first point at the beginning, in the subsection on
the inflationary background. Then the subsection on perturbations
discusses the second and third points.

\subsection{The Background Geometry}\label{prim}

\begin{figure}[ht]
\vspace{-3cm}
\hspace{-2cm} \includegraphics[width=14.0cm,height=9cm]{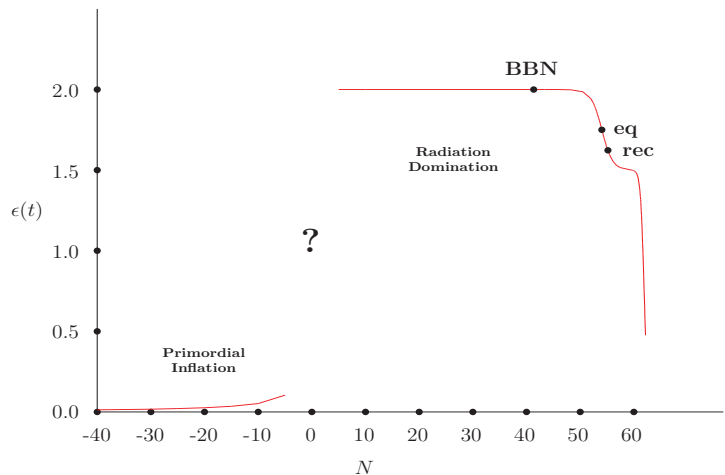} 
\caption{\label{epsilon} \tiny The red curve shows the first 
slow roll parameter 
$\epsilon(t)\equiv -\dot{H}/{H^2}$ as a function of the number 
of e-foldings $N$ since the end of primordial inflation. A 
question mark stands for the phase of reheating between the 
epochs of primordial inflation and radiation domination, because 
there are many models for this period. Significant events marked 
on the graph are Big Bang Nucleosynthesis ({\bf BBN}) when the 
seven lightest isotopes were produced, matter-radiation equality 
({\bf eq}) when the energy density was composed of equal amounts 
of relativistic and non-relativistic matter, and recombination 
({\bf rec}) when neutral Hydrogen formed and the cosmic microwave 
radiation began free-streaming. Observable cosmological 
perturbations experience first horizon crossing near the lower 
left hand corner of the graph.}
\end{figure}

On scales larger than about 100 Mpc the observable universe is 
approximately homogeneous, isotropic and spatially flat. The 
invariant element for such a geometry can be put in the form,
\begin{equation}
ds^2 = -c^2 dt^2 + a^2(t) d\vec{x} \!\cdot\! d\vec{x} \; . \label{FLRW}
\end{equation}
Two derivatives of the scale factor $a(t)$ have great significance,
the Hubble parameter $H(t)$ and the first slow roll parameter
$\epsilon(t)$,
\begin{equation}
H(t) \equiv \frac{\dot{a}}{a} \qquad , \qquad \epsilon(t) \equiv 
-\frac{\dot{H}}{H^2} \; . \label{cparms}
\end{equation}

Inflation is defined as $H(t)>0$ with $\epsilon(t) < 1$. One can see 
that it is possible from the current values of the cosmological 
parameters (denoted by a subscript zero) \cite{Ade:2013zuv},
\begin{equation}
H_0 \approx 2.2 \times 10^{-18}~{\rm Hz} \qquad , \qquad 
\epsilon_0 \approx 0.47 \; . \label{paramsnow}
\end{equation}
However, the important phase of inflation for my purposes is {\it
Primordial Inflation}, which is conjectured to have occurred during
the first $10^{-32}$ seconds of existence. If the BICEP2 detection 
of primordial B-mode polarization is accepted then we finally know 
the values of $H(t)$ and $\epsilon(t)$ near the end of primordial 
inflation \cite{Ade:2014xna},
\begin{equation}
H_i \approx 1.8 \times 10^{+38}~{\rm Hz} \qquad , \qquad 
\epsilon_i \approx 0.013 \; . \label{paramsthen}
\end{equation}
I will comment later on the significance of $H_i$. Let us here 
note that $\epsilon_i$ is very near the de Sitter limit of 
$\epsilon = 0$ at which the Hubble parameter becomes constant. 
This is a very common background to use when estimating quantum
effects during primordial inflation.

We have direct observational evidence that both the scale factor 
and its logarithmic time derivative $H(t)$ have changed over
many orders of magnitude during cosmic history. In contrast, the
deceleration parameter only varies over the small range $0 \leq
\epsilon(t) \leq 2$. Figure \ref{epsilon} shows what we think we
 know about $\epsilon(t)$ as a function of the number of e-foldings 
since the end of primordial inflation at $t = t_e$,
\begin{equation}
N(t) \equiv \ln\Bigl[ \frac{a(t)}{a(t_e)} \Bigr] \; . \label{N}
\end{equation}
It is a tribute to decades of observational work that only a small
portion of this figure is really unknown, corresponding to the
phase of re-heating at the end of inflation.

Primordial inflation was advanced in the late 1970's and early 
1980's to explain the absence of observed relics (primordial black
holes, magnetic monopoles, cosmic strings) and the initial 
conditions (homogeneous, isotropic and spatially flat) for the 
long epoch of radiation domination which is visible on Figure 
\ref{epsilon}. After some notable precursors \cite{Brout:1977ix,
Starobinsky:1980te,Kazanas:1980tx,Sato:1980yn}, the paper of Guth
\cite{Guth:1980zm} focussed attention on the advantages of 
a early epoch of inflation and, incidentally, coined the name.
Important additional work concerned finding an acceptable way to
commence inflation and to make it end \cite{Linde:1981mu,
Albrecht:1982wi}. The first completely successful model was 
Linde's ``Chaotic Inflation'' \cite{Linde:1983gd}. 

One of the most powerful motivations for primordial inflation is 
that it explains the {\it Horizon Problem} of why events far back 
in our past light-cone seem so uniform. I will review the argument 
here because the same analysis is useful for the next subsection. 
From the cosmological geometry (\ref{FLRW}) we can easily compute 
the coordinate distance $R(t_2,t_1)$ traversed by a light ray 
whose trajectory obeys $ds^2 = 0$,
\begin{equation}
R(t_2,t_1) \equiv \int_{t_2}^{t_1} \!\!\! \frac{c dt}{a(t)} \; .
\label{light-cone}
\end{equation}
Now note the relation,
\begin{equation}
\frac{d}{dt} \Biggl[ \frac1{(\epsilon \!-\! 1) H a}\Biggr] = 
\frac1{a} \Biggl[1 - \frac{\dot{\epsilon}}{(\epsilon \!-\! 1)^2 H}
\Biggr] \; . \label{exact}
\end{equation}
One can see from Figure \ref{epsilon} that $\epsilon(t)$ was nearly
constant over long periods of cosmic evolution, in particular 
during the epoch of radiation domination, which would extend back 
to the beginning if it were not for primordial inflation. So we can 
drop the second term of (\ref{exact}) to conclude,
\begin{equation}
R(t_2,t_1) \approx \frac{c}{(\epsilon_1 \!-\! 1) H_1 a_1} -
\frac{c}{(\epsilon_2 \!-\! 1) H_2 a_2} \; . \label{Rapprox}
\end{equation}

One additional exact relation brings the horizon problem to focus,
\begin{equation}
\frac{d}{dt} \Bigl[ H(t) a(t) \Bigr] = - \Bigl[ \epsilon(t) \!-\! 1
\Bigr] H^2(t) a(t) \; . \label{horizon}
\end{equation}
Combining equation (\ref{horizon}) with (\ref{Rapprox}) reveals a 
crucial distinction between inflation ($\epsilon(t) < 1$) and 
deceleration ($\epsilon(t) > 1$): {\it during deceleration the 
radius of the light-cone is dominated by its upper limit, whereas 
the lower limit dominates during inflation.} The horizon problem 
derives from assuming that there was no phase of primordial inflation 
so that the epoch of radiation domination extends back to the 
beginning of the universe. Suppose that the universe began at $t = 
t_2$ and we view some early event such as recombination ({\bf rec} 
on Fig. \ref{epsilon}) or big bang nucleosynthesis ({\bf BBN} on 
Fig. \ref{epsilon}). At time $t = t_1$ we can see things out to the 
radius of our past light-cone $R(t_1,t_0)$ which is vastly larger 
than the radius of the forward light-cone $R(t_2,t_1) \approx 
c/[(\epsilon_1 - 1) H_1 a_1]$ that anything can have travelled 
from the beginning of time. For example, the cosmic microwave 
radiation is uniform to one part in $10^5$, which is far better 
thermal equilibrium than the air of the room in which you are 
sitting. Without a phase of primordial inflation we are seeing
about 2200 different patches of the sky which have not even had 
time to exchange a single photon, much less achieve a high degree
of thermal equilibrium \cite{Woodard:2009ns}. Of course the problem
just gets worse the further back we look. At the time of big bang
nucleosynthesis we are seeing about $10^{15}$ causally disconnected
regions, which are nonetheless in rough thermal equilibrium
\cite{Woodard:2009ns}.

Without inflation the radius $R(t_2,t_1)$ of the forward light-cone
is almost independent of the beginning of time $t_2$. No matter how 
early we make $t_2$ it is not possible to increase $R(t_2,t_1)$ more 
than about $c/[(\epsilon_1 -1) H_1 a_1]$. Hence the high degree 
of uniformity we observe in the early universe would have to be a 
spectacularly unlikely accident. Primordial inflation solves the 
problem neatly by making the lower limit of the forward light-cone
dominate, $R(t_2,t_1) \approx c/[(1-\epsilon_2) H_2 a_2]$. We
can make the radius of the forward light-cone much larger than the 
radius of the past light-cone, so that causal processes would have
had plenty of time to achieve the high degree of equilibrium that 
is observed.

Before closing this subsection I want to return to the numerical
values quoted for $H_0$ and $H_i$ in relations 
(\ref{paramsnow}-\ref{paramsthen}). The loop counting parameter of 
quantum gravity can be expressed in terms of the square of the 
Planck time, $T^2_{\rm Pl} \equiv \hbar G/c^5 \approx 2.9 \times 
10^{-87}~{\rm sec}^2$. Quantum gravitational effects from a process 
whose characteristic frequency is $\omega$ are typically of order 
$\omega^2 T^2_{\rm Pl}$. For inflationary particle production the 
characteristic frequency is of course the Hubble parameter, so we
can easily compare the strengths of quantum gravitational effects
during the current phase of inflation and from the epoch of 
primordial inflation,
\begin{equation}
\frac{\hbar G H_0^2}{c^5} \approx 1.4 \times 10^{-122}
\qquad , \qquad \frac{\hbar G H_i^2}{c^5} \approx 9.4 \times 
10^{-11} \; . \label{loopcount}
\end{equation}
The minuscule first number is why we will never detect quantum 
gravitational effects from the current phase of inflation. Although 
the second number is tiny, it is not so small as to preclude 
detection, if only the signal can persist until the present day. In 
the next subsection I will explain how that can happen.

The loop counting parameter $\hbar G H^2/c^5$ is the quantum 
gravitational analog of the quantum electrodynamic fine structure
constant $\alpha \equiv e^2/4\pi\epsilon_0\hbar c \approx 7.3 
\times 10^{-3}$. Both parameters control the strength of perturbative
corrections. Recall that a result in quantum electrodynamics
--- for example, the invariant amplitude of Compton scattering --- 
typically consists of a lowest, tree order contribution of strength
$\alpha$, then each additional loop brings an extra factor of 
$\alpha$. In the same way, the lowest, tree order quantum gravity
effects from inflationary particle production have strength $\hbar G
H^2/c^5$, and each addition loop brings an extra factor of $\hbar G
H^2/c^5$. Because the quantum gravitational loop counting parameter
from primordial inflation is so much smaller than its quantum
electrodynamics cousin, we expect that quantum gravitational 
perturbation theory should be wonderfully accurate. In fact, all 
that can be resolved with current data is the tree order effect,
although I will argue in section \ref{21cm} that the one loop 
correction may eventually be resolved. Beyond that there is no hope.

\subsection{Inflationary Particle Production}\label{pert}

The phenomenon of polarization in a medium is covered in undergraduate
electrodynamics. The medium contains a vast number of bound charges. 
The application of an electric field makes positive charges move with the 
field and the negative changes move opposite. That charge separation 
polarizes the medium and tends to reduce the electric field strength.

One of the amazing predictions of quantum field theory is that virtual 
particles are continually emerging from the vacuum, existing for a brief
period, and then disappearing. How long these virtual particles can exist
is controlled by the energy-time uncertainty principle, which gives the 
minimum time $\Delta t$ needed to resolve and energy difference $\Delta E$,
\begin{equation}
\Delta t \Delta E \gtwid \hbar \; . \label{ET}
\end{equation}
If one imagines the emergence of a pair of positive and negatively charged 
particles of mass $m$ and wave vector $\pm \vec{k}$ then the energy went 
from zero to $E = 2 [m^2 c^4 + \hbar^2 c^2 k^2]^{\frac12}$. To {\it not} 
resolve a violation of energy conservation, the energy-time uncertainty 
principle requires the pair to disappear after a time $\Delta t$ given by,
\begin{equation}
\Delta t \sim \frac{\hbar}{\sqrt{m^2 c^4 + \hbar^2 c^2 k^2}} \; . 
\label{flatDt}
\end{equation}
The rest is an exercise is classical (that is, non-quantum) physics. If we
ignore the change in the particles' momentum then their positions obey,
\begin{equation}
\frac{d^2}{c dt^2} \Bigl( \sqrt{m^2 c^2 + \hbar^2 k^2} \, \Delta \vec{x}_{\pm}
\Bigr) = \pm e \vec{E} \quad \Longrightarrow \quad \Delta 
\vec{x}_{\pm}(\Delta t) = \frac{\pm \hbar^2 e \vec{E}}{2 c [m^2 c^2 + 
\hbar^2 k^2]^{\frac32}} \; .
\end{equation}
Hence the polarization induced by wave vector $\vec{k}$ is,
\begin{equation}
\vec{p} = +e \Delta \vec{x}_+(\Delta t) - e \Delta \vec{x}_-(\Delta t) =
\frac{\hbar^2 e^2 \vec{E}}{c [m^2 c^2 + \hbar^2 k^2]^{\frac32}} \; .
\end{equation}
The full vacuum polarization density comes from integrating $d^3k/(2\pi)^3$.

The simple analysis I have just sketched gives pretty nearly the prediction 
from one loop quantum electrodynamics, which is in quantitative agreement
with experiment. It allows us to understand two features of vacuum 
polarization which would be otherwise obscure:
\begin{itemize}
\item{That the largest effect derives from the lightest charged particles 
because they have the longest persistence times $\Delta t$ and therefore
induce the greatest polarization; and}
\item{That the electrodynamic interaction becomes stronger at short distances
because the longest wave length (hence smallest $k$) virtual particles could 
induce more polarization than is allowed by the travel time between two very
close sources.}
\end{itemize}

Cosmological expansion can strengthen quantum effects because it causes the
virtual particles which drive them to persist longer. This is easy to see 
from the geometry (\ref{FLRW}). Because spatial translation invariance is
unbroken, particles still have conserved wave numbers $\vec{k}$. However,
because the physical distance is the coordinate distance scaled by $a(t)$,
the physical energy of a particle with mass $m$ and wave number $k = 
2\pi/\lambda$ becomes time dependent,
\begin{equation}
E(t,\vec{k}) = \sqrt{m^2 c^4 + \frac{\hbar^2 c^2 k^2}{a^2(t)} } \; .
\end{equation}
Hence the relation for the persistence time $\Delta t$ of a virtual pair 
which emerges at time $t$ changes from (\ref{flatDt}) to,
\begin{equation}
\int_{t}^{t+\Delta t} \!\!\!\! dt' E(t',\vec{k}) \sim \hbar \; . 
\label{expDt}
\end{equation}
Massless particles persist the longest, just as they do in flat space. 
However, for inflation it is the lower limit of (\ref{expDt}) which 
dominates, so that even taking $\Delta t$ to infinity does not cause the
integral to grow past a certain point. One can see this from the de Sitter 
limit,
\begin{equation}
\int_{t}^{t+\Delta t} \!\!\!\! dt' \frac{\hbar c k}{a(t')} = 
\frac{\hbar c k}{H_i a(t)} \Bigl[ 1 - e^{-H_i \Delta t}\Bigr] \; . 
\label{inflDt}
\end{equation}
A particle with $c k < H(t) a(t)$ is said to be super-horizon, and we have
just shown that {\it any massless virtual particle which emerges from the 
vacuum with a super-horizon wave number during inflation will persist forever.}

It turns out that almost all massless particles possess a symmetry known as 
{\it conformal invariance} which suppresses the rate at which they emerge from
the vacuum. This keeps the density of virtual particles small, even though any 
that do emerge can persist forever. One can see the problem by specializing 
the Lagrangian of a massless, conformally coupled scalar $\psi(t,\vec{x})$ to 
the cosmological geometry (\ref{FLRW}),
\begin{equation}
\mathcal{L} = -\frac12 \partial_{\mu} \psi \partial_{\nu} \psi g^{\mu\nu} 
\sqrt{-g} -\frac{R}{12} \psi^2 \sqrt{-g} \longrightarrow \frac{a^3}{2} \Bigl[
\frac{\dot{\psi}^2}{c^2} - \frac{\partial_i \psi \partial_i \psi}{a^2} - 
\frac{(\dot{H} + 2 H^2) \psi^2}{c^2} \Bigr] .
\end{equation}
The equation for a canonically normalized, spatial plane wave of the form 
$\psi(t,\vec{x}) = v(t,k) e^{i\vec{k} \cdot \vec{x}}$ can be solved for a 
general scale factor $a(t)$,
\begin{equation}
\ddot{v} + 3 H \dot{v} + \Bigl[ \frac{c^2 k^2}{a^2} + \dot{H} + 2 H^2\Bigr]
v = 0 \; \Longrightarrow \; v(t,k) = \sqrt{\frac{\hbar}{2c k}}
\frac{\exp[ -i c k \! \int_{t_i}^t \! \frac{dt'}{a(t')}]}{a(t)} \; . 
\label{vmode}
\end{equation}
The factor of $1/a(t)$ in (\ref{vmode}) suppresses the emergence rate,
even though destructive interference from the phase dies off, just as the 
energy-time uncertainty principle (\ref{inflDt}) predicts. The stress-energy 
contributed by this field is,
\begin{equation}
T_{\mu\nu} = \Bigl[\delta^{\rho}_{\mu} \delta^{\sigma}_{\nu} \!-\! \frac12 
g_{\mu\nu} g^{\rho\sigma} \Bigr] \partial_{\rho} \psi \partial_{\sigma} \psi
+ \frac16 \Bigl[ R_{\mu\nu} \!-\! \frac12 g_{\mu\nu} R \!+\! g_{\mu\nu} 
\square \!-\! D_{\mu} D_{\nu} \Bigr] \psi^2 \; ,
\end{equation}
where $D_{\mu}$ is the covariant derivative and $\square$ is the covariant 
d'Alembertian. We can get the 0-point energy of a single wave vector $\vec{k}$
by specializing $T_{00}$ to the cosmological geometry (\ref{FLRW}) and multipling 
by a factor of $a^3(t)$,
\begin{equation}
\mathcal{E}(t,\vec{k}) = \frac{a^3}{2} \Bigl[ \vert \dot{v}\vert^2 + \Bigl( 
\frac{c^2 k^2}{a^2} \!+\! H^2\Bigr) \vert v\vert^2 + H \Bigl( v \dot{v}^* \!+\! 
\dot{v} v^*\Bigr) \Bigr] = \frac{\hbar c k}{2 a(t)} \; . \label{confE}
\end{equation}
This is just the usual $\frac12 \hbar \omega$ term which is not strengthened 
but rather weakened by the cosmological expansion.

Only gravitons and massless, minimally coupled scalars are both massless and 
not conformally invariant so that they can engender significant quantum effects 
during inflation. Because they obey the same mode equation 
\cite{Lifshitz:1945du,Grishchuk:1974ny} it will suffice to specialize the 
scalar Lagrangian to the cosmological geometry (\ref{FLRW}),
\begin{equation}
\mathcal{L} = -\frac12 \partial_{\mu} \phi \partial_{\nu} \phi g^{\mu\nu} 
\sqrt{-g} \longrightarrow \frac12 a^3 \Bigl[\frac{\dot{\phi}^2}{c^2} - 
\frac1{a^2} \partial_i \phi \partial_i \phi \Bigr] \; .
\end{equation} 
The equation for a canonically normalized, spatial plane wave of the form 
$\phi(t,\vec{x}) = u(t,k) e^{i\vec{k} \cdot \vec{x}}$ is simpler than that
of its conformally coupled cousin (\ref{vmode}) but more difficult to solve,
so I will specialize the solution to de Sitter,
\begin{equation}
\ddot{u} + 3 H \dot{u} + \frac{c^2 k^2}{a^2} \, u = 0 \; \Longrightarrow \; 
u(t,k) = \sqrt{\frac{\hbar}{2c k}} \Bigl[1 + \frac{i H_i a(t)}{c k}\Bigr] 
\frac{\exp[ -ic k \! \int_{t_i}^t \! \frac{dt'}{a(t')}]}{a(t)} \; . 
\label{umode}
\end{equation}
The minimally coupled mode function $u(t,k)$ has the same phase factor
as the conformal mode function (\ref{vmode}), and they both fall off
like $1/a(t)$ in the far sub-horizon regime of $ck \gg H_i a(t)$. 
However, they disagree strongly in the super-horizon regime during which
$v(t,k)$ continues to fall off whereas $u(t,k)$ approaches a phase times
$H_i \sqrt{\hbar/2 c^3 k^3}$. One can see from the equation on the left 
of (\ref{umode}) that $u(t,k)$ approaches a constant for any inflating
geometry. 

The 0-point energy in wave vector $\vec{k}$ is,
\begin{equation}
\mathcal{E}(t,\vec{k}) = \frac12 a^3 \Bigl[ \vert \dot{u}\vert^2 +
\frac{c^2 k^2}{a^2} \vert u\vert^2 \Bigr] = \frac{\hbar ck}{a(t)} 
\Bigl[ \frac12 + \Bigl( \frac{H_i a(t)}{2 ck} \Bigr)^2 \Bigr] 
\; . \label{E/k}
\end{equation}
Because each wave vector is an independent harmonic oscillator with 
mass proportional to $a^3(t)$ and frequency $k/a(t)$ we can read off
the occupation number from expression (\ref{E/k}),
\begin{equation}
N(t,\vec{k}) = \Bigl[ \frac{H_i a(t)}{2 ck} \Bigr]^2 \; . \label{N/k}
\end{equation}
As one might expect, this number is small in the sub-horizon regime.
It becomes of order one at the time $t_k$ of horizon crossing, $c k =
H(t_k) a(t_k)$, and $N(t,\vec{k})$ grows explosively afterwards.
This is crucial because it means that {\it inflationary particle 
production is an infrared effect}. That means we can study it 
reliably using quantum general relativity, even though that theory
is not perturbatively renormalizable. 

The final point I wish to make is that the mode function $u(t,k)$
becomes constant after first horizon crossing. For de Sitter this
constant is calculable,
\begin{equation}
u(t,k) \Biggl\vert_{\rm dS} \longrightarrow i H_i 
\sqrt{\frac{\hbar}{2 c^3 k^3}} \exp\Bigl[ -\frac{i ck}{H_i a_i}\Bigr] 
\; . \label{freeze}
\end{equation}
However, one can see from the mode equation on the left hand side of
(\ref{umode}) that the approach to a constant happens for any 
inflating geometry. Recall from equation (\ref{horizon}) that the
inverse horizon length $c^{-1} H(t) a(t)$ grows during inflation and 
decreases during the later phase of deceleration which encompasses
so much of the cosmological history depicted in Figure \ref{epsilon}. 
Hence we can give the following rough summary of the ``life cycle of
a mode'' of wave number $k$:
\begin{itemize} 
\item{At the onset of primordial inflation the mode has $ck \gg H(t) 
a(t)$ so the mode function oscillates and falls off like $1/a(t)$;}
\item{If inflation lasts long enough the mode will eventually 
experience first horizon crossing $ck = H(t_k) a(t_k)$, after which 
mode function becomes approximately constant; and}
\item{During the phase of deceleration which follows primordial 
inflation, modes which experienced first horizon crossing near the 
end of inflation re-enter the horizon $ck = H(T_k) a(T_k)$, after 
which they begin participating in dynamical processes with amplitude 
larger by a factor of $a(T_k)/a(t_k)$ than they would have had 
without first horizon crossing.}
\end{itemize}
This is how quantum gravitational effects from the epoch of 
primordial inflation become fossilized so that they can be 
detected now.

\section{Tree Order Power Spectra}\label{tree}

Although the evidence for primordial inflation is overwhelming, there
is not yet any compelling mechanism for causing it. The simplest 
class of successful models is based on general relativity plus a 
scalar inflaton $\varphi(x)$ whose potential $V(\varphi)$ is
regarded as a free function \cite{Linde:1983gd},
\begin{equation}
\mathcal{L} = \Bigl[ \frac{c^4 \mathbf{R}}{16 \pi G} - \frac12 
\varphi_{,\mu} \varphi_{,\nu} \mathbf{g}^{\mu\nu} - V(\varphi)\Bigr]
\sqrt{-\mathbf{g}} \; . \label{GR+phi}
\end{equation}
Here $\mathbf{g}_{\mu\nu}(x)$ is the $D$-dimensional, spacelike 
metric with Ricci curvature $\mathbf{R}$. (I will work in $D$ 
spacetime dimensions to facilitate the use of dimensional 
regularization, even though the $D=4$ limit must eventually be 
taken for physical results.) The purpose of this section is to 
show how this simple model can not only drive primordial inflation 
but also the quantum gravitational fluctuations whose imprint on 
the cosmic background radiation has been imaged with stunning 
accuracy \cite{Hinshaw:2012aka,Hou:2012xq,Sievers:2013ica,
Ade:2013zuv}.

I first demonstrate that the potential $V(\varphi)$ can be chosen to 
support the cosmological geometry (\ref{FLRW}) with any scale factor 
$a(t)$ for which the Hubble parameter is monotonically decreasing. 
I also comment on the many problems of plausibility which seem to 
point to the need for a better model. I then decompose perturbations 
about the background (\ref{FLRW}) into a scalar $\zeta(x)$ and a 
transverse-traceless tensor $h_{ij}(x)$. Owing to the particle 
production mechanism adumbrated in section \ref{pert}, certain modes 
of $\zeta$ and $h_{ij}$ become hugely excited during primordial 
inflation, and then freeze in so that they can survive to much later 
times. The strength of this effect is quantified by the {\it 
primordial scalar and tensor power spectra}, which I define and
compute at tree order, along with associated observables. I then
discuss the controversy which has arisen concerning an alternate 
definition of the tree order power spectra. The section closes with 
an explanation of why the tree order power spectra are the first 
quantum gravitational effects ever to have been resolved. 

I will adopt the notation employed in recent studies by Maldacena 
\cite{Maldacena:2002vr} and by Weinberg \cite{Weinberg:2005vy}, 
however, the original work for tensors was done in 1979 by 
Starobinsky \cite{Starobinsky:1979ty}, and for scalars in 1981 by 
Mukhanov and Chibisov \cite{Mukhanov:1981xt}. Important subsequent 
work was done over the course of the next several years by Hawking 
\cite{Hawking:1982cz}, by Guth and Pi \cite{Guth:1982ec}, by 
Starobinsky \cite{Starobinsky:1982ee}, by Bardeen, Steinhardt and 
Turner \cite{Bardeen:1983qw}, and by Mukhanov \cite{Mukhanov:1985rz}. 
Some classic review articles on the subject are 
\cite{Mukhanov:1990me,Liddle:1993fq,Lidsey:1995np}.

\subsection{The Background for Single-Scalar Inflation}\label{single}

There is no question that a minimally coupled scalar potential model of
the form (\ref{GR+phi}) can support inflation because there is a 
constructive procedure for finding the potential $V(\varphi)$ given the
expansion history $a(t)$ \cite{Tsamis:1997rk,Saini:1999ba,
Capozziello:2005mj,Guo:2006ab}. For the geometry (\ref{FLRW}) the scalar
depends just on time $\varphi_0(t)$ and only two of Einstein's equations 
are nontrivial,
\begin{eqnarray}
\frac12 (D\!-\!2) (D\!-\!1) H^2 & = & \frac{8\pi G}{c^2} \Bigl[ 
\frac{\dot{\varphi}^2_0}{c^2} + V(\varphi_0)\Bigr] \; , \label{E1} \\
-(D\!-\!2) \dot{H} - \frac12 (D\!-\!2) (D\!-\!1) H^2 & = & 
\frac{8\pi G}{c^2} \Bigl[ \frac{\dot{\varphi}^2_0}{c^2} - 
V(\varphi_0)\Bigr] \; . \label{E2}
\end{eqnarray}
By adding (\ref{E1}) and (\ref{E2}) one obtains the relation,
\begin{equation} \nonumber
(D\!-\!2) \dot{H} = \frac{8 \pi G}{c^4} \, \dot{\varphi}^2_0 \; .
\label{back1}
\end{equation}
Hence one can reconstruct the scalar's evolution provided the Hubble 
parameter is monotonically decreasing, and that relation can be
inverted (numerically if need be) to solve for time as a function of
\begin{equation}
\varphi_0(t) = \varphi_0(t_2) \pm \int_{t_2}^{t} \!\! dt' 
\sqrt{\frac{-(D\!-\!2) c^4 \dot{H}(t')}{8\pi G}} 
\qquad \Longleftrightarrow \qquad t = T(\varphi_0) \; . \label{back2}
\end{equation}
One then determines the potential by subtracting (\ref{E2}) from 
(\ref{E1}) and evaluating the resulting function of time at $t = 
T(\varphi_0)$,
\begin{equation}
V(\varphi_0) = \frac{(D\!-\!2) c^2}{16 \pi G} \Biggl[ 
\dot{H}\Bigl(T(\varphi_0)\Bigr) + 3 H^2\Bigl(T(\varphi_0)\Bigr) 
\Biggr] \; .
\end{equation}

Just because scalar potential models (\ref{GR+phi}) can be adjusted 
to work does not mean they are particularly plausible. They suffer 
from six sorts of sometimes contradictory fine-tuning problems:
\begin{enumerate}
\item{{\it Initial Conditions} --- Inflation must begin with the 
inflaton approximately homogeneous, and potential-dominated, over 
more than a Hubble volume \cite{Vachaspati:1998dy};}
\item{{\it Duration} --- The inflaton potential must be flat enough 
to make inflation last for at least 50 e-foldings 
\cite{Linde:1981mu,Albrecht:1982wi};}
\item{{\it Scalar Perturbations} --- Getting the right magnitude 
for the scalar power spectrum requires $\hbar G^3/c^{11} \times
V^3/{V'}^2 \sim 10^{-11}$ \cite{Mukhanov:1990me};}
\item{{\it Tensor Perturbations} --- Getting the right magnitude
for the tensor power spectrum requires $c^4/G \times (V'/V)^2
\sim 1$ \cite{Mukhanov:1990me};}
\item{{\it Reheating} --- The inflaton must couple to ordinary 
matter (its gravitational couplings do not suffice) so that its 
post-inflationary kinetic energy produces a hot, radiation 
dominated universe \cite{Allahverdi:2010xz};} 
\item{{\it Cosmological Constant} --- The minimum of the scalar 
potential must have the right value $\hbar G^2 V_{\rm min}/c^7 
\approx 10^{-123}$ to contribute the minuscule vacuum energy we 
observe today \cite{Riess:1998cb,Perlmutter:1998np,Wang:2006ts,
Alam:2006kj}.}
\end{enumerate}
Note that adding the matter couplings required to produce 
reheating puts 2-4 at risk because matter loop effects induce 
Coleman-Weinberg corrections to the inflaton effective potential. 
Nor does fundamental theory provide any explanation for why the
cosmological constant is so small \cite{Weinberg:1988cp,
Carroll:2000fy}. The degree of fine-tuning needed to enforce
these six conditions strains credulity, and disturbs even those 
who devised the early models \cite{Ijjas:2013vea,Guth:2013sya,
Linde:2014nna,Ijjas:2014nta}.

Opinions differ, but I feel it is a mistake to make too much of the
defects of single-scalar inflation. The evidence for an early phase 
of accelerated expansion is overwhelming and really incontrovertible, 
irrespective of what caused it. Further, all that we know about low 
energy effective field theory confirms that the general relativity 
portion of Lagrangian (\ref{GR+phi}) must be valid, even at the 
scales of primordial inflation. That suffices to establish the 
quantum gravitational character of primordial perturbations, even 
without a compelling model for what caused inflation. So I will go 
forward with the analysis on the basis of the single-scalar model 
(\ref{GR+phi}), firm in the belief that whatever eventually 
supplants it must exhibit many of the same features.

\subsection{Gauge-Fixed, Constrained Action}\label{gauge}

We decompose $\mathbf{g}_{\mu\nu}$ into lapse, shift and spatial metric
according to Arnowitt, Deser and Misner (ADM) \cite{Deser:1959zza,
Arnowitt:1960es,Arnowitt:1962hi},
\begin{equation}
\mathbf{g}_{\mu\nu} dx^{\mu} dx^{\nu} = -N^2 c^2 dt^2 + g_{ij} (dx^i \!-\!
N^i c dt) (dx^j \!-\! N^j c dt) \; .
\end{equation}
ADM long ago showed that the Lagrangian has a very simple dependence upon 
the lapse \cite{Deser:1959zza,Arnowitt:1960es,Arnowitt:1962hi},
\begin{equation}
\mathcal{L} = \Bigl({\rm Surface\ Terms}\Bigr) - 
\frac{c^4 \sqrt{g}}{16 \pi G} \Bigl[N \cdot A + \frac{B}{N}\Bigr] \; . 
\label{simple}
\end{equation}
The quantity $A$ is a potential energy,
\begin{equation}
A = - R + \frac{16 \pi G}{c^4} \Bigl[V(\varphi) + \frac12 \varphi_{,i} 
\varphi_{,j} g^{ij} \Bigr] \; , \label{A}
\end{equation}
where $R$ is the $(D-1)$-dimensional Ricci scalar formed from $g_{ij}$.
The quantity $B$ in (\ref{simple}) is a sort of kinetic energy,
\begin{equation}
B = (E^i_{~i})^2 - E^{ij} E_{ij} - \frac{8\pi G}{c^4} \Bigl(
\frac{\dot{\varphi}}{c} - \varphi_{,i} N^i\Bigr)^2  \; , \label{B}
\end{equation}
where $E_{ij}/N$ is the extrinsic curvature,
\begin{equation}
E_{ij} \equiv \frac12 \Bigl[ N_{i ; j} + N_{j ; i} - c^{-1} \dot{g}_{ij}
\Bigr] \; ,
\end{equation}
and a semi-colon denotes spatial covariant differentiation using the 
connection compatible with $g_{ij}$. 

ADM fix the gauge by specifying $N(t,\vec{x})$ and $N^i(t,\vec{x})$,
however, Maldacena \cite{Maldacena:2002vr} and Weinberg 
\cite{Weinberg:2005vy} instead impose the conditions,
\begin{eqnarray}
G_0(t,\vec{x}) & \equiv & \varphi(t,\vec{x}) - \varphi_0(t) = 0 \; , 
\label{G0} \\
G_i(t,\vec{x}) & \equiv & \partial_j h_{ij}(t,\vec{x}) = 0 \; . \label{Gi}
\end{eqnarray}
The transverse-traceless graviton field $h_{ij}(t,\vec{x})$ is defined by 
decomposing the spatial metric into a conformal part and a unimodular part 
$\widetilde{g}_{ij}$,
\begin{equation}
g_{ij} = a^2(t) e^{2 \zeta(t,\vec{x})} \widetilde{g}_{ij}(t,\vec{x}) 
\; \Longrightarrow \; \sqrt{g} = a^{D-1} e^{(D-1) \zeta} \; . \label{conf}
\end{equation}
The unimodular part is obtained by exponentiating the transverse-traceless
graviton field $h_{ij}(t,\vec{x})$,
\begin{equation}
\widetilde{g}_{ij} \equiv \Bigl[e^h\Bigr]_{ij} = \delta_{ij} + h_{ij} + 
\frac12 h_{ik} h_{jk} + O(h^3) \qquad , \qquad h_{ii} = 0 \; .
\end{equation}
The Faddeev-Popov determinant associated with (\ref{G0}-\ref{Gi}) depends 
only on $h_{ij}$, and becomes singular for $\epsilon = 0$.

Of course no gauge can eliminate the physical scalar degree of freedom 
which is evident in (\ref{GR+phi}). With condition (\ref{G0}) the inflaton 
degree of freedom resides in $\zeta(t,\vec{x})$ and linearized gravitons 
are carried by $h_{ij}(t,\vec{x})$. In this gauge the lapse and shift are
constrained variables which mediate important interactions between the
dynamical fields but contribute no independent degrees of freedom. Varying 
(\ref{simple}) with respect to $N$ produces an algebraic equation for $N$,
\begin{equation}
A - \frac{B}{N^2} = 0 \qquad \Longrightarrow \qquad 
N = \sqrt{ \frac{B}{A}} \;
\end{equation}
This gives the constrained Lagrangian a ``virial'' form \cite{Kahya:2010xh},
\begin{equation}
\mathcal{L}_{\rm const} = \Bigl({\rm Surface\ Terms}\Bigr) -
\frac{c^4 \sqrt{g}}{8 \pi G} \, \sqrt{A B} \; . \label{virial}
\end{equation}
From relations (\ref{back1}-\ref{back2}) one can see that the background 
values of the potential and kinetic terms are equal, $A_0 = B_0 = (D \!-\!2) 
c^{-2} [\dot{H} + (D \!-\!1) H^2]$. Hence the background value of the lapse 
is unity.

There is unfortunately no nonperturbative solution for the shift 
$N^i(t,\vec{x})$ in terms of $\zeta$ and $h_{ij}$, so its constraint 
equation must be solved perturbatively. One first employs (\ref{conf}) to 
exhibit how the potential (\ref{A}) depends on $\zeta$ and $h_{ij}$,
\begin{equation}
A = A_0 - R \equiv A_0 (1 + \alpha) \; . \label{Aexp}
\end{equation}
Here the spatial Ricci scalar is,
\begin{equation}
R = \frac{e^{-2\zeta}}{a^2} \Biggl[ \widetilde{R} - 2 (D \!-\!2) 
\widetilde{\nabla}^2 \zeta - (D\!-\!2) (D\!-\!3) \zeta^{,k} 
\widetilde{g}^{k\ell} \zeta_{,\ell} \Biggr] \; ,
\end{equation}
where $\widetilde{R} = O(h^2)$ is the Ricci scalar formed from 
$\widetilde{g}_{ij}$ and $\widetilde{\nabla}^2 \equiv \partial_i 
\widetilde{g}^{ij} \partial_j$ is the conformal scalar Laplacian. The
full scalar Laplacian is,
\begin{equation}
\nabla^2 \equiv \frac1{\sqrt{g}} \partial_i \Bigl[\sqrt{g} \,
g^{ij} \partial_j\Bigr] \; .
\end{equation}
At this stage one can recognize that the dimensionless 3-curvature 
perturbation $\mathcal{R}$ is just $\zeta$, in $D=4$ dimensions and to 
linearized order \cite{Liddle:1993fq},
\begin{equation}
\mathcal{R}(t,\vec{x}) \equiv -\frac{1}{4 \nabla^2} \, R 
= \Bigl(\frac{D\!-\!2}{2}\Bigr) \zeta(t,\vec{x}) + O\Bigl(\zeta^2,\zeta h, 
h^2\Bigr) \; . \label{Rdef}
\end{equation}
The kinetic energy (\ref{B}) can be expressed as,
\begin{eqnarray}
\lefteqn{B \equiv A_0 (1 \!+\! \beta) = A_0 + 2 (D\!-\!2) c^{-1} H 
\Bigl[ (D\!-\!1) (c^{-1} \dot{\zeta} \!-\! \zeta_{,k} \widetilde{N}^k) 
\!-\! \widetilde{N}^k_{~ ,k} \Bigr] -\widetilde{E}^{k\ell} 
\widetilde{E}_{k\ell} } \nonumber \\
& & \hspace{.5cm} + (D\!-\!2) (D\!-\!1) \Bigl(\frac{\dot{\zeta}}{c} 
\!-\! \zeta_{,k} \widetilde{N}^k \Bigr)^2 \!\!\!- 2 (D\!-\! 2) \Bigl( 
\frac{\dot{\zeta}}{c} \!-\! \zeta_{,k} \widetilde{N}^k\Bigr) 
\widetilde{N}^k_{~ ,k} + (\widetilde{N}^{k}_{~ ,k})^2 \; . \qquad 
\label{Bexp}
\end{eqnarray}
Here we define $\widetilde{N}^i \equiv N^i$, $\widetilde{N}_i \equiv
\widetilde{g}_{ij} \widetilde{N}^j$ and $\widetilde{E}_{ij} \equiv 
\frac12 [ \widetilde{N}_{i ; j} + \widetilde{N}_{j ; i} - c^{-1}
\dot{h}_{ij} ]$.

The next step is to expand the volume part of the constrained 
Lagrangian in powers of $\alpha$ and $\beta$,
\begin{eqnarray}
- \frac{c^4 \sqrt{g}}{8 \pi G} \, \sqrt{A B} & = & -\frac{c^4 a^{D-1} 
e^{(D-1) \zeta}}{8\pi G} \, A_0 \sqrt{ (1 \!+\! \alpha) (1 + \beta)} 
\; , \\
& = & -\frac{c^4 a^{D-1} e^{(D-1)\zeta}}{8\pi G} \, A_0 
\Biggl\{1 \!+\! \frac{(\alpha \!+\! \beta)}{2} \!-\! \frac{(\alpha \!-\! 
\beta)^2}{8} \!+\! \dots \Biggr\} . \qquad \label{expansion}
\end{eqnarray}
As Weinberg noted, the terms involving no derivatives of $\zeta$ or 
$h_{ij}$ sum up to a total derivative \cite{Weinberg:2005vy},
\begin{equation}
a^{D-1} A_0 = \frac{\partial}{\partial t} \Bigl[ (D\!-\! 2) H a^{D-1}
\Bigr] \; .
\end{equation}
Another important fact is that quadratic mixing between $\widetilde{N}^i$ 
and $\zeta$ can be eliminated with the covariant field redefinition
\cite{Kahya:2010xh},
\begin{equation}
\widetilde{S}^k \equiv \widetilde{N}^k + \widetilde{g}^{k\ell} 
\partial_{\ell} \frac1{\widetilde{\nabla}^2} \Bigl[ 
\frac{ce^{-2\zeta}}{H a^2} \widetilde{\nabla}^2 \zeta - \epsilon 
(c^{-1} \dot{\zeta} \!-\! \zeta_{,i} \widetilde{N}^i)\Bigr] \; .
\end{equation}
After much work the quadratic Lagrangians emerge,
\begin{eqnarray}
\mathcal{L}_{S^2} &\! =\! & \frac{c^4 a^{D-1}}{32 \pi G} \Biggl\{ 
\partial_{\ell} \widetilde{S}^k \partial_{\ell} \widetilde{S}^k \!\!+\! 
\Bigl( \frac{D \!-\! 3 \!+\! \epsilon}{D \!-\! 1 \!-\! \epsilon}\Bigr) 
\partial_{\ell} \widetilde{S}^k \partial_k \widetilde{S}^{\ell} \Biggr\} 
\; , \qquad \label{freeN} \\
\mathcal{L}_{\zeta^2} &\! = \!& \frac{(D \!-\!2) c^4 \epsilon \, 
a^{D-1}}{16\pi G} \Biggl\{ \frac{\dot{\zeta}^2}{c^2} - \frac{\partial_k 
\zeta \partial_k \zeta}{a^2} \Biggr\} , \label{freeZ} \\
\mathcal{L}_{h^2} &\! = \!& \frac{c^4 a^{D-1}}{64\pi G} \Biggl\{ 
\frac{\dot{h}_{ij} \dot{h}_{ij}}{c^2} - \frac{\partial_k h_{ij} 
\partial_k h_{ij}}{a^2} \Bigr\} . \label{freeh}
\end{eqnarray}
These results suffice for the analysis of this section. To consider loop
corrections (or non-Gaussiantity) one must solve the constraint equation 
for $\widetilde{S}^i$,
\begin{equation}
\partial_j \partial_j \widetilde{S}^i + \Bigl( \frac{D \!-\! 3 \!+\! 
\epsilon}{D \!-\! 1 \!-\! \epsilon}\Bigr) \partial_i \partial_j 
\widetilde{S}^j = O\Bigl(\zeta^2,\zeta h,h^2,S \zeta\Bigr) \; .
\label{constraint}
\end{equation}
That is a tedious business which has only been carried out to a few
orders. I will review what is known in section \ref{howto}.

\subsection{Tree Order Power Spectra}\label{subtree}

As we will see in section \ref{nonlinear}, there is not yet general 
agreement on how to define the primordial power spectra when loop
corrections are included \cite{Miao:2012xc,Miao:2013oko}. At tree 
order we can dispense with dimensional regularization, and also 
forget about the distinction between $\zeta$ and the dimensionless 
3-curvature perturbation (\ref{Rdef}). The following definitions
suffice:
\begin{eqnarray}
\Delta^2_{\mathcal{R}}(k,t) & \equiv & \frac{k^3}{2 \pi^2} \int 
\!\! d^3x \, e^{-i \vec{k} \cdot \vec{x}} \Bigl\langle \Omega 
\Bigl\vert \zeta(t,\vec{x}) \zeta(t,\vec{0}) \Bigr\vert \Omega 
\Bigr\rangle \; , \label{Deltaz} \\
\Delta^2_{h}(k,t) & \equiv & \frac{k^3}{2 \pi^2} \int \!\! d^3x \,
e^{-i \vec{k} \cdot \vec{x}} \Bigl\langle \Omega \Bigl\vert 
h_{ij}(t,\vec{x}) h_{ij}(t,\vec{0}) \Bigr\vert \Omega \Bigr\rangle 
\; . \label{Deltah}
\end{eqnarray}
Although it is useful to retain the time dependence in expressions
(\ref{Deltaz}-\ref{Deltah}), the actual predictions of primordial
inflation are obtained by evaluating the time-dependent power
spectra safely between the first and second horizon crossing times 
$t_k$ and $T_k$ described in section \ref{pert},
\begin{equation}
\Delta^2_{\mathcal{R}}(k) \equiv \Delta^2_{\mathcal{R}}(k,t) 
\Bigl\vert_{t_k \ll t \ll T_k} \qquad , \qquad \Delta^2_h(k) 
\equiv \Delta^2_h(k,t) \Bigl\vert_{t_k \ll t \ll T_k} \; .
\label{exacttree}
\end{equation}

The state $\vert \Omega \rangle$ in expressions 
(\ref{Deltaz}-\ref{Deltah}) is annihilated by $\alpha(\vec{k})$
and $\beta(\vec{k},\lambda)$ in the free field expansions of
$\zeta$ and $h_{ij}$,
\begin{eqnarray}
\zeta(t,\vec{x}) & = & \int \!\! \frac{d^3k}{(2\pi)^3} \Biggl\{ 
v(t,k) e^{i\vec{k} \cdot \vec{x}} \alpha(\vec{k}) 
+ v^*(t,k) e^{-i \vec{k} \cdot \vec{x}} \alpha^{\dagger}(\vec{k})
\Biggr\} \; , \qquad \label{freezeta} \\
h_{ij}(t,\vec{x}) & = & \int \!\! \frac{d^3k}{(2\pi)^3} 
\sum_{\lambda = \pm} \Biggl\{ u(t,k) e^{i\vec{k} \cdot \vec{x}} 
\epsilon_{ij}(\vec{k},\lambda) \beta(\vec{k},\lambda) 
+ {\rm Conjugate} \Biggr\} \; . \qquad \label{freegrav} 
\end{eqnarray}
The polarization tensors $\epsilon_{ij}(\vec{k},\lambda)$ are the
same as those of flat space. If we adopt the usual normalizations 
for the creation and annihilation operators,
\begin{equation}
\Bigl[ \alpha(\vec{k}) , \alpha^{\dagger}(\vec{k}') \Bigr] = 
(2\pi)^3 \delta^3( \vec{k} \!-\! \vec{k}') \quad , \quad
\Bigl[ \beta(\vec{k},\lambda) , \alpha^{\dagger}(\vec{k}',\lambda') 
\Bigr] = \delta_{\lambda \lambda'} (2\pi)^3 \delta^3( \vec{k} \!-\! 
\vec{k}') \; . \label{CCRs}
\end{equation}
then canonical quantization of the free Lagrangians 
(\ref{freeZ}-\ref{freeh}) implies that the mode functions obey,
\begin{eqnarray}
\ddot{v} + \Bigl[3 H \!+\! \frac{\dot{\epsilon}}{\epsilon} \Bigr]
\dot{v} + \frac{c^2 k^2}{a^2} v = 0 & , & v \dot{v}^* - \dot{v} v^*
= \frac{i 4 \pi \hbar G}{c^2 \epsilon a^3} \; , \label{vdef} \\
\ddot{u} + 3 H \dot{u} + \frac{c^2 k^2}{a^2} u = 0 & , & 
u \dot{u}^* - \dot{u} u^* = \frac{i 32 \pi \hbar G}{c^2 a^3} \; . 
\label{u2def}
\end{eqnarray}
It has long been known that the graviton mode function $u(t,k)$ obeys 
the same equation (\ref{umode}) as that of a massless, minimally coupled 
scalar \cite{Lifshitz:1945du,Grishchuk:1974ny}. Only their normalizations
differ by the square root of $32\pi G/c^2$.
  
By substituting the free field expansions (\ref{freezeta}-\ref{freegrav})
into the definitions (\ref{Deltaz}-\ref{Deltah}) of the power spectra,
and then making use of the canonical commutation relations (\ref{CCRs}), 
one can express the tree order power spectra in terms of scalar and 
tensor mode functions,
\begin{eqnarray}
\Delta^2_{\mathcal{R}}(k,t) & = & \frac{k^3 \vert v(,kt)\vert^2}{2 
\pi^2} \; , \label{Deltav} \\
\Delta^2_h(k,t) & = & \frac{k^3 \vert u(t,k)\vert^2}{2 \pi^2}
\sum_{\lambda = \pm} \epsilon_{ij} \epsilon_{ij}^* =
\frac{k^3 \vert u(k,t)\vert^2}{\pi^2} \; . \label{Deltau}
\end{eqnarray}
One of the frustrating things about primordial inflation is that we
don't know what $a(t)$ is so we need results which are valid for
any reasonable expansion history. This means that even tree order 
expressions such as (\ref{Deltav}-\ref{Deltau}) can only be evaluated 
approximately because there are no simple expressions for the mode 
functions for a general scale factor $a(t)$ \cite{Tsamis:2002qk,
Tsamis:2003zs,Tsamis:2003px}.

One common approximation is setting $\epsilon(t)$ to a constant, 
the reliability of which can be gauged by studying the region (at 
$N \approx -40$) of Figure 1 at which currently observable 
perturbations freeze in. (The necessity of nonconstant $\epsilon(t)$
later is not relevant for the validity of assuming constant 
$\epsilon(t)$ to estimate the amplitude at freeze-in.) For constant 
$\epsilon(t)$ both mode functions are proportional to a Hankel 
function of the first kind,
\begin{eqnarray}
\lefteqn{ \dot{\epsilon} = 0 \qquad \Longrightarrow \qquad v(t,k) = 
\frac{u(t,k)}{\sqrt{8\epsilon}} \; , } \label{constv} \\
& & \hspace{-.5cm} u(t,k) = \sqrt{\frac{8 \pi^2 \hbar G}{(1 \!-\! 
\epsilon) c^2 H(t) a^3(t)}} \, H^{(1)}_{\nu}\Bigl( \frac{c k}{(1 
\!-\! \epsilon) H(t) a(t)}\Bigr) \; , \; \nu = \frac12 \Bigl( 
\frac{3 \!-\! \epsilon}{1 \!-\! \epsilon}\Bigr) \; . \qquad 
\label{constu}
\end{eqnarray}
Between first and second horizon crossing ($t_k \ll t \ll T_k$) we
can take the small argument limit of the Hankel function,
\begin{eqnarray}
\lefteqn{ \dot{\epsilon}(t) = 0 \quad , \quad t_k \ll t \ll T_k} 
\nonumber \\
& \Longrightarrow & u(t,k) \approx \sqrt{\frac{8 \pi^2 \hbar G}{
(1 \!-\! \epsilon) c^2 H(t) a^3(t)}} \times 
\frac{-i \Gamma(\nu)}{\pi} \Bigl[ \frac{ 2 (1 \!-\! \epsilon) H(t)
a(t)}{c k}\Bigr]^{\nu} \; , \qquad \label{smallk} \\
& & \hspace{1cm} = \sqrt{\frac{16 \pi^2 \hbar G}{c^5 k^3}} \times
\frac{-i \Gamma(\nu)}{\pi} \Bigl[ \frac{2 (1 \!-\! \epsilon) H(t)
a^{\epsilon}(t)}{(c k)^{\epsilon}} \Bigr]^{\frac1{1 - \epsilon}} 
\; . \qquad
\end{eqnarray}
Constant $\epsilon(t)$ also implies $H(t) a^{\epsilon}(t)$ is a
constant, which we may as well evaluate at the time of first
horizon crossing, $H(t) a^{\epsilon}(t) = H^{1-\epsilon}(t_k)
(ck)^{\epsilon}$. With the doubling formula ($\Gamma(2x) = 
2^{2x-1}/\sqrt{\pi} \times \Gamma(x) \Gamma(x+\frac12)$) we at 
length obtain,
\begin{eqnarray}
\lefteqn{ \dot{\epsilon}(t) = 0 \quad , \quad t_k \ll t \ll T_k} 
\nonumber \\
& \Longrightarrow & u(t,k) \approx -i \Bigl[ \frac{1 \!-\! 
\epsilon}{2^{\epsilon}} \Bigr]^{\frac1{1-\epsilon}} 
\frac{\Gamma(\frac2{1-\epsilon})}{\Gamma(\frac1{1-\epsilon})}
\times \sqrt{\frac{16 \pi \hbar G H^2(t_k)}{c^5 k^3}} \; . 
\qquad \label{finalconst}
\end{eqnarray} 
The factor multiplying the square root has nearly unit modulus
for small $\epsilon$ --- and the BICEP2 result is $\epsilon_i =
0.013$ \cite{Ade:2014xna}, while previous data sets give the
even smaller bound of $\epsilon_i < 0.007$ at $95\%$ confidence
\cite{Hinshaw:2012aka,Hou:2012xq,Sievers:2013ica}. Hence it 
should be reliable to drop this factor, resulting in the 
approximate forms,
\begin{equation}
\Delta^2_{\mathcal{R}}(k) \approx 
\frac{ \hbar G H^2(t_k)}{\pi c^5 \epsilon(t_k)} \qquad , \qquad 
\Delta^2_h(k) \approx \frac{16 \hbar G H^2(t_k)}{\pi c^5} \; .
\label{treepower}
\end{equation}

The WKB approximation is another common technique for estimating 
the freeze-in amplitudes of $v(t,k)$ and $u(t,k)$ which appear
in expressions (\ref{Deltav}-\ref{Deltau}) for the tree order
power spectra. Recall that the method applies to differential
equations of the form $\ddot{f} + \omega^2(t) f = 0$. From 
expression (\ref{u2def}) one can see that reaching this form for 
the tensor mode function requires the rescaling $f(t,k) = 
a^{\frac32}(t) \times u(t,k)$. It is then simple to recognize the 
correctly normalized WKB solution and its associated frequency as,
\begin{eqnarray}
u(t,k) & \approx & \sqrt{ \frac{16 \pi \hbar G}{c^2 a^3(t) \vert
\omega(t,k)\vert} } \exp\Biggl[ -i \!\! \int_{t_2}^t \!\! dt'
\omega(t',k) \Biggr] \; , \label{WKBu} \qquad \\
\omega^2(t,k) & \equiv & \frac{c^2 k^2}{a^2(t)} - \Bigl[\frac94 
\!-\! \frac32 \epsilon(t) \Bigr] H^2(t) \; . \label{omegau}
\end{eqnarray}
In the sub-horizon regime of $ck \gg H(t) a(t)$ the frequency is
real $\omega(t,k) \approx ck/a(t)$ and the solution (\ref{WKBu})
both oscillates and falls off like $1/a(t)$. Freeze-in occurs in
the super-horizon regime of $ck \ll H(t) a(t)$ during which the
frequency is imaginary $\omega(t,k) \approx i H(t) [\frac32 -
\frac12 \epsilon(t)]$. We can estimate the freeze-in amplitude 
by computing the real part of the exponent,
\begin{equation}
-i \!\! \int_{t_k}^t \!\! dt' \omega(t',k) \approx \int_{t_k}^t 
\!\! dt' \Biggl\{ \frac32 \frac{\dot{a}(t')}{a(t')} + \frac12
\frac{\dot{H}(t')}{H(t')} \Biggr\} = 
\ln\Biggl[\frac{a^{\frac32}(t) H^{\frac12}(t)}{a^{\frac32}(t_k)
H^{\frac12}(t_k)}\Biggr] \; . \label{uexponent}
\end{equation}
Substituting (\ref{uexponent}) in (\ref{WKBu}) and using 
$H(t_k) a(t_k) = c k$ gives the following result for the 
freeze-in modulus,
\begin{equation}
t_k \ll t \ll T_k \qquad \Longrightarrow \qquad \Bigl\vert
u(t,k) \Bigr\vert_{\rm WKB} \approx \sqrt{\frac{16 \pi 
\hbar G H^2(t_k)}{c^5 k^3 \vert \frac32 \!-\! \frac12 \epsilon(t)
\vert}} \; . \label{normuWKB}
\end{equation}
If we drop the order one factor of $\vert \frac32 - \frac12 
\epsilon(t) \vert$ the result is (\ref{treepower}).

From expression (\ref{vdef}) we see that reaching the WKB form
for the scalar mode function requires the rescaling $f(t,k) = 
\epsilon^{\frac12}(t) \times a^{\frac32}(t) \times v(t,k)$. Its 
frequency is simpler to express if I first introduce the (Hubble
form of the) second slow roll parameter $\eta(t)$,
\begin{equation}
\eta(t) \equiv \epsilon(t) - \frac{\dot{\epsilon}(t)}{2 H(t) 
\epsilon(t)} \; . \label{etadef}
\end{equation}
The correctly normalized WKB approximation for the scalar 
mode function and its frequency is,
\begin{eqnarray}
v(t,k) & \approx & \sqrt{ \frac{2 \pi \hbar G}{c^2 \epsilon(t) 
a^3(t) \vert \nu(t,k)\vert} } \exp\Biggl[ -i \!\! \int_{t_2}^t 
\!\! dt' \nu(t',k) \Biggr] \; , \label{WKBv} \qquad \\
\nu^2(t,k) & \equiv & \frac{c^2 k^2}{a^2(t)} - \Bigl[\Bigl(
\frac32 \!+\! \epsilon(t) \!-\! \eta(t)\Bigr) \Bigl(\frac32
\!-\! \eta(t)\Bigr) + \frac{\dot{\epsilon} \!-\! \dot{\eta}}{
H(t)} \Bigr] H^2(t) \; . \label{omegav} \qquad
\end{eqnarray}
Freeze-in occurs as one evolves from the sub-horizon regime of
$\nu(t,k) \approx ck/a(t)$ to the super-horizon regime of 
$\nu(t,k) \approx i H(t) [\frac32 + \frac12 \epsilon(t) - 
\eta(t)]$. The real part of the exponent of (\ref{WKBv}) is,
\begin{equation}
-i \!\! \int_{t_k}^t \!\! dt' \nu(t',k) \approx \!\!\int_{t_k}^t 
\!\! dt' \Biggl\{ \frac32 \frac{\dot{a}(t')}{a(t')} + \frac12
\frac{\dot{H}(t')}{H(t')} + \frac12 \frac{\dot{\epsilon}(t')}{
\epsilon(t')} \Biggr\} = \ln\Biggl[\frac{a^{\frac32}(t) 
H^{\frac12}(t) \epsilon^{\frac12}(t)}{a^{\frac32}(t_k)
H^{\frac12}(t_k) \epsilon^{\frac12}(t_k)}\Biggr] . 
\label{vexponent}
\end{equation}
Substituting in (\ref{WKBv}) gives a freeze-in modulus which
is again roughly consistent with (\ref{treepower}),
\begin{equation}
t_k \ll t \ll T_k \qquad \Longrightarrow \qquad \Bigl\vert
v(t,k) \Bigr\vert_{\rm WKB} \approx \sqrt{\frac{2 \pi 
\hbar G H^2(t_k)}{c^5 k^3 \epsilon(t_k) \vert \frac32 \!+\! 
\frac12 \epsilon(t) \!-\! \eta(t)\vert}} \; . \label{normvWKB}
\end{equation}

It should be obvious from the discordant factors of order one 
in expressions (\ref{finalconst}), (\ref{normuWKB}) and
(\ref{normvWKB}), that the results (\ref{treepower}) for the 
tree order power spectra are only approximate. This is confirmed 
by numerical integration of explicit models \cite{Wang:1997cw,
Martin:1999wa}. In addition to order one factors depending on 
the geometry at $t = t_k$ there is also a nonlocal ``memory 
factor'' which depends on the precise manner in which the mode 
evolves up to first horizon crossing \cite{Tsamis:2003px}. Most
of the ambiguity derives from not having a definitive model for
what caused inflation. Once the expansion history is known it
is possible to derive wonderfully accurate results by 
numerically integrating either the mode functions 
\cite{Easther:2010qz}. It is even more efficient to numerically 
evolve the time-dependent power spectra directly, without the 
irrelevant phase information \cite{Romania:2012tb}.

The tree order power spectra (\ref{exacttree}) give the 
tensor-to-scalar ratio $r(k)$, the scalar spectral index $n_s(k)$,
and the tensor spectral index $n_t(k)$,
\begin{eqnarray}
r(k) & \equiv & \frac{\Delta^2_h(k)}{\Delta^2_{\mathcal{R}}(k)} 
\approx 16 \epsilon(t_k) \; , \label{rdef} \\
n_s(k) & \equiv & 1 + \frac{\partial \ln(\Delta^2_{\mathcal{R}}(k)
}{\partial \ln(k)} \approx 1 - 4 \epsilon(t_k) + 2 \eta(t_k) \; .
\label{nsdef} \\
n_t(k) & \equiv & \frac{\partial \ln(\Delta^2_h(k)}{\partial 
\ln(k)} \approx - 2\epsilon(t_k) \; . \label{ntdef}
\end{eqnarray}
In each case the definition is exact, and the approximate result
derives from expressions (\ref{treepower}) with an additional
approximation to relate $dk$ to $dt_k$,
\begin{equation}
c dk = [1 \!-\! \epsilon(t_k)] H^2(t_k) a(t_k) dt_k \approx
ck \times H(t_k) dt_k \; . \label{dkdt}
\end{equation}
Comparison of (\ref{rdef}) and (\ref{ntdef}) implies an important 
test on single-scalar inflation which is violated in more general
models \cite{Polarski:1995zn,GarciaBellido:1995fz,Sasaki:1995aw},
\begin{equation}
r \approx -8 n_t \; . \label{Sscalar}
\end{equation}
Certain general trends are also evident from the approximate 
results (\ref{treepower}):
\begin{itemize}
\item{$r < 1$ because $\epsilon \ll 1$;}
\item{$n_t < 0$ because $H(t)$ decreases; and} 
\item{$n_s - 1 < n_t$ because $\epsilon(t)$ tends to increase.}
\end{itemize}

To anyone who works in quantum gravity it is breath-taking that 
we have {\it any} data, so it seems petulant to complain that the
some of the parameters, and particularly their dependences upon $k$,
are still poorly constrained. The scalar power spectrum can be
inferred from the measurements of the intensity and the $E$-mode of
polarization in cosmic microwave radiation which originates at the
time of recombination ({\bf rec} on Figure \ref{epsilon}) and then
propagates through the fossilized metric perturbations left over
from the epoch of primordial inflation. The latest full-sky results 
come from the Planck satellite and are fit to the form 
\cite{Ade:2013zuv},
\begin{equation}
\Delta^2_{\mathcal{R}}(k) \approx A_s \Bigl( \frac{k}{k_0}
\Bigr)^{n_s -1 + \frac{d n_s}{d \ln(k)} \ln(\frac{k}{k_0})} \; .
\end{equation}
The fiducial wave number is $k_0 = 0.050~{\rm Mpc}^{-1}$, and the
quantities ``$n_s$'' and ``$dn_s/d\ln(k)$'' are the scalar spectral
index and its derivative evaluated at $k = k_0$. When combined with
the polarization data from the WMAP satellite \cite{Hinshaw:2012aka}
the Planck team reports \cite{Ade:2013zuv},
\begin{equation}
10^{9} \times A_s = 2.196^{+0.051}_{-0.060} \; , \; n_s = 0.9603 
\pm 0.0073 \; , \; \frac{d n_s}{d \ln(k)} = -0.013 \pm 0.018 \; . 
\label{Plancknumbers}
\end{equation}

The tensor-to-scalar ratio is reported at a much smaller wave number 
of $k = 0.002~{\rm Mpc}^{-1}$. Bounds on $r_{0.002}$ can be derived 
from analyzing how the intensity and the $E$-mode of polarization 
of the cosmic microwave background radiation depend upon $k$. 
Because this sort of $k$ dependence might also indicate ``running''
of the scalar spectral index ($dn_s/d\ln(k) \neq 0$) the limits on 
$r_{0.002}$ become significantly weaker if one allows for running.
Combining Planck and previous data sets gives the following bounds 
at $95\%$ confidence \cite{Ade:2013zuv},
\begin{eqnarray}
r_{0.002} & < & 0.11 \qquad \frac{d n_s}{d \ln(k)} = 0 \; , \\
r_{0.002} & < & 0.26 \qquad \frac{d n_s}{d \ln(k)} \neq 0 \; .
\end{eqnarray}
Direct detection requires a measurement of the $B$-mode of 
polarization. The BICEP2 team have done this and they report a 
result consistent with,
\begin{equation}
r_{0.002} \approx 0.20 \; . \label{BICEP2}
\end{equation}

Although the tensor power spectrum is still poorly known, and
controversial \cite{Flauger:2014qra,Paul}, resolving it is 
terrifically important because it tests single-scalar inflation 
through relation (\ref{Sscalar}) and incidentally fixes the scale 
of primordial inflation. If only $\Delta^2_{\mathcal{R}}(k)$ is
resolved one can always construct a single-scalar potential 
$V(\varphi)$ which will explain it. To see this, suppose we have 
measured the scalar power spectrum for some range of wave numbers 
$k_2 < k < k_1$ and use the approximate formula (\ref{treepower}), 
along with the small $\epsilon$ relation (\ref{dkdt}) between 
$dt_k$ and $dk$, to reconstruct the inflationary Hubble parameter,
\begin{equation}
\frac1{H^2(t_k)} - \frac1{H^2_2} = \frac{2 \hbar G}{\pi c^5} \!\!
\int_{k_2}^k \! \frac{dk'}{k' \, \Delta^2_{\mathcal{R}}(k')} \; . 
\label{HgivenDeltaR}
\end{equation}
The Hubble parameter $H_2 \equiv H(t_2)$ is an integration 
constant which we can choose to make the tensor power spectrum 
smaller than any bound. Now use (\ref{HgivenDeltaR}), with 
(\ref{dkdt}), to reconstruct the relation between $t$ and $k$,
\begin{equation}
H_2 (t \!-\! t_2) = \int_{k_2}^k \!\! \frac{dk'}{k'} \sqrt{1 +
\frac{2 \hbar G H_2^2}{\pi c^5} \! \int_{k_2}^{k'} \! \frac{dk''
}{k'' \, \Delta^2_{\mathcal{R}}(k'')} } \; . \label{tfromk}
\end{equation}
This expression can always be inverted numerically, and the rest
of the construction is the same as that given in section 
\ref{single}.

\subsection{The Controversy over Adiabatic Regularization}
\label{adiabatic}

It is obvious from their free Lagrangians (\ref{freezeta}-\ref{freeh}) 
that the 2-point correlators of $\zeta(t,\vec{x})$ and 
$h_{ij}(t,\vec{x})$ diverge quadratically when the two fields are 
evaluated at the same spacetime point. This is not enough to induce any 
tree order divergence in my definitions (\ref{Deltaz}-\ref{Deltah}).
However, it is problematic for the more common definition which is based 
on a spectral resolution of the coincident 2-point function,
\begin{eqnarray}
\Bigl\langle \Omega \Bigl\vert \zeta(t,\vec{x}) \zeta(t,\vec{x})
\Bigr\vert \Omega \Bigr\rangle & = & \int_0^{\infty} \!\! 
\frac{dk}{k} \, \Delta^2_{\mathcal{R}}(k,t) \; , \label{altz} \\ 
\Bigl\langle \Omega \Bigl\vert h_{ij}(t,\vec{x}) h_{ij}(t,\vec{x}) 
\Bigr\vert \Omega \Bigr\rangle & = & \int_0^{\infty} \!\! 
\frac{dk}{k} \, \Delta^2_{h}(k,t) \; . \label{alth}
\end{eqnarray}
In 2007 Leonard Parker \cite{Parker:2007ni} pointed out that removing 
this divergence with the standard technique of adiabatic regularization 
\cite{Parker:1969au,Parker:1974qw,Fulling:1974pu,Bunch:1980vc,
Anderson:1987yt} can change the power spectra by several orders of 
magnitude. 

Subsequent work by Parker and collaborators showed that adiabatic 
regularization of the scalar and tensor power spectra would alter the 
single-scalar consistency relation (\ref{Sscalar}) and would also 
reconcile the conflict between even WMAP data \cite{Hinshaw:2012aka} 
and a quartic inflaton potential $V(\varphi) = \lambda \varphi^4$ 
\cite{Agullo:2008ka,Agullo:2009vq}. Such profound changes in the 
labour of three decades provoked the natural objection that no 
technique for addressing ultraviolet divergences ought to affect the
infrared regime in which inflationary particle production takes 
place \cite{Durrer:2009ii,Marozzi:2011da}. Parker and his 
collaborators replied that consistency of renormalization theory 
requires adiabatic subtractions which affect all modes, including
those in the infrared \cite{Agullo:2011qg}. 

I find this debate fascinating because it is an example of how 
inflationary cosmology is challenging the way we think about hitherto
abstract issues in quantum gravity and vice versa. I don't
know the answer but I have encountered the same problem when trying 
to work out the pulse of gravitons which would be produced by a very 
peculiar model in which $H(t)$ oscillates from positive to negative 
at the end of inflation \cite{Romania:2010zq,Romania:2011ez}. The
resolution may not lie with any change in the way we renormalize but 
rather with greater care in how we connect theory to observation,
for example, defining the tree order power spectra from expressions
(\ref{Deltaz}-\ref{Deltah}) rather than from spectral resolutions of 
the coincident 2-point functions (\ref{altz}-\ref{alth})
\cite{Bastero-Gil:2013nja}. Whatever we find, it is worthwhile to
reflect on the wonder of what is taking place. These are the same 
problems which the men of genius who founded flat space quantum 
field theory had to puzzle out when they settled on non-coincident 
one-particle-irreducible functions as the basis for renormalization 
and computation of the S-matrix. It is a privilege to reprise their 
roles.  

\subsection{Why These Are Quantum Gravitational Effects}\label{whyQG}

The factors of $\hbar G$ in expressions (\ref{treepower}) ought to 
establish that both power spectra are legitimate quantum 
gravitational, the first ever detected. Unfortunately, three 
objections seem to be delaying general recognition of this simple but 
revolutionary fact:
\begin{itemize}
\item{Expressions (\ref{treepower}) are tree order results;}
\item{There is still debate over whether or not the graviton signal
$\Delta^2_{h}(k)$ has been resolved \cite{Ade:2014xna,
Flauger:2014qra,Paul}; and}
\item{There is not yet a compelling model for what caused primordial 
inflation.}
\end{itemize}
I will argue below that all three objections result from imposing 
unreasonably high standards on what qualifies as a quantum 
gravitational effect.

The first objection might be re-stated as, ``it's not quantum gravity
if it's only tree order.'' This is applied to no other force. For 
example, both the photo-electric effect and beta decay occur tree 
order, yet no one disputes that they are quantum manifestations of the 
electro-weak interaction. The same thing could be said of Planck's 
black-body spectrum, and any number of other tree order effects such 
as Bhabha scattering.

The second objection might be restated as, ``it's not quantum gravity
if it doesn't involve gravitons.'' This is also silly because 
$\Delta^2_{\mathcal{R}}(k)$ has certainly been resolved and it is 
just as certainly a quantum gravitational effect in view of the 
factor of $\hbar G$ evident in expression (\ref{treepower}). We saw
in sections \ref{gauge} and \ref{subtree} that the scalar power 
spectrum derives from the gravitational response to quantum matter, 
the same way that all the solar system tests of general relativity 
derive from the gravitational response to classical matter. Were we
to insist that only gravitons can test quantum gravity then logical 
consistency would imply that only gravitational radiation tests 
classical gravity, at which point we are left with only the binary 
pulsar data!

Indeed, a little reflection on the problem of perturbative quantum 
gravity \cite{Woodard:2009ns} reveals that the lowest order problem 
is not from gravitons --- which cause no uncontrollable divergences 
until two loop order \cite{Goroff:1985th,vandeVen:1991gw} --- but 
rather from exactly the same gravitational response to quantum matter 
which the scalar power spectrum tests. All experimentally confirmed 
matter theories engender quantum gravitational divergences at just 
one loop order \cite{'tHooft:1974bx,Deser:1974zzd,Deser:1974cz,
Deser:1974cy,Deser:1974nb,Deser:1974xq}. If a sensible quantum 
gravity expert was told he could only know one of the two power 
spectra and then asked to choose which one, he ought to pick 
$\Delta^2_{\mathcal{R}}(k)$ because it tells him about the lowest 
order problem. Fortunately, we will know both power spectra, and 
probably sooner rather than later. It is even possible we will 
eventually resolve one loop corrections.

The final objection could be re-stated as, ``it isn't quantum 
gravity if we can't make a unique prediction for it.'' This seems
as ridiculous as trying to argue that galactic rotation curves don't
necessarily derive from gravity just because we are not yet certain 
whether their shapes are explained by Newtonian gravity with dark 
matter or by some modification of gravity. Which is not to deny how
wonderfully improved the situation would be with a compelling model 
for inflation. If we had one then the two power spectra would 
provide a definitive test of quantum gravity, the same way that the 
photo-electric effect and Bhabha scattering test quantum 
electrodynamics.

Sceptics are free to accuse me of unwarranted optimism but I believe
that working out what drove primordial inflation is just a matter of 
time in the data-rich environment which is developing. Measurements
of $n_s$ with increasingly tight upper bounds on $r_{0.002}$ have
already ruled out some potentials such as $V(\varphi) \sim 
\varphi^4$ \cite{Hinshaw:2012aka}, and all models with constant 
$\epsilon(t)$ \cite{Hou:2012xq,Sievers:2013ica}. This process is 
bound to continue, and even accelerate, as the data gets better. 
There are plans to reduce the errors on $n_s$ by a factor of five 
using galaxy surveys \cite{IPSIG}. (This will begin filling in the 
question mark region of reheating on Figure \ref{epsilon}.) If the 
BICEP2 detection really means $r_{0.002} \approx 0.20$ it would 
rule out a host of models with small $r_{0.002}$ 
\cite{Ade:2014xna}. We will know within the next five years by 
checking if the BICEP2 signal possesses the key frequency and 
angular dependences needed to distinguish it as primordial 
gravitons. If so then it should be possible to reduce the errors 
on $r_{0.002}$ to the percent level within the next decade. As 
higher resolution polarization measurements are made over the 
course of the next 15 years it should be possible to remove the 
gravitational lensing signal (known as ``de-lensing'') to reach 
the sensitivity needed to measure $n_t$. It is inconceivable to 
me that theorists will remain idle while these events transpire. 
Past experience shows that theory and experiment develop 
synergistically. The data are not going to run out any time soon, 
and I believe fundamental theorists will eventually receive 
enough guidance to develop a truly compelling a model for 
primordial inflation.

Let me close this section by pointing out that just the fact of
observing scalar perturbations from primordial inflation tells us
two significant things about quantum gravity \cite{Woodard:2009ns}:
\begin{itemize}
\item{It is no longer viable to avoid quantizing gravity; and}
\item{The problem of ultraviolet divergences cannot be explained
by making spacetime discrete.}
\end{itemize}
The first point is obvious from the fact that the scalar power 
spectrum represents the gravitational response to quantum 
fluctuations of matter, which would be absent if the source of
classical gravity were taken to be the expectation value of the
matter stress tensor in some state. To see the second point note 
that although discretization at any scale makes quantum 
gravitational loop integrals finite, it will not keep them {\it 
small} unless the discretization length is larger than  
$\sqrt{\hbar G/c^3} \sim 1.6 \times 10^{-34}~{\rm m}$. 
But primordial inflation posits that the universe has expanded 
by the staggering factor of about $e^{100} \sim 10^{43}$ from a 
time when quantum gravitational effects were small. Hence the 
current co-moving scale of discretization would correspond to 
about a million kilometers! 

\section{Loop Corrections to the Power Spectra}\label{loop}

From (\ref{Plancknumbers}) one can see that the scalar power
spectrum is currently measured with an accuracy of more than 
two significant figures. However, resolving the one loop 
correction would require about ten significant figures because
the loop counting parameter of inflationary quantum gravity is
no larger than $\hbar G H_i^2/c^5 \sim 10^{-10}$. Although 
there is no hope of achieving this precision within the next 
two decades, the data is potentially recoverable and theorists 
have begun thinking about how to predict the results when 
(and if) one loop corrections are resolved in the far future. 
This section describes the basic formalism and the significant
issues. I close by adumbrating a process through which the 
missing eight significant figures might be made up.

\subsection{How to Make Computations}\label{howto}

I will return later in this section to the issue of precisely 
what theoretical quantities correspond to the observed scalar
and tensor power spectra. For now let me assume that the tree 
order definitions (\ref{Deltaz}-\ref{Deltah}) remain valid.
One striking fact about these expressions is that neither of 
them is an S-matrix element. Nor is either the matrix element 
of some product of noncoincident local operators (because both
are at the same time) between an ``in'' state which is free 
vacuum at asymptotically early times and an ``out'' state 
which is free vacuum at asymptotically late times. One can 
define a formal S-matrix for the simplest cosmologies 
\cite{Marolf:2012kh} but it calls for measurements which are 
precluded by causality. More generally, the entire formalism 
of in-out matrix elements --- which is all most of us were 
taught to calculate --- is inappropriate for cosmology 
because the universe began with an initial singularity 
\cite{Borde:2001nh} and no one knows how it will end. 
Persisting with in-out quantum field theory would make loop 
corrections possess two highly undesirable features:
\begin{itemize}
\item{They would be dominated by assumptions about the 
``out'' vacuum owing to vast expansion of spacetime; and}
\item{The matrix elements of even Hermitian operators would
be complex numbers because the ``in'' and ``out'' vacua must 
differ due to inflationary particle production.}
\end{itemize}

The more appropriate quantity to study in cosmology is the
expectation value of some operator in the presence of a 
prepared state which is released at a finite time. Of course
one could always employ the canonical formalism to make such
computations, but particle physicists yearn for a technique
that is as simple as the Feynman rules are for in-out matrix
elements. Julian Schwinger devised such a formalism for 
quantum mechanics in 1961 \cite{Schwinger:1960qe}. Over the 
next two years it was generalized to quantum field by
Mahanthappa \cite{Mahanthappa:1962ex} and by Bakshi and
Mahanthappa \cite{Bakshi:1962dv,Bakshi:1963bn}. Keldysh
applied it to statistical field theory in 1964
\cite{Keldysh:1964ud} where the technique has become routine. 
Until very recently its use in quantum field theory was limited 
to a handful of people working on phase transitions and gravity
\cite{Chou:1984es,Jordan:1986ug,Calzetta:1986ey,
Calzetta:1986cq}. Most particle theorists were majestically 
ignorant of the technique and so attached to the in-out 
formalism that they dismissed as mistakes what are significant 
and deliberate deviations of the Schwinger-Keldysh formalism, 
such as the absence of an imaginary part. The stifling 
atmosphere which prevailed is well conveyed by the lofty 
disdain in the words of a referee I had for a 2003 grant 
renewal proposal to the Department of Energy:
\begin{quotation}
In his work with Tsamis, Woodard has focused on what they 
interpret as an instability of de Sitter space due to a 
two-loop infrared divergence associated with long-wavelength, 
virtual gravitons. They describe this as the accumulation of 
gravitational attraction of ``large-wavelength virtual 
gravitons.'' That is a puzzling statement in itself--the
accumulation that they describe would build up only if 
gravitons were really being produced. In fact, they think 
these virtual gravitons are rendered real as they are 
``pulled apart by rapid expansion of spacetime.'' I believe 
that there is absolutely no evidence for this. Real particle 
production should show up as an imaginary contribution to
the graviton vacuum polarization tensor, at least if 
unitarity in de Sitter space resembles flat space.
\end{quotation}

The thinking of particle theorists underwent a radical
transformation in 2005 when Nobel laureate Steven Weinberg
undertook a study of loop corrections to the power spectra
\cite{Weinberg:2005vy}. He quickly realized that the in-out
formalism was inappropriate and, because he did not then 
know of the Schwinger-Keldysh formalism, he independently 
discovered a version of it which is better suited to this
problem than the usual one. (His student Bua Chaicherdsakul
told Weinberg of the older technique, and he gave full credit
to Schwinger in his paper.) I well recall the day Weinberg's
paper appeared on the arXiv. I chanced to be visiting the
University of Utrecht then and a very knowledgeable and not 
unsympathetic colleague commented, ``I guess I will finally 
have to learn the Schwinger-Keldysh formalism.'' Weinberg's 
words on the general problem of computing loop effects in 
primordial inflation are also worth quoting in defence of 
the intellectual curiosity which is sometimes lacking in 
particle theory:
\begin{quotation}
This paper will discuss how calculations of cosmological
correlations can be carried to arbitrary orders of 
perturbation theory, including the quantum effects 
represented by loop graphs. So far, loop corrections to
correlation functions appear to be much too small ever to 
be observed. The present work is motivated by the opinion
that we ought to understand what our theories entail, even 
where in practice its predictions cannot be verified
experimentally, just as field theorists in the 1940's and
1950's took pains to understand quantum electrodynamics to
all orders of perturbation theory, even though it was only
possible to verify results in the first few orders.  
\end{quotation}

The best way to understand the Schwinger-Keldysh formalism is
by relating its functional integral representation to the 
canonical formalism. Recall how this goes for the in-out 
formalism in the context of a real scalar field $\phi(t,\vec{x})$
whose Lagrangian is the spatial integral of its Lagrangian
density,
\begin{equation}
L[\phi(t)] \equiv \int \!\! d^{D-1}x \, \mathcal{L}[\phi(t,\vec{x})]
\; . 
\end{equation}
The in-out formalism gives matrix elements of $T^{*}$-ordered
products of operators, which means that any derivatives are taken
outside the time-ordering symbol. The usual relation is adapted
to asymptotic scattering problems but, for our purposes, it is
better to consider the matrix element between a state $\vert \Psi
\rangle$ whose wave functional at time $t = t_2$ is 
$\Psi[\phi(t_2)]$ and a state $\langle \Phi\vert$ with wave 
functional $\Phi[\phi(t_1)]$. The well-known functional integral
expression for the matrix element of the $T^*$-ordered product of
some operator $\mathcal{O}_a[\phi]$ is,
\begin{equation}
\Bigl\langle \Phi \Bigl\vert T^*\Bigl(\mathcal{O}_a[\phi]\Bigr) 
\Bigr\vert \Psi \Bigr\rangle = \Fint [d\phi] \, \mathcal{O}_a[\phi] 
\, \Phi^*[\phi(t_1)] \, e^{\frac{i}{\hbar} \int_{t_2}^{t_1} \! dt 
L[\phi(t)]} \, \Psi[\phi(t_2)] \; . \label{1stint}
\end{equation}
We can use (\ref{1stint}) to obtain a similar expression for the 
matrix element of the anti-$T^*$-ordered product of some 
operator $\mathcal{O}_b[\phi]$ in the presence of the conjugate
states,
\begin{eqnarray}
\Bigl\langle \Psi \Bigl\vert \overline{T}^*\Bigl(\mathcal{O}_b[\phi]
\Bigr) \Bigr\vert \Phi \Bigr\rangle & = & \Bigl\langle \Phi 
\Bigl\vert T^*\Bigl(\mathcal{O}_b^{\dagger}[\phi]\Bigr) \Bigr\vert 
\Psi \Bigr\rangle^* \; , \\
& = & \Fint [d\phi] \, \mathcal{O}_b[\phi] \, \Phi[\phi(t_1)] \,
e^{-\frac{i}{\hbar} \int_{t_2}^{t_1} \! dt L[\phi(t)]} \, 
\Psi^*[\phi(t_2)] \; . \label{2ndint}
\end{eqnarray}
Summing over a complete set of wavefunctionals $\Phi[\phi(t_1)]$ 
gives a delta functional,
\begin{equation}
\sum_{\Phi} \Phi\Bigl[\phi_-(t_1)\Bigr] \, \Phi^*\Bigl[\phi_+(t_1)
\Bigr] = \delta\Bigl[\phi_-(t_1) \!-\! \phi_+(t_1) \Bigr]
\; . \label{sum}
\end{equation}
Multiplying (\ref{1stint}) by (\ref{2ndint}), and using (\ref{sum}),
gives a functional integral expression for the expectation value of 
any anti-$T^*$-ordered operator $\mathcal{O}_b$ multiplied by any 
$T^*$-ordered operator $\mathcal{O}_a$,
\begin{eqnarray}
\lefteqn{\Bigl\langle \Psi \Bigl\vert \overline{T}^*\Bigl(\mathcal{O}_b[
\phi]\Bigr) T^*\Bigl(\mathcal{O}_a[\phi]\Bigr) \Bigr\vert \Psi
\Bigr\rangle = \Fint [d\phi_+] [d\phi_-] \, \delta\Bigl[\phi_-(\ell)
\!-\! \phi_+(\ell)\Bigr] } \nonumber \\
& & \hspace{1.5cm} \times \mathcal{O}_b[\phi_-] \mathcal{O}_a[\phi_+]
\Psi^*[\phi_-(t_2)] e^{\frac{i}{\hbar} \int_s^{\ell} dt \Bigl\{L[\phi_+(t)] 
- L[\phi_-(t)]\Bigr\}} \Psi[\phi_+(t_2)] \; . \qquad \label{fund}
\end{eqnarray}
This is the fundamental Schwinger-Keldysh relation between the canonical 
operator formalism and the functional integral formalism.

What we might term the ``Feynman rules'' of the Schwinger-Keldysh formalism
follow from (\ref{fund}) in close analogy to those for in-out matrix elements. 
Because the same field operator is represented by two different dummy fields,
$\phi_{\pm}(x)$, the endpoints of lines carry a $\pm$ polarity. External 
lines associated with the operator $\mathcal{O}_b[\phi]$ have the $-$ 
polarity while those associated with the operator $\mathcal{O}_a[\phi]$ have 
the $+$ polarity. Interaction vertices are either all $+$ or all $-$. The
same is true for counterterms, which means that mixed-polarity diagrams 
cannot harbor primitive divergences. Vertices with $+$ polarity are the same 
as in the usual Feynman rules whereas vertices with the $-$ polarity have an 
additional minus sign. Propagators can be $++$, $-+$, $+-$ and $--$.

\begin{figure}[ht]
\vspace{-4cm}
\hspace{2cm} \includegraphics[width=4cm,height=4cm]{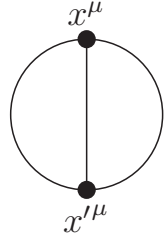} 
\vspace{3cm}
\caption{\label{SKdiagram} \tiny A typical Schwinger-Keldysh loop.
The vertex at $x^{\mu}$ connects to an amputated $+$ leg. The
vertex at ${x'}^{\mu}$ is internal and must be summed over $+$
and $-$ polarities. The cancellation between polarities makes the
integral (\ref{SKint}) pure imaginary and restricts the 
integration to points ${x'}^{\mu}$ on or within the past 
light-cone of $x^{\mu}$.}
\end{figure}

The four propagators can be read off from the fundamental relation 
(\ref{fund}) when the free Lagrangian is substituted for the full one. 
I denote canonical expectation values in the free theory with a 
subscript $0$. With this convention one sees that the $++$ propagator 
is the ordinary Feynman result,
\begin{equation}
i\Delta_{\scriptscriptstyle ++}(x;x') = \Bigl\langle \Omega \Bigl\vert
T\Bigl(\phi(x) \phi(x') \Bigr) \Bigr\vert \Omega \Bigr\rangle_0 =
i\Delta(x;x') \; . \label{++}
\end{equation}
The other cases are simple to read off and to relate to (\ref{++}),
\begin{eqnarray}
i\Delta_{\scriptscriptstyle -+}(x;x') & = &  
\Bigl\langle \Omega \Bigl\vert \phi(x) \phi(x') \Bigr\vert \Omega 
\Bigr\rangle_0 \!\!=\! \theta(t\!-\!t') i\Delta(x;x') \!+\! 
\theta(t'\!-\!t) \Bigl[i\Delta(x;x')\Bigr]^* , \qquad \label{-+} \\
i\Delta_{\scriptscriptstyle +-}(x;x') & = &  
\Bigl\langle \Omega\Bigl\vert \phi(x') \phi(x) \Bigr\vert \Omega 
\Bigr\rangle_0 \!\!=\! \theta(t\!-\!t') \Bigl[i\Delta(x;x')\Bigr]^* 
\!+\! \theta(t'\!-\!t) i\Delta(x;x') , \qquad \label{+-} \\
i\Delta_{\scriptscriptstyle --}(x;x') & = & 
\Bigl\langle \Omega\Bigl\vert \overline{T}\Bigl(\phi(x) \phi(x') 
\Bigr) \Bigr\vert \Omega \Bigr\rangle_0 = \Bigl[i\Delta(x;x')
\Bigr]^* . 
\label{--}
\end{eqnarray}
The close relations between the various propagators and the minus 
signs from $-$ vertices combine to enforce causality and reality in 
the Schwinger-Keldysh formalism. For example, in a diagram with 
the topology depicted in Figure \ref{SKdiagram}, suppose the 
vertex at $x^{\mu}$ is connected to an amputated external $+$ line. 
If the vertex at ${x'}^{\mu}$ is internal then we must sum over
$+$ and $-$ variations and integrate to give a result proportional
to,
\begin{equation}
\int \!\! d^Dx' \Biggl\{ \Bigl[ i\Delta_{\scriptscriptstyle 
++}(x;x')\Bigr]^3 - \Bigl[ i\Delta_{\scriptscriptstyle +-}(x;x')
\Bigr]^3 \Biggr\} = 2 i \!\! \int_{t_2}^t \!\! dct' \! \int \!\!
d^{D-1}x' \, {\rm Im}\Biggl( \Bigl[ i\Delta(x;x') \Bigr]^3 
\Biggr) . \label{SKint}
\end{equation}

Although expression (\ref{1stint}) is simple to derive from the 
canonical formalism, few particle theorists would have recognized it 
before Weinberg's paper for two reasons:
\begin{itemize}
\item{The action integral runs between the finite times
$t_2 \leq t \leq t_1$; and}
\item{It contains state wave functionals $\Psi[\phi(t_2)]$ and
$\Phi^*[\phi(t_1)]$.}
\end{itemize}
The over-specialization of quantum field theory to asymptotic 
scattering problems led to generations of particle theorists being 
inculcated with the dogma that it is irrelevant to consider any 
state but ``the'' vacuum (often defined as ``the unique, 
normalizable energy eigenstate''), and that this state is 
automatically selected by extending the temporal integration to 
$-\infty < t < +\infty$. This was always nonsense, but it sufficed 
for asymptotic scattering theory as long as infrared problems were 
treated using the Bloch-Nordsieck technique \cite{Bloch:1937pw}, 
the universal applicability of which also became dogma in spite of 
simple counter-examples \cite{Veneziano:1972rs}. 

Inflationary cosmology has forced us to consider releasing the 
universe in a prepared state at a finite time. When this is done 
one realizes that the state wave functional $\Psi[\phi(t_2)]$ can 
be broken up into a free part, whose logarithm is quadratic in the 
perturbation field, and a series of perturbative corrections 
involving higher powers of the field,
\begin{equation}
\Psi\Bigl[\phi(t_2)\Bigr] = \Psi_0\Bigl[\phi(t_2)\Bigr] 
\Biggl\{ 1 + O\Bigl( \phi^3(t_2)\Bigr) \Biggr\} \; . \label{state}
\end{equation}
For example, the free vacuum state wave functional of a massive 
scalar in flat space is,
\begin{equation}
\Omega_0\Bigl[\phi(t_2)\Bigr] \propto \exp\Biggl[ -\frac1{2 c
\hbar} \! \int \!\! d^{D-1}x \, \phi(t,\vec{x}) \sqrt{-\nabla^2 
\!+\! \frac{m^2 c^2}{\hbar^2} } \, \phi(t,\vec{x}) \Biggr] \; . 
\label{freevac}
\end{equation}
It can be shown that the free part of the vacuum wave functional 
combines with the quadratic surface variations of the action to
enforce Feynman boundary conditions \cite{Tsamis:1984kx}. Rather
than the usual hand-waving, that is how inverting the kinetic 
operator gives a unique solution for the propagator. The 
perturbative correction terms (\ref{state}), which must be present 
even to recover the flat space limit, correspond to nonlocal 
interactions on the initial value surface \cite{Kahya:2009sz}.

\subsection{$\epsilon$-Suppression and Late Time Growth}\label{epssup}

Making exact computations requires the $\zeta(t,\vec{x})$ and
$h_{ij}(t,\vec{x})$ propagators and their interaction vertices.
From the free Lagrangian (\ref{freeZ}), and the appropriate 
$D$-dimensional generalization of the scalar mode function 
(\ref{vdef}), one can give a formal expression for the $\zeta$ 
Feynman propagator,
\begin{equation}
i\Delta_{\zeta}(x;x') = \int \!\! \frac{d^{D-1}k}{(2\pi)^{D-1}} 
\Biggl\{\theta(t \!-\! t') v(t,k) v^*(t',k) + \theta(t' \!-\! t) 
v^*(t,k) v(t',k) \Biggr\} e^{i \vec{k} \cdot \vec{x}} .
\end{equation}
Expressions (\ref{freeh}) and (\ref{u2def}) give a similar result
for the graviton Feynman propagator,
\begin{eqnarray}
\lefteqn{ i\Bigl[\mbox{}_{ij} \Delta_{k\ell}\Bigr](x;x') = 
\Bigl[ \Pi_{i (k} \Pi_{\ell ) j} - \frac1{D \!-\! 2} \Pi_{ij} 
\Pi_{k\ell} \Bigr] } \nonumber \\
& & \hspace{.5cm} \times \! \int \!\! \frac{d^{D-1}k}{(2\pi)^{D-1}} 
\Biggl\{\theta(t \!-\! t') u(t,k) u^*(t',k) + \theta(t' \!-\! t) 
u^*(t,k) u(t',k) \Biggr\} e^{i \vec{k} \cdot \vec{x}} , \qquad
\end{eqnarray}
where the transverse projection operator is $\Pi_{ij} \equiv 
\delta_{ij} - \partial_i \partial_j/\partial_k \partial_k$. 
Unfortunately, we do not possess simple expressions for either 
the scalar or tensor mode functions for a general expansion history 
$a(t)$, nor are all the gauge-fixed and constrained interactions
yet known to the order required for a full one loop computation, 
and nothing has been done about renormalization. I will therefore 
concentrate on characterizing how loop corrections behave with 
respect to the two most important issues which control their 
strength:
\begin{itemize}
\item{Enhancement by inverse factors of the slow roll
parameter $\epsilon(t)$; and}
\item{Enhancement by secular growth from infrared effects.}
\end{itemize}

\begin{figure}[ht]
\vspace{-5cm}
\hspace{2cm} \includegraphics[width=6cm,height=4cm]{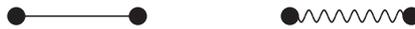} 
\vspace{2cm}
\caption{\label{treegraphs} \tiny 
Diagrammatic representation for the tree order power spectra. A 
straight line represents the $\zeta$ propagator while the graviton
propagator is wavy.}
\end{figure}

To understand the issue of $\epsilon$-enhancement let us first note 
from the free Lagrangians (\ref{freeZ}-\ref{freeh}) that the scalar
and tensor propagators have the following dependences upon 
$\epsilon(t)$ and the various fundamental constants, 
\begin{equation}
i \Delta_{\zeta} \sim \frac{\hbar G}{c^5 \epsilon} \times 
{\rm Frequency}^2 \qquad , \qquad i \Delta_h \sim  
\frac{\hbar G}{c^5} \times {\rm Frequency}^2 \; .
\end{equation}
For the effects of inflationary particle production the relevant 
frequency is the Hubble parameter $H(t)$. (Of course it could be at 
any time in the past, as could the factor of $1/\epsilon(t)$ in the 
$\zeta$ propagator.) This offers a very simple explanation for the
approximate forms (\ref{treepower}) I derived for the tree order
power spectra in section \ref{subtree}. The relevant diagrams are
given in Figure \ref{treegraphs}.

To find the gauge-fixed and constrained interactions one must solve 
the constraint equation (\ref{constraint}) for $S^i[\zeta,h]$, then
substitute back into (\ref{expansion}). There are many terms, even 
at the lowest orders, and they generally combine (sometimes after
partial temporal integrations) so that the final result is suppressed
by crucial powers of $\epsilon$. Each term has two net derivatives,
however, this counting must include $-1$ derivatives from factors of
$1/H$, and $-2$ derivatives from factors of $1/\partial_k \partial_k$
which arise in solving the constraint equation (\ref{constraint}). 
The $\zeta^3$ interaction was derived in 2002 by Maldacena 
\cite{Maldacena:2002vr}, and simple results were obtained in 2006
for the $\zeta^4$ terms by Seery, Lidsey and Sloth \cite{Seery:2006vu}.
At the level of detail I require these two interactions take the 
form,
\begin{equation}
\frac1{\hbar} \mathcal{L}_{\zeta^3} \sim \frac{c^4 \epsilon^2
a^{D-1}}{16 \pi \hbar G} \, \zeta \partial \zeta \partial \zeta 
\qquad , \qquad \frac1{\hbar} \mathcal{L}_{\zeta^4} \sim 
\frac{c^4 \epsilon^2 a^{D-1}}{16 \pi \hbar G} \, \zeta^2 
\partial \zeta \partial \zeta \; . \label{z34}
\end{equation}
In 2007 Jarhus and Sloth discussed the next two interactions 
\cite{Jarnhus:2007ia},
\begin{equation}
\frac1{\hbar} \mathcal{L}_{\zeta^5} \sim \frac{c^4 \epsilon^3 
a^{D-1}}{16 \pi \hbar G} \, \zeta^3 \partial \zeta \partial \zeta
\qquad , \qquad \mathcal{L}_{\zeta^6} \sim \frac{c^4 \epsilon^3 
a^{D-1}}{16 \pi \hbar G} \, \zeta^4 \partial \zeta \partial \zeta 
\; . \label{z56}
\end{equation}
Results for the lowest $\zeta$-$h_{ij}$ interactions were reported
in 2012 by Xue, Gao and Brandenberger \cite{Xue:2012wi}. Making no
distinction between which fields are differentiated, these 
interactions take the general form,
\begin{equation}
\frac1{\hbar} \Biggl[\mathcal{L}_{\zeta h^2} + 
\mathcal{L}_{\zeta^2 h} + \mathcal{L}_{\zeta^2 h^2} \Biggr] \sim 
\frac{c^4 \epsilon a^{D-1}}{16 \pi \hbar G} \Biggl[ \zeta \partial 
h \partial h + h \partial \zeta \partial \zeta + h^2 \partial 
\zeta \partial \zeta \Biggr] \; . \label{zhs}
\end{equation}
And because they persist even in the de Sitter limit of $\epsilon=0$
it is obvious that the purely graviton interactions are not 
$\epsilon$-suppressed,
\begin{equation}
\frac1{\hbar} \Biggl[\mathcal{L}_{h^3} + \mathcal{L}_{h^4} + 
\dots \Biggr] \sim \frac{c^4 a^{D-1}}{16 \pi \hbar G} 
\Biggl[ h \partial h \partial h + h^2 \partial h \partial h + 
\dots \Biggr] \; . \label{pureh}
\end{equation}

\begin{figure}[ht]
\vspace{-3cm}
\hspace{1cm} \includegraphics[width=10cm,height=5cm]{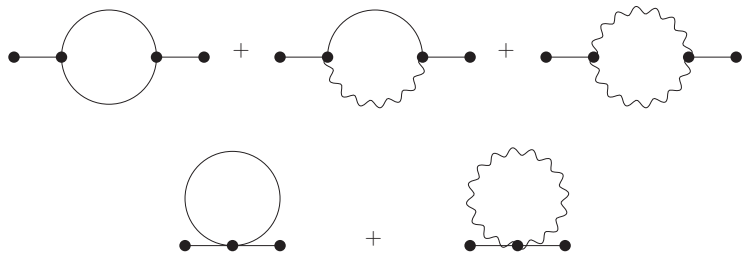} 
\vspace{1cm}
\caption{\label{zetaloops} \tiny 
One loop corrections to the scalar power spectrum. Straight lines 
represent the $\zeta$ propagator while the graviton propagator is 
wavy.}
\end{figure}

The various diagrams which contribute to the one loop correction
to $\Delta^2_{\mathcal{R}}(k)$ are depicted in Figure 
\ref{zetaloops}. In each case the leftmost point is fixed at
$x^{\mu} = (t,\vec{x})$ and the rightmost point is fixed at 
${x'}^{\mu} = (t,\vec{0})$. Interior points are integrated. For
example, the leftmost diagram on the first line has the general
form,
\begin{equation}
\int \!\! d^Dy \, i\Delta_{\zeta}(x;y) V_{\zeta^3}(y) \!\! \int \!\!
d^Dy' \, i\Delta_{\zeta}(x';y') V_{\zeta^3}(y') \Bigl[ 
i\Delta_{\zeta}(y;y') \Bigr]^2 \; , \label{leftgraph}
\end{equation}
where $V_{\zeta^3}(y)$ and $V_{\zeta^3}(y')$ denote the vertex
operators one can read off from the $\zeta^3$ interaction. To recover
the ordering in (\ref{Deltaz}) the $x^{\mu}$ line must have $-$
polarity and the ${x'}^{\mu}$ must be $+$, while the $y^{\mu}$ and
${y'}^{\mu}$ vertices would be summed over all $\pm$ variations. 

I will return to the possibility that vertex integrations lead to
temporal growth but for now let me assume that the two net derivatives 
in each vertex combine with the associated integral to produce a
factor of $c/H^2$. Under this assumption one can estimate the strength
of any diagram by combining:
\begin{itemize}
\item{A factor of $\hbar G H^2/c^5\epsilon$ for each $\zeta$ 
propagator;}
\item{A factor of $\hbar G H^2/c^5$ for each $h_{ij}$ propagator; and}
\item{A factor of $c^5 \epsilon^N/\hbar G H^2$ for each vertex with
either $2N-1$ or $2N$ $\zeta$ fields and any number of $h_{ij}$ 
fields.}
\end{itemize} 
For example, the estimated result for the leftmost diagram on the
first line of Figure \ref{zetaloops} is,
\begin{equation}
\Bigl( \frac{\hbar G H^2}{c^5 \epsilon} \Bigr)^4 \times 
\Bigl( \frac{\hbar G H^2}{c^5} \Bigr)^0 \times \Bigl( \frac{c^5 
\epsilon^2}{\hbar G H^2} \Bigr)^2 = \Bigl( \frac{\hbar G H^2}{c^5
\epsilon} \Bigr)^2 = \Bigl( \frac{\hbar G H^2}{c^5 \epsilon}\Bigr)
\times \Bigl( \frac{\hbar G H^2 \epsilon}{c^5} \Bigr) \; . 
\label{leftestimate}
\end{equation}
In the final expression of (\ref{leftestimate}) I have extracted the 
tree order result (\ref{treepower}), so one sees that this one loop
correction is down by the factor of $\hbar G H^2/c^5$ (which was
inevitable on dimensional grounds) times an extra factor of 
$\epsilon$. Neither of the two diagrams on the bottom line of Figure 
\ref{zetaloops} has this extra suppression,
\begin{eqnarray}
\Bigl( \frac{\hbar G H^2}{c^5 \epsilon} \Bigr)^3 \times 
\Bigl( \frac{\hbar G H^2}{c^5} \Bigr)^0 \times \Bigl( \frac{c^5 
\epsilon^2}{\hbar G H^2} \Bigr)^1 & = & \Bigl( \frac{\hbar G H^2}{c^5 
\epsilon}\Bigr) \times \Bigl( \frac{\hbar G H^2}{c^5} \Bigr) \; , \\
\Bigl( \frac{\hbar G H^2}{c^5 \epsilon} \Bigr)^2 \times 
\Bigl( \frac{\hbar G H^2}{c^5} \Bigr)^1 \times \Bigl( \frac{c^5 
\epsilon}{\hbar G H^2} \Bigr)^1 & = & \Bigl( \frac{\hbar G H^2}{c^5 
\epsilon}\Bigr) \times \Bigl( \frac{\hbar G H^2}{c^5} \Bigr) \; . 
\label{bottomestimate}
\end{eqnarray}

\begin{figure}[ht]
\vspace{-3cm}
\hspace{1cm} \includegraphics[width=10cm,height=5cm]{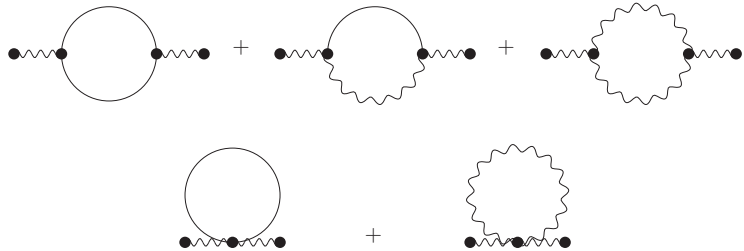} 
\vspace{1cm}
\caption{\label{gravloops} \tiny 
One loop corrections to the tensor power spectrum. Straight lines 
represent the $\zeta$ propagator while the graviton propagator is 
wavy.}
\end{figure}

The one loop corrections to $\Delta^2_h(k)$ are depicted in Figure
\ref{gravloops}. The same rules suffice to estimate the strengths
of these corrections, although one must recall that the tree order
result (\ref{treepower}) has no $\epsilon$ enhancement. For example,
the leftmost diagram on the first line of Figure \ref{gravloops} 
contributes,
\begin{equation}
\Bigl( \frac{\hbar G H^2}{c^5 \epsilon} \Bigr)^2 \times 
\Bigl( \frac{\hbar G H^2}{c^5} \Bigr)^2 \times \Bigl( \frac{c^5 
\epsilon}{\hbar G H^2} \Bigr)^2 = \Bigl( \frac{\hbar G H^2}{c^5}
\Bigr) \times \Bigl( \frac{\hbar G H^2}{c^5} \Bigr) \; . 
\label{lefth}
\end{equation}
None of the one loop corrections to $\Delta^2_h(k)$ is any stronger 
than (\ref{lefth}); the central diagram on the first line is 
actually suppressed by an additional factor of $\epsilon$. We 
therefore conclude that one loop corrections to each of the power
spectra are generically suppressed from the tree results 
(\ref{treepower}) by a factor of $\hbar G H^2/c^5 \sim 10^{-10}$. 

In these estimates it will be noted that I have not specified {\it 
when} the various factors of $H(t)$ and $\epsilon(t)$ are evaluated. 
Both quantities are thought to be nearly constant during much of 
primordial inflation --- in which case, it does not matter much 
when they are evaluated. However, it is well to recall that the 
actual loop corrections are integrals of sometimes differentiated 
propagators, like expression (\ref{leftgraph}). Weinberg noted the 
possibility for these integrations to grow with the co-moving time
\cite{Weinberg:2005vy}. That this can happen is associated with 
infrared divergences of the $\zeta$ and $h_{ij}$ propagators
which are evident from the small $k$ limiting form (\ref{smallk})
of both mode functions for constant $\epsilon$ \cite{Ford:1977in}.
This leads to two sources of possible secular growth:
\begin{itemize}
\item{A coincident propagator --- such as the four diagrams on
the bottom lines of Figures \ref{zetaloops} and \ref{gravloops}
--- grows like $\hbar G H^2/c^5 \times H t$ 
\cite{Vilenkin:1982wt,Linde:1982uu,Starobinsky:1982ee}; and}
\item{When the analogous in-out expression would be infrared 
divergent, a Schwinger-Keldysh integration such as (\ref{SKint}) 
is finite but grows with time \cite{Tsamis:1994ca}.}
\end{itemize}
The physical origin of both effects is that even the long wave 
length parts of the $\zeta$ and $h_{ij}$ effective actions are 
affected by the on-going process of inflationary particle 
production.

Weinberg proved an important theorem which limits the growth of
loop corrections to the primordial power spectra for the 
single-scalar model (\ref{GR+phi}) plus an arbitrary number of 
free scalars which are minimally coupled to gravity 
\cite{Weinberg:2006ac}. His result is that the largest possible 
secular enhancement to the depressingly small estimates 
(\ref{bottomestimate}) and (\ref{lefth}) consists of powers of 
the number of inflationary e-foldings. His student Bua 
Chaicherdsakul extended the result to cover fermions and gauge 
particles \cite{Chaicherdsakul:2006ui}. However, the situation 
changes radically if one allows matter couplings to the inflaton 
because the resulting Coleman-Weinberg corrections to its 
effective potential can induce important changes in the 
expansion history. For example, if an $m^2 \varphi^2$ inflaton 
were coupled to a massless fermion the resulting negative energy 
Coleman-Weinberg correction would cause the universe to end in a 
Big Rip singularity \cite{Miao:2005am}. Because the gauge 
(\ref{G0}) forces the inflaton to agree with its classical 
trajectory, changes in the physical expansion history manifest 
in secular growth of $\zeta(t,\vec{x})$ correlators. It should 
also be noted that Weinberg's theorem is limited to the 
inflationary power spectra. Explicit computations show that 
loop corrections to other correlators such as the vacuum 
polarization \cite{Prokopec:2002uw} and the fermion self-energy 
\cite{Prokopec:2003qd} can grow like powers of the inflationary
scale factor. 

\subsection{Nonlinear Extensions}\label{nonlinear}

No one disputes Weinberg's bound, but some cosmologists disagree
that there can be any secular corrections. Weinberg gave two 
examples \cite{Weinberg:2005vy}, which other authors confirmed 
\cite{Adshead:2008gk}. However, Senatore and Zaldarriaga 
identified a problem with the use of dimensional regularization 
in one of these examples, and went on to argue that no secular 
enhancements are possible under any circumstances 
\cite{Senatore:2009cf}. It seems very clear that models can be
devised for which quantum corrections to the {\it naive} 
correlators (\ref{Deltaz}-\ref{Deltah}) grow with time like 
powers of the number of e-foldings, just as Weinberg stated 
\cite{Kahya:2010xh}. Close examination of claims to the 
contrary \cite{Senatore:2012nq,Senatore:2012wy,Pimentel:2012tw}
reveals that the authors are not actually disputing this, but 
rather arguing that the naive correlators 
(\ref{Deltaz}-\ref{Deltah}) should be replaced with other
theoretical quantities which fail to show secular growth.
That brings up the fascinating and crucial issue of what 
operators represent the measured power spectra.

The problem with trying to overcome the loop suppression 
through secular enhancements is that the growth begins at 
first horizon crossing and terminates with the end of inflation. 
But observable modes experienced first horizon crossing at most 
50 e-foldings before the end of inflation, which means the 
enhancement can be at most some small power of 50.
The issue which focussed people's attention on modifying 
the naive observables was not secular growth but rather the 
closely associated problem of sensitivity to the infrared 
cutoff. Ford and Parker showed in 1977 that the propagator 
of a massless, minimally coupled scalar has an infrared 
divergence for any constant $\epsilon(t)$ cosmology in the 
range $0 \leq \epsilon \leq \frac32$ \cite{Ford:1977in}.
In view of relations (\ref{constv}-\ref{constu}) this same 
problem afflicts both the $\zeta$ and $h_{ij}$ propagators. 
Like all infrared divergences, this one derives from posing 
an unphysical question. The problem in this case is 
arranging large correlations for super-horizon modes which 
no local observer can control. There are two fixes which 
have been suggested:
\begin{itemize}
\item{Either arrange for the initially super-horizon modes
to be in some less highly correlated state 
\cite{Vilenkin:1983xp}; or else}
\item{Work on a spatially compact manifold such as $T^{D-1}$ 
whose coordinate radius is such that there are no initially
super-horizon modes \cite{Tsamis:1993ub}.}
\end{itemize}
In practice each fix amounts to cutting off the Fourier mode
sum at some minimum value $k = L^{-1}$. If infrared 
divergences could be shown to afflict loop corrections to 
the power spectra, and if the cutoff $L$ were large enough, 
then loop corrections might be significant.

I recommend the review article by Seery on
infrared loop corrections to inflationary correlators
\cite{Seery:2010kh}. Important work was done by a number
of authors \cite{Sloth:2006az,Sloth:2006nu, Bilandzic:2007nb,
vanderMeulen:2007ah,Lyth:2007jh,Seery:2007we,Seery:2007wf,
Bartolo:2007ti,Urakawa:2008rb,Riotto:2008mv,Enqvist:2008kt,
Seery:2009hs,Kumar:2009ge,Burgess:2009bs}. In 2010 Giddings
and Sloth were able to give a convincing argument that 
graviton loop corrections to $\Delta^2_{\mathcal{R}}(k)$
are indeed sensitive to the infrared cutoff $L$
\cite{Giddings:2010nc,Giddings:2010ui}, and hence able to
make significant corrections. This disturbed people who think 
about gauge invariance in gravity because the actual infrared
{\it divergence} --- as opposed to the closely associated 
secular growth factor --- is a constant in space and time,
and a constant field configuration $h_{ij}(t,\vec{x})$ ought 
to be gauge equivalent to zero. Even before the work of
Giddings and Sloth the possibility of such corrections had
prompted Urakawa and Tanaka to argue for modifying the 
original definition (\ref{Deltaz}) so that the spatial 
argument of the first field $\zeta(t,\vec{x})$ is replaced 
by the metric-dependent geodesic $\vec{X}[g](\vec{x})$ 
which is a constant invariant length $\Vert \vec{x} \Vert$ 
from the other point $\vec{0}$ in the direction $\vec{x}$ 
\cite{Urakawa:2009my,Urakawa:2009gb},
\begin{equation}
\Delta^2_{\mathcal{R}}(k,t) \longrightarrow \frac{k^3}{2 \pi^2} 
\int \!\! d^3x \, e^{-i \vec{k} \cdot \vec{x}} \Bigl\langle 
\Omega \Bigl\vert \zeta\Bigl(t,\vec{X}[g](\vec{x}\Bigr) 
\zeta(t,\vec{0}) \Bigr\vert \Omega \Bigr\rangle \; .
\label{newaltz}
\end{equation}
After the paper by Giddings and Sloth it was quickly established 
that these sorts of partially invariant observables are free 
of the infrared divergence \cite{Byrnes:2010yc,Urakawa:2010it,
Urakawa:2010kr,Gerstenlauer:2011ti,Tanaka:2011aj,Chialva:2011bg,
Urakawa:2011fg}. In subsequent work Giddings and Sloth have
sought to identify invariant observables which still show the
enhancement \cite{Giddings:2011zd,Giddings:2011ze}.
Tanaka and Urakawa have also continued their work on the problem,
\cite{Tanaka:2012wi,Tanaka:2013xe,Tanaka:2013caa,Tanaka:2014ina}

The discussion of infrared effects attracted me because I had for 
years been working on these in de Sitter background. I was also 
fascinated by the struggle to identify physical observables in 
quantum gravity because my long-neglected doctoral thesis dealt 
with that very subject \cite{Tsamis:1989yu}. In fact had I 
considered corrections involving precisely the same sort of 
geodesics as in (\ref{newaltz})! One thing I discovered is that they 
introduce new ultraviolet divergences associated with integrating 
graviton fields over the 1-dimensional background geodesic 
\cite{Tsamis:1989yu}. These new divergences change the power 
spectrum into the expectation value of a nonlocal composite 
operator which no one currently understands how to renormalize 
\cite{Miao:2012xc}. Shun-Pei Miao and I also demonstrated that 
(\ref{newaltz}) disturbs the careful pattern of $\epsilon$-suppression
which we saw section \ref{epssup}; one loop corrections to 
(\ref{newaltz}) go like the tree order result (\ref{treepower}) 
times $\hbar G H^2/c^5 \epsilon$ \cite{Miao:2012xc}. For the very
same reason, non-Gaussianity would also be unsuppressed
\cite{Miao:2012xc}. So changing what theoretical quantity we
identify with the scalar power spectrum from (\ref{Deltaz}) to 
(\ref{newaltz}) in order to avoid sensitivity on the infrared cutoff 
would come at the high price of introducing uncontrollable 
ultraviolet divergences and observable non-Gaussianity. It seems a 
bad bargain, and I mean no disrespect to colleagues who are 
struggling to puzzle out the truth, as am I, when I say we must do 
better.

It seems clear to me that we need new ideas. One radical and 
thought-provoking proposal is the suggestion by Miao and Park to 
abandon correlators altogether and instead quantum-correct the 
mode function relations (\ref{Deltav}-\ref{Deltau}) 
\cite{Miao:2013oko}. Among other things, this would avoid the new 
ultraviolet divergence which Fr\"ob, Roura and Verdaguer have 
found in one loop corrections to the tensor power spectrum because 
the two times coincide \cite{Frob:2012ui}.

It also seems to me that too few physicists appreciate the wondrous 
opportunity which has befallen us to shape a new discipline
by defining its observables. The debate on this vital subject is 
sometimes confused, and too often degenerates into shouting matches. 
In an effort to clarify matters Shun-Pei Miao and I laid out ten 
principles which are worth repeating here \cite{Miao:2012xc}:
\begin{enumerate}
\item{IR divergence differs from IR growth;}
\item{The leading IR logs might be gauge independent;}
\item{Not all gauge dependent quantities are unphysical;}
\item{Not all gauge invariant quantities are physical;}
\item{Nonlocal ``observables'' can null real effects;}
\item{Renormalization is crucial and unresolved;}
\item{Extensions involving $\zeta$ must be $\epsilon$-suppressed;}
\item{It is important to acknowledge approximations;}
\item{Sub-horizon modes cannot have large IR logs; and}
\item{Spatially constant quantities are observable.}
\end{enumerate}

\subsection{The Promise of 21cm Radiation}\label{21cm}

Particle physicists are familiar with the saying, ``yesterday's
discovery is tomorrow's background.'' Cosmologists are today 
witnessing the final stages of this process in the context of 
observations of the cosmic microwave background, as interlocking 
developments in technology and understanding of astrophysical 
processes have permitted fundamental theory to be probed more and
more deeply. A brief survey of the history is instructive:
\begin{itemize}
\item{{\bf 1964} --- discovery of the monopole, for which Penzias
and Wilson received the 1978 Nobel Prize;}
\item{{\bf 1970's} --- discovery of the dipole, which gives the
Earth's motion relative to the CMB;}
\item{{\bf 1992} --- discovery of lowest higher multipoles in the
temperature-temperature correlator by COBE, for which Mather and
Smoot received the 2006 Nobel Prize;}
\item{{\bf 1999} --- detection of the first Doppler peak by 
BOOMERanG and MAXIMA, supporting inflation and not cosmic
strings as the primary source of structure formation;}
\item{{\bf early 2000's} --- detection of $E$-mode polarization 
by DASI and CBI, and demonstration by WMAP of the $T$-$E$ 
anti-correlation predicted by inflation;}
\item{{\bf 2003-2010} --- full sky maps of temperature and $E$-mode
correlators by WMAP, and their use for precision determinations
of cosmological parameters;}
\item{{\bf 2013} --- full sky map of Planck resolves seven Doppler
peaks and give tighter bounds on $\Lambda$CDM parameters;}
\item{{\bf 2013} --- First detection of $B$-mode polarization from
gravitational lensing by the South Pole Telescope;}
\item{{\bf 2014} --- Detection of primordial $B$-mode polarization
claimed by BICEP2, confirming another key prediction of primordial 
inflation, fixing the inflationary energy scale to be $\sim 2 \times 
10^{16}~{\rm GeV}$, and incidentally establishing the existence and 
quantization of gravitons; and}
\item{{\bf 2014} --- Resolution of six acoustic peaks of $E$-mode
polarization by the Atacama Cosmology Telescope Polarimeter, which
provides an independent determination of $\Lambda$CDM parameters.}
\end{itemize}

The first steps are even now being taken in what could be an equally 
fruitful evolution, whose full realization will consume decades as it 
yields a steady series of discoveries. I refer to the project of 
surveying large volumes of the Universe using the 21 cm line 
\cite{Furlanetto:2006jb}. The discovery potential is obvious from the 
comparison between an x-ray and a CT-scan: {\it all that has been 
learned from the cosmic microwave background derives from the surface 
of last scattering, whereas 21 cm radiation allows us to make a 
tomograph of the universe.} 

Current and planned projects probe two regimes of cosmic redshift:
\begin{itemize}
\item{$0 \ltwid z \ltwid 4$ --- in which the radiation from unresolved 
galaxies is observed to probe baryon acoustic oscillations 
\cite{Pober:2012zz,Ansari:2012vn,Battye:2012tg,CHIME,Ali:2013rfa,
Chen:2012xu}.}
\item{$6 \ltwid z \ltwid 10$ --- in which intergalactic Hydrogen is 
observed to probe the epoch of reionization \cite{Pober:2013jna,
vanHaarlem:2013dsa,Lonsdale:2009cb,Zheng:2013tpz,Parsons:2009in}.}
\end{itemize}
The first of these provides important information for understanding the 
mysterious physics which is causing the current universe to accelerate, 
the discovery of which earned Perlmutter, Schmidt and Riess the 2011 
Nobel Prize. The second is crucial to understanding the first generation
of stars, and will eventually be an important foreground in future 
observations.

As technology and engineering improve, and as astrophysical effects 
are better understood, it is possible to foresee a time (decades from 
now) when redshifts as high as $z \sim 50$ are observed to measure the 
matter power spectrum with staggering accuracy. There is enough 
potentially recoverable data in the 21 cm radiation to resolve one 
loop corrections \cite{Masui:2010cz}. Current measurements of 
$\Delta^2_{\mathcal{R}}(k)$ do not test fundamental theory because we 
lack a compelling mechanism for driving primordial inflation, but that 
is bound to change over the decades required for the full maturity of 
21 cm cosmology. And when we do understand the driving mechanism, it 
will be possible to untangle the one loop correction from the tree 
order effect, which will test quantum gravity. {\bf This could be for 
quantum gravity what the measurement of} $g-2$ {\bf was for quantum 
electrodynamics.} The data is there, and people will be working for 
decades to harvest it.

\section{Other Quantum Gravitational Effects}\label{other}

As I have explained, the driving force for quantum gravitational 
effects during inflation is the production of nearly massless,
minimally coupled scalars (if there are any) and gravitons. The
presence of these particles is quantified by the scalar and tensor
power spectra. Because Einstein + anything is an interacting 
quantum field theory, the newly created particles must interact, at
some level, both with themselves and with other particles. This 
section describes how to study those interactions. I first list the 
various linearized effective field equations, then I describe the 
propagators and how to represent the tensor structure of the 
associated one-particle-irreducible (1PI) 2-point functions. The 
section closes with a review of results and open problems. However, 
the issue of back-reaction is so convulted and contentious that it 
merits its own subsection.

\subsection{Linearized Effective Field Equations}\label{lineqns}

We want to study how the propagation of a single particle is
affected by the vast sea of infrared gravitons and scalars 
produced by inflation. That can be done by computing the 
1PI 2-point function of the particle in question and then using it 
to quantum-correct the linearized effective field equation. Recall 
from (\ref{loopcount}) that quantum gravitational loop corrections 
from inflationary particle production are suppressed by $\hbar G 
H_i^2/c^5 \sim 10^{-10}$. Because this number is so small it is 
seldom necessary include nonlinear effects or to go beyond one 
loop order. The usual unit conventions of relativistic quantum field 
theory apply in which time is measured so that $c = 1$, and mass is 
measured so that $\hbar = 1$. The loop-counting parameter of quantum 
gravity is $\kappa^2 \equiv 16 \pi G (\times \hbar/c^3) \approx 1.3 
\times 10^{-69}~{\rm m}^2$.
 
Because $\epsilon_i \approx 0.013$ is so small, most work is done
on de Sitter background, for which $\epsilon(t) = 0$ and $H(t)$ is
a constant. Computations are done on a portion of the full de 
Sitter manifold which is termed ``the cosmological patch'' 
in the recent literature, and sometimes ``open conformal 
coordinates'' in the older literature,
\begin{equation}
ds^2 = a^2(\eta) \Bigl[-d\eta^2 + d\vec{x} \!\cdot\! d\vec{x}
\Bigr] \qquad , \qquad a(\eta) \equiv -\frac1{H \eta} = e^{H t}
\; . \label{deSitter}
\end{equation}
The $D-1$ spatial coordinates exist in the same range $-\infty < 
x^i < +\infty$ as Minkowski space, but the conformal time $\eta$
is limited to the range $-\infty < \eta \leq 0$. The de Sitter
metric is $g_{\mu\nu} = a^2 \eta_{\mu\nu}$, where $\eta_{\mu\nu}$
is the Minkowski metric. In contrast to section \ref{loop}, metric
fluctuations are characterized by the conformally rescaled and 
canonically normalized graviton field $h_{\mu\nu}$,
\begin{equation}
g^{\rm full}_{\mu\nu}(x) \equiv a^2 \Bigl[\eta_{\mu\nu} + 
\kappa h_{\mu\nu}(x) \Bigr] \; . \label{gravitondef}
\end{equation}
Graviton indices are raised and lowered with the Minkowski metric,
$h^{\mu}_{~\nu} \equiv \eta^{\mu\rho} h_{\rho\nu}$, $h^{\mu\nu} 
\equiv \eta^{\mu\rho} \eta^{\nu\sigma} h_{\rho\sigma}$, $h \equiv
\eta^{\mu\nu} h_{\mu\nu}$. Fermion fields are also conformally 
rescaled,
\begin{equation}
\psi^{\rm full}_i(x) \equiv a^{\frac{D-1}2} \times \psi_i(x) \; .
\end{equation}

The various 1PI 2-point functions are evaluated using dimensional
regularization, then fully renormalized with the appropriate 
counterterms in the sense of Bogoliubov, Parasiuk 
\cite{Bogoliubov:1957gp}, Hepp \cite{Hepp:1966eg} and Zimmermann
\cite{Zimmermann:1968mu,Zimmermann:1969jj} (BPHZ). After this the
unregulated limit of $D=4$ is taken. As explained in section 
\ref{howto}, quantum corrections to the in-out effective field 
equations at spacetime point $x^{\mu}$ are dominated by 
contributions from points ${x'}^{\mu}$ in the infinite future when 
the 3-volume has been expanded to infinity. Quantum corrections to 
the in-out matrix elements of field operators are also generally 
complex, even for real fields. These results are correct for 
in-out scattering theory, but they have no physical relevance for 
cosmology where the appropriate question is what happens to the 
expectation value of the field operator in the presence of a 
prepared state which is released at some finite time. One solves 
that sort of problem using the Schwinger-Keldysh formalism 
\cite{Chou:1984es,Jordan:1986ug,Calzetta:1986ey,Calzetta:1986cq}. 
In this technique each of the 1PI $N$-point functions of the 
in-out formalism gives rise to $2^N$ Schwinger-Keldysh $N$-point 
functions. It is the sum of the $++$ and $+-$ 1PI 2-point 
functions which appears in the linearized Schwinger-Keldysh 
effective field equation. This combination is both real and 
causal.
 
The 1PI 2-point function for a scalar is known as its
``self-mass-squared'', $-i M^2(x;x')$. The quantum-corrected,
linearized field equation for a minimally coupled scalar with
mass $m$ is,
\begin{equation}
a^2 \Bigl(-\partial^2_0 - 2 H a \partial_0 + \nabla^2\Bigr)
\varphi(x) - m^2 a^4 \varphi(x) - \int \!\! d^4x' M^2(x;x') 
\varphi(x') = 0 \; . \label{scalareqn}
\end{equation}
The (conformally rescaled) fermion's 1PI 2-point function is called 
its ``self-energy'', $-i [\mbox{}_i \Sigma_j](x;x')$. If 
$\gamma^{\mu}_{ij}$ stands for the usual gamma matrices then the 
quantum-corrected, linearized field equation for a fermion with
mass $m$ is,
\begin{equation}
i \gamma^{\mu}_{ij} \partial_{\mu} \psi_j(x) - m a \psi_i(x) - 
\int \!\! d^4x' \Bigl[ \mbox{}_i \Sigma_j\Bigr](x;x') \psi_j(x') 
= 0 \; . \label{spinoreqn}
\end{equation}
The 1PI 2-point function for a photon has the evocative name ``vacuum 
polarization'', $+i [\mbox{}^{\mu} \Pi^{\nu}](x;x')$. If $F^{\nu\mu}
\equiv \eta^{\nu\rho} \eta^{\mu\sigma} (\partial_{\rho} A_{\sigma} -
\partial_{\sigma} A_{\rho})$ is the usual field strength tensor then
the quantum-corrected Maxwell equation can be written,
\begin{equation}
\partial_{\nu} F^{\nu\mu}(x) + \int \!\! d^4x' \Bigl[\mbox{}^{\mu}
\Pi^{\nu} \Bigr](x;x') A_{\nu}(x') = J^{\nu}(x) \; , \label{vectoreqn}
\end{equation}
where $J^{\mu}(x)$ is the current density. Finally, the 1PI 2-point 
function for a (conformally rescaled and canonically normalized) 
graviton is termed the ``graviton self-energy'', $-i[\mbox{}^{\mu\nu} 
\Sigma^{\rho\sigma}](x;x')$. It is used to quantum correct the 
linearized Einstein equation as,
\begin{equation}
\partial_{\alpha} \Bigl[a^2 \mathcal{L}^{\mu\nu\alpha\beta\rho\sigma} 
\partial_{\beta} h_{\rho\sigma}(x)\Bigr] - \int \!\! d^4x' \Bigl[ 
\mbox{}^{\mu\nu} \Sigma^{\rho\sigma}\Bigr](x;x') h_{\rho\sigma}(x') = 
-\frac{\kappa a^2}{2} \eta^{\mu\rho} \eta^{\nu\sigma} 
T_{\rho\sigma}(x) \; , \label{tensoreqn}
\end{equation}
where $T_{\rho\sigma}$ is the linearized stress tensor and I define,
\begin{eqnarray}
\lefteqn{\mathcal{L}^{\mu\nu\rho\sigma\alpha\beta} \equiv \frac12
\eta^{\alpha\beta} \Bigl[ \eta^{\mu (\rho} \eta^{\sigma) \nu} \!-\!
\eta^{\mu\nu} \eta^{\rho\sigma} \Bigr] } \nonumber \\
& & \hspace{3.5cm} + \frac12 \eta^{\mu\nu} \eta^{\rho (\alpha}
\eta^{\beta) \sigma} + \frac12 \eta^{\rho\sigma} \eta^{\mu (\alpha}
\eta^{\beta) \nu} - \eta^{\alpha) (\rho} \eta^{\sigma) (\mu}
\eta^{\nu) (\beta} \; . \qquad
\end{eqnarray}

The scalar equation (\ref{scalareqn}) and its spinor counterpart
(\ref{spinoreqn}) of course describe the propagation of scalars and
spinors, respectively. The vector and tensor equations 
(\ref{vectoreqn}-\ref{tensoreqn}) can be similarly used to study 
the propagation of dynamical photons and gravitons, but they also 
describe modifications of the electrodynamic and gravitational 
forces. Dynamical quanta show no modification in flat space quantum
field theory (enforcing that is what typically fixes the field
strength renormalization) so it is at least possible that nothing
happens as well during inflation. However, the force laws are 
guaranteed to show an effect during inflation because they do so
in flat space background \cite{BjerrumBohr:2002sx,
Donoghue:1993eb,Donoghue:1994dn}.

\subsection{Propagators and Tensor 1PI Functions}\label{props}

The symmetries of the general cosmological geometry (\ref{FLRW}) 
are homogeneity and isotropy. However, the de Sitter limit of 
$\epsilon(t) = 0$ results in the appearance of two additional 
symmetries. Although it obvious to cosmologists that these extra
symmetries can be at best approximate for inflationary cosmology,
they have exerted a powerful influence on mathematical physicists
owing to the expectation that the full de Sitter group should 
play the same role in organizing and simplifying quantum field 
theory on de Sitter that Poincar\'e invariance has played for 
flat space. That expectation has remained unfulfilled owing to 
the time dependence intrinsic to inflationary particle production.

In our $D$-dimensional conformal coordinate system 
(\ref{deSitter}) the $\frac{1}{2}D(D+1)$ de Sitter 
transformations can be decomposed as follows:
\begin{itemize}
\item{$(D-1)$ spatial tranlations,}
\begin{equation}
\eta' = \eta\;, ~~ x'^{i} = x^{i} + \epsilon^{i} \;. \label{sym1}
\end{equation}
\item{$\frac{1}{2}(D-1)(D-2)$ spatial rotations,}
\begin{equation}
\eta' = \eta\;, ~~ x'^{i} = R^{ij}x^{j}\;. \label{sym2}
\end{equation}
\item{One dilatation,}
\begin{equation}
\eta' = k\eta\;, ~~ x'^{i} = kx^{j}\;. \label{sym3}
\end{equation}
\item{$(D-1)$ spatial special conformal transformations,}
\begin{equation}
\eta' = \frac{\eta}{1 - 2\vec{\theta}\cdot\vec{x}
+ \parallel\! \vec{\theta} \!\parallel^2 x\cdot x}\;, ~~ x' =
\frac{x^{i} - \theta^{i}x\cdot x}{1 - 2\vec{\theta}\cdot\vec{x} +
\parallel\! \vec{\theta} \!\parallel^2 x\cdot x}\;. \label{sym4}
\end{equation}
\end{itemize}
Homogeneity is (\ref{sym1}) and isotropy is (\ref{sym2}). The
two additional symmetries which appear in the de Sitter limit of
$\epsilon(t) = 0$ are (\ref{sym3}-\ref{sym4}).

Although infrared divergences induce de Sitter breaking, 
they do so in a limited way that leaves the largest part of the 
result de Sitter invariant. For dimensional regularization 
computations it is best to express this de Sitter invariant part 
in terms of the length function $y(x;z)$,
\begin{equation}
y(x;x') \equiv a a' H^2 \Bigl[ \Bigl\Vert \vec{x} \!-\! \vec{x}'
\Bigr\Vert^2 - \Bigl(\vert \eta \!-\! \eta'\vert \!-\! i
\varepsilon\Bigr)^2 \Bigr]\; . \label{ydef}
\end{equation}
Except for the factor of $i\varepsilon$ (whose purpose is to 
enforce Feynman boundary conditions) the function $y(x;x)$ is 
related to the invariant length $\ell(x;x')$ from 
$x^{\mu}$ to ${x'}^{\mu}$,
\begin{equation}
y(x;x') = 4 \sin^2\Bigl( \frac12 H \ell(x;x')\Bigr) \; .
\end{equation}
The four Schwinger-Keldysh polarity variations 
(\ref{++}-\ref{--}) never affect the de Sitter breaking terms.
They can all be obtained by making simple changes of the 
$i\varepsilon$ term in (\ref{ydef}),
\begin{eqnarray}
y_{\scriptscriptstyle ++}(x;x') & = & a a' H^2 \Bigl[ 
\Bigl\Vert \vec{x} \!-\! \vec{x}' \Bigr\Vert^2 - \Bigl(
\vert \eta \!-\! \eta'\vert \!-\! i\varepsilon\Bigr)^2
\Bigr] \; , \label{y++} \\
y_{\scriptscriptstyle -+}(x;x') & = & a a' H^2 \Bigl[ 
\Bigl\Vert \vec{x} \!-\! \vec{x}' \Bigr\Vert^2 - \Bigl(
\eta \!-\! \eta' \!-\! i\varepsilon\Bigr)^2 \Bigr] \; , 
\label{y-+} \\
y_{\scriptscriptstyle +-}(x;x') & = & a a' H^2 \Bigl[ 
\Bigl\Vert \vec{x} \!-\! \vec{x}' \Bigr\Vert^2 - \Bigl(
\eta \!-\! \eta' \!+\! i\varepsilon\Bigr)^2 \Bigr] \; , 
\label{y+-} \\
y_{\scriptscriptstyle --}(x;x') & = & a a' H^2 \Bigl[ 
\Bigl\Vert \vec{x} \!-\! \vec{x}' \Bigr\Vert^2 - \Bigl(
\vert \eta \!-\! \eta'\vert \!+\! i\varepsilon\Bigr)^2
\Bigr] \; . \label{y--}
\end{eqnarray}

The best way of expressing higher spin propagators on de Sitter,
as on flat space, is by acting differential operators on scalar
propagators. I work with a general scalar propagator 
$i\Delta_b(x;x')$ which obeys the equation,
\begin{equation}
\Bigl[ \square + (b^2 \!-\! b_A^2) H^2 \Bigr] i\Delta_b(x;x') =
\frac{i \delta^D(x \!-\! x')}{\sqrt{-g}} \; .
\end{equation}
Here and henceforth the index $b_A$ is $b_A \equiv \frac{D-1}2$
and $\square$ stands for the covariant scalar d'Alembertian,
\begin{equation}
\square \equiv \frac1{\sqrt{-g}} \, \partial_{\mu} \Bigl( \sqrt{-g}
\, g^{\mu\nu} \partial_{\nu} \Bigr) = \frac1{a^2} \Bigl( -
\partial_0^2 - (D\!-\! 2) H a \partial_0 + \nabla^2 \Bigr) \; .
\label{square}
\end{equation}
For the case of $b < b_A$ the propagator has a positive 
mass-squared $m^2 = (b_A^2 - b^2) H^2$ and its propagator is a de 
Sitter invariant Hypergeometric function of $y(x;x')$. Its expansion
for $b = \nu$ is,
\begin{eqnarray}
\lefteqn{i\Delta^{\rm dS}_{\nu}(x;x') = \frac{H^{D-2}}{(4
\pi)^{\frac{D}2}} \Biggl\{ \Gamma\Bigl(\frac{D}2 \!-\! 1\Bigr)
\Bigl(\frac{4}{y}\Bigr)^{\frac{D}2-1} - \frac{\Gamma(\frac{D}2)
\Gamma(1 \!-\! \frac{D}2)}{ \Gamma(\frac12 \!+\! \nu) \Gamma(\frac12
\!-\! \nu)} \sum_{n=0}^{\infty} } \nonumber \\
& & \hspace{-.7cm} \times \Biggl[ \frac{\Gamma(\frac32 \!+\! \nu
\!+\! n) \Gamma(\frac32 \!-\! \nu \!+\! n)}{ \Gamma(3 \!-\!
\frac{D}2 \!+\! n) \, (n \!+\! 1)!} \Bigl(\frac{y}4 \Bigr)^{n -
\frac{D}2 +2} \!\!\!-\! \frac{\Gamma(b_A \!+\! \nu \!+\! n)
\Gamma(b_A \!-\! \nu \!+\! n)}{\Gamma(\frac{D}2 \!+\! n) \, n!}
\Bigl(\frac{y}4\Bigr)^n \Biggr] \! \Biggr\} . \qquad
\label{naivescal}
\end{eqnarray}

When $b \geq b_A$ the naive mode sum is infrared divergent for
the same reason as the problem discovered by Ford and Parker 
\cite{Ford:1977in} which I described in section \ref{nonlinear}. 
One sometimes encounters contrary statements in the mathematical 
physics literature \cite{Faizal:2011iv}, but close examination 
reveals that the authors admit they are constructing a formal
solution to the propagator equation which is not a true propagator
by the illegitimate technique of adding negative norm states to 
the theory \cite{Miao:2013isa}. The proper technique 
\cite{Vilenkin:1983xp,Tsamis:1993ub} of cutting off the mode sum 
amounts to adding to (\ref{naivescal}) a de Sitter breaking, 
infrared correction \cite{Allen:1987tz,Miao:2010vs},
\begin{eqnarray}
\lefteqn{\Delta^{\rm IR}_{\nu}(x;x') =
\frac{H^{D-2}}{(4\pi)^{\frac{D}2}} \frac{\Gamma(\nu)
\Gamma(2\nu)}{\Gamma(b_A)
\Gamma(\nu \!+\! \frac12)} \times \theta(\nu \!-\! b_A) } \nonumber \\
& & \hspace{.7cm} \times \sum_{N=0}^{[\nu - b_A]} \frac{(a a'
)^{\nu - b_A - N}}{\nu \!-\! b_A \!-\! N} \sum_{n=0}^N \Bigl(
\frac{a}{a'} \!+\! \frac{a'}{a}\Bigr)^n \sum_{m=0}^{
[\frac{N-n}2]} C_{Nnm} (y \!-\!2)^{N-n-2m} \; , \qquad
\label{series}
\end{eqnarray}
where the coefficients $C_{Nnm}$ are,
\begin{eqnarray}
\lefteqn{C_{Nnm} = \frac{(-\frac14)^N}{m! n! (N \!-\!n \!-\! 2m)!}
\times \frac{\Gamma(b_A \!+\! N \!+\! n \!-\!
\nu)}{\Gamma(b_A \!+\! N \!-\! \nu)} } \nonumber \\
& & \hspace{2cm} \times \frac{\Gamma(b_A)}{\Gamma(b_A \!+\! N\!-\!
2m)} \times \frac{\Gamma(1 \!-\!\nu)}{\Gamma(1 \!-\! \nu \!+\! n
\!+\! 2m)} \times \frac{\Gamma(1 \!-\! \nu)}{\Gamma(1 \!-\! \nu
\!+\! m)} \; . \qquad \label{cdef}
\end{eqnarray}
The full propagator is therefore,
\begin{equation}
i\Delta_b(x;x') = \lim_{\nu \rightarrow b} \Bigl[ i\Delta^{\rm
dS}_{\nu}(x;x') + \Delta^{\rm IR}_{\nu}(x;x') \Bigr] \; . 
\label{fullprop}
\end{equation}

It is often useful to discuss integrated propagators which obey,
\begin{equation}
\Bigl[ \square + (b^2 \!-\! b_A^2) H^2 \Bigr] i \Delta_{bc}(x;z) =
i\Delta_c(x;z) \; . \label{Int1}
\end{equation}
The solution is easily seen to be \cite{Miao:2010vs,Miao:2009hb},
\begin{equation}
i\Delta_{bc}(x;z) = \frac1{(b^2 \!-\! c^2) H^2} \Bigl[
i\Delta_c(x;z) \!-\! i\Delta_b(x;z)\Bigr] = i\Delta_{cb}(x;z) \; .
\label{Int2}
\end{equation}
I also employ a doubly integrated propagator which obeys,
\begin{equation}
\Bigl[ \square + (b^2 \!-\! b_A^2) H^2 \Bigr] i \Delta_{bcd}(x;z) =
i\Delta_{cd}(x;z) \; . \label{Int4}
\end{equation}
The solution can be written in a form which is manifestly symmetric
under any interchange of the three indices $a$, $b$ and $c$,
\begin{eqnarray}
i\Delta_{bcd}(x;z) &\!\! = \!\!& \frac{ i\Delta_{bd}(x;z) \!-\!
i\Delta_{bc}(x;z)}{(c^2 \!-\! d^2) H^2} \; , \qquad \label{Int5a} \\
& \!\!=\!\! & \frac{ (d^2 \!-\! c^2) i\Delta_{b}(x;z) \!+\! (b^2
\!-\! d^2) i\Delta_{c}(x;z) \!+\! (c^2 \!-\! b^2) i\Delta_{d}(x;z)
}{(b^2 \!-\! c^2) (c^2 \!-\! d^2) (d^2 \!-\! b^2) H^4} \; . \qquad
\label{Int5b}
\end{eqnarray}

The propagator for a (conformally rescaled) fermion of mass $m$ 
was constructed by Candelas and Raine \cite{Candelas:1975du}
in terms of the scalar propagator (\ref{naivescal}) with
$\nu = -\frac12 + i \frac{m}{H}$,
\begin{eqnarray}
\lefteqn{i \Bigl[\mbox{}_i S_j\Bigr](x;x') = \Bigl[ i \gamma^{\mu}
\partial_{\mu} + a m\Bigr] (a a')^{\frac{D}2 -1} } \nonumber \\
& & \hspace{3.5cm} \times \Biggl\{ \Bigl( \frac{I \!-\! \gamma^0}{2} 
\Bigr) i\Delta^{\rm dS}_{\nu}(x;x') + \Bigl( \frac{I \!+\! 
\gamma^0}{2}\Bigr) i \Delta^{\rm dS}_{\nu^*}(x;x') \Biggr\} . 
\label{fermprop} \qquad
\end{eqnarray}
Except for the conformal rescaling, expression (\ref{fermprop}) is
de Sitter invariant as long as the mass is real. Tachyonic fermions 
with $m^2 \leq -D H^2$ break de Sitter invariance the same way that 
tachyonic scalars do.

Vector and tensor fields raise the issue of gauge fixing. Normally
we would accomplish this by adding to the action gauge fixing terms 
which respect the coordinate isometries of the background, however, 
there is an obstacle to doing this on any background such as de 
Sitter which possesses a linearization instability \cite{Miao:2009hb}. 
This leaves two alternatives:
\begin{itemize}
\item{One either add a de Sitter breaking gauge fixing term which 
cannot be extended to the full de Sitter manifold; or}
\item{One can impose some gauge condition on the field operators.}
\end{itemize}
After some discussion of the problem I will give results for both 
alternatives.

The obstacle to adding invariant gauge fixing terms came as a surprise 
because one can actually derive them in flat space quantum field 
theory by starting from an exact gauge and making functional changes 
of variables \cite{Coleman:1974hr}. In 2009 Miao, Tsamis and I 
identified precisely where this procedure breaks down when a 
linearization instability is present \cite{Miao:2009hb} but one can 
see the physics problem quite simply by considering flat space 
electrodynamics on the manifold $T^3 \times R$. Because the spatial 
sections are compact, both sides of the spatially averaged, 
$\mu = 0$ Maxwell equation must vanish separately,
\begin{equation}
\partial_{\nu} F^{\nu\mu} = J^{\mu} \qquad \Longrightarrow \qquad
\int_{T^3} \!\! d^3x \, \partial_i F^{i0}(t,\vec{x}) = \int_{T^3} 
\!\! d^3x \, J^0(t,\vec{x}) = 0 \; .
\end{equation}
This zero charge constraint follows from the invariant field
equations so it must be true in any valid gauge. However, adding
a Poincar\'e invariant gauge fixing term result in a very different 
theory. The field equations of Feynman gauge are,
\begin{equation}
\Bigl[-\partial_t^2 + \nabla^2\Bigr] A^{\mu}(t,\vec{x}) =
J^{\mu}(t,\vec{x}) \; . \label{Feynman}
\end{equation}
These equations can be solved for any total charge so there cannot
be any question that the theory has been changed.

The use of covariant gauge fixing terms is so ingrained that some
people's first reaction to the obstacle on de Sitter is, ``let's 
just go ahead and do it anyway!'' As it happens, Emre Kahya and I
had stumbled upon what happens if that is done in 2005. We used two
different gauges to compute and fully renormalize the one loop 
self-mass-squared $-iM^2(x;x')$ for a charged, massless, minimally 
coupled scalar on de Sitter \cite{Kahya:2005kj,Kahya:2006ui}. 
There was no problem with a de Sitter-breaking gauge fixing term 
\cite{Woodard:2004ut} --- nor is there any problem 
\cite{Prokopec:2006ue} when Lorentz gauge is enforced as a strong 
operator condition \cite{Tsamis:2006gj} --- but we found on-shell 
singularities when using the de Sitter-Feynman gauge propagator 
\cite{Allen:1985wd}. The origin of these singularities seems to be 
that integrating the self-mass-squared against the scalar wave 
function measures the $A_0 J^0$ interaction of the particle with 
its own field, and one can see from equation (\ref{Feynman}) that 
the solution for $A_0(t,\vec{x})$ must grow like $Q t^2/2$ in 
Feynman gauge.

My favorite de Sitter-breaking gauge fixing term for electromagnetism 
is \cite{Woodard:2004ut},
\begin{equation}
\mathcal{L}_{GF} = -\frac12 a^{D-4} \Bigl(\eta^{\mu\nu} A_{\mu , \nu}
- (D \!-\!4) H a A_0\Bigr)^2 \; . \label{EM}
\end{equation}
Because space and time components are treated differently it is useful to
have an expression for the purely spatial part of the Minkowski metric,
\begin{equation}
\overline{\eta}_{\mu\nu} \equiv \eta_{\mu\nu} + \delta^0_{\mu} \delta^0_{\nu}
\; .
\end{equation}
In this gauge the photon propagator takes the form of a sum of constant tensor 
factors times scalar propagators,
\begin{equation}
i\Bigl[{}_{\mu} \Delta_{\nu}\Bigr](x;x') = \overline{\eta}_{\mu\nu} \, a a'
i\Delta_B(x;x') - \delta^0_{\mu} \delta^0_{\nu} \, a a' i\Delta_C(x;x') \; .
\label{photonnoncov}
\end{equation}
The B-type and $C$-type propagators are special cases of (\ref{naivescal})
with $\nu = (D-3)/2$ and $\nu = (D-5)/2$, respectively,
\begin{eqnarray}
\lefteqn{i \Delta_B(x;x') =  \frac{H^{D-2}}{(4\pi)^{\frac{D}2}} \Biggl\{
\Gamma\Bigl(\frac{D}2 \!-\! 1\Bigr) \Bigl(\frac4{y} \Bigr)^{\frac{D}2-1} }
\nonumber \\
& & \hspace{3cm} + \sum_{n=0}^{\infty} \Biggl[ \frac{\Gamma(n \!+\! 
\frac{D}2)}{\Gamma(n \!+\! 2)} \Bigl( \frac{y}4 \Bigr)^{n +2 - \frac{D}2} 
- \frac{\Gamma(n \!+\! D \!-\! 2)}{\Gamma(n \!+\! \frac{D}2)} 
\Bigl(\frac{y}4 \Bigr)^n \Biggr] \Biggr\} , \qquad \label{DeltaB} \\
\lefteqn{i \Delta_C(x;x') =  \frac{H^{D-2}}{(4\pi)^{\frac{D}2}} \Biggl\{
\Gamma\Bigl(\frac{D}2 \!-\! 1\Bigr) \Bigl(\frac4{y} \Bigr)^{\frac{D}2-1} }
\nonumber \\
& & \hspace{.3cm} - \sum_{n=0}^{\infty} \Biggl[ \frac{(n \!-\! \frac{D}2 
\!+\!  3) \Gamma(n \!+\! \frac{D}2 \!-\! 1)}{\Gamma(n \!+\! 2)} 
\Bigl(\frac{y}4 \Bigr)^{n +2 - \frac{D}2} \!\!\!- \frac{(n\!+\!1) \Gamma(n 
\!+\! D \!-\! 3)}{\Gamma(n \!+\! \frac{D}2)} \Bigl(\frac{y}4 \Bigr)^n
\Biggr] \Biggr\} . \qquad \label{DeltaC}
\end{eqnarray}
Note that the infinite sums in (\ref{DeltaB}) and (\ref{DeltaC}) vanish 
for $D=4$, which means they only need to be included when multiplied by
a divergence, and even then only the lowest terms of the sums are 
required. In fact the $B$-type and $C$ type propagators agree in $D=4$, 
and the photon propagator in this gauge is the same for $D=4$ as it is 
in flat space! Despite the noncovariant gauge, this propagator shows no
physical breaking of de Sitter invariance.

My favorite de Sitter breaking gauge fixing term is \cite{Tsamis:1992xa,
Woodard:2004ut},
\begin{equation}
\mathcal{L}_{GF} = -\frac12 a^{D-2} \eta^{\mu\nu} F_{\mu} F_{\nu} \; , \;
F_{\mu} \equiv \eta^{\rho\sigma} \Bigl(\psi_{\mu\rho , \sigma}
- \frac12 \psi_{\rho \sigma , \mu} + (D \!-\! 2) H a \psi_{\mu \rho}
\delta^0_{\sigma} \Bigr) . \label{GR}
\end{equation}
Note that it breaks (\ref{sym4}) but preserves (\ref{sym1}-\ref{sym3}),
just like its electromagnetic cousin (\ref{EM}). In this gauge the 
graviton propagator takes the form of a sum of three constant index 
factors times a scalar propagator,
\begin{equation}
i\Bigl[{}_{\mu\nu} \Delta_{\rho\sigma}\Bigr](x;x') = \sum_{I=A,B,C}
\Bigl[{}_{\mu\nu} T^I_{\rho\sigma}\Bigr] i\Delta_I(x;x') \; . \label{gprop}
\end{equation}
The constant index factors are,
\begin{eqnarray}
\Bigl[{}_{\mu\nu} T^A_{\rho\sigma}\Bigr] & = & 2 \, \overline{\eta}_{\mu (\rho}
\overline{\eta}_{\sigma) \nu} - \frac2{D\!-\! 3} \overline{\eta}_{\mu\nu}
\overline{\eta}_{\rho \sigma} \; , \label{TA} \\
\Bigl[{}_{\mu\nu} T^B_{\rho\sigma}\Bigr] & = & -4 \delta^0_{(\mu}
\overline{\eta}_{\nu) (\rho} \delta^0_{\sigma)} \; , \label{TB} \\
\Bigl[{}_{\mu\nu} T^C_{\rho\sigma}\Bigr] & = & \frac2{(D \!-\!2) (D \!-\!3)}
\Bigl[(D \!-\!3) \delta^0_{\mu} \delta^0_{\nu} + \overline{\eta}_{\mu\nu}\Bigr]
\Bigl[(D \!-\!3) \delta^0_{\rho} \delta^0_{\sigma} + \overline{\eta}_{\rho
\sigma}\Bigr] \; . \label{TC}
\end{eqnarray}
The de Sitter invariant $B$-type and $C$-type propagators appear as well 
in the photon propagator (\ref{photonnoncov}). The $A$-type propagator 
corresponds to (\ref{fullprop}) with $b = (D-1)/2$, and it is better known
as the propagator of a massless, minimally coupled scalar \cite{Onemli:2002hr,
Onemli:2004mb},
\begin{eqnarray}
\lefteqn{i \Delta_A(x;x') =  \frac{H^{D-2}}{(4\pi)^{\frac{D}2}} \Biggl\{
\frac{\Gamma(\frac{D}2)}{(\frac{D}2 \!-\! 1)} \Bigl(\frac4{y} \Bigr)^{
\frac{D}2-1} \!\!\! + \frac{\Gamma(\frac{D}2 \!+\! 1)}{(\frac{D}2 \!-\! 2)} 
\Bigl(\frac4{y} \Bigr)^{\frac{D}2 -2} \!\!\! + 
\frac{\Gamma(D \!-\! 1)}{\Gamma(\frac{D}2)} \, \ln(a a') } \nonumber \\
& & \hspace{.7cm} + A_1 - \sum_{n=1}^{\infty} \Biggl[
\frac{\Gamma(n \!+\!  \frac{D}2 \!+\! 1)}{(n \!-\! \frac{D}2 \!+\! 2) 
\Gamma(n \!+\! 2)} \Bigl(\frac{y}4 \Bigr)^{n +2 - \frac{D}2} \!\!\! - 
\frac{\Gamma(n \!+\! D \!-\! 1)}{n \, \Gamma(n \!+\! \frac{D}2)} 
\Bigl(\frac{y}4 \Bigr)^n \Biggr] \Biggr\} , \qquad \label{DeltaA}
\end{eqnarray}
where the constant $A_1$ is,
\begin{equation}
A_1 = \frac{\Gamma(D\!-\!1)}{\Gamma(\frac{D}2)} \Biggl\{
-\psi\Bigl(1 \!-\! \frac{D}2\Bigr) + \psi\Bigl(\frac{D\!-\!1}2\Bigr)
+ \psi(D \!-\!1) + \psi(1) \Biggr\} .
\end{equation}

The de Sitter breaking factor of $\ln(a a')$ on the first line of 
(\ref{DeltaA}) is intensely disturbing to mathematical physicists who 
insist that it must be an artefact of the de Sitter breaking gauge 
(\ref{GR}) \cite{Higuchi:2011vw}. However, there are several ways of 
seeing the gravitons show real de Sitter breaking \cite{Miao:2011ng}. 
The simplest and most physical is to note that linearized 
gravitons in transverse-traceless gauge obey the same equation as the 
massless, minimally coupled scalar \cite{Lifshitz:1945du,
Grishchuk:1974ny}, whose propagator is admitted to break de Sitter 
invariance \cite{Allen:1987tz}. In fact the breaking of de Sitter 
invariance is a consequence of the time dependence of inflationary 
particle production and is embedded in the primordial power spectra 
\cite{Miao:2013isa}. One should also note that the gauge fixing term 
(\ref{GR}) preserves dilatation invariance (\ref{sym3}) while the 
propagator (\ref{DeltaA}) breaks this symmetry. Finally, one can show 
that the propagator is not invariant even when naive de Sitter 
transformations are augmented to include the compensating gauge 
transformation needed to restore (\ref{GR}) \cite{Kleppe:1993fz}.

It is also possible to enforce gauge conditions --- even de Sitter
invariant ones --- as strong operator equations. When this is done,
the resulting propagators can no longer be expressed using constant
tensor factors. The most economical representation is to write them
as differential projection operators acting on scalar structure 
functions. The differential operators are constructed using the
covariant derivative operator $D_{\mu}$ based on the de Sitter
connection,
\begin{equation}
\Gamma^{\rho}_{~ \mu\nu} = a H \Bigl( \delta^{\rho}_{~ \mu} 
\delta^0_{~\nu} + \delta^{\rho}_{~\nu} \delta^0_{~\mu} - 
\eta^{0\rho} \eta_{\mu\nu} \Bigr) \; .
\end{equation}
We raise and lower indices using the de Sitter metric, $D^{\mu} 
\equiv g^{\mu\nu} D_{\nu}$.

The Lorentz gauge condition $D^{\mu} A_{\mu} = 0$ implies that 
the photon propagator obeys,
\begin{equation}
D^{\mu} i \Bigl[\mbox{}_{\mu} \Delta_{\nu}\Bigr](x;x') = 0 =
{D'}^{\nu} i \Bigl[\mbox{}_{\mu} \Delta_{\nu}\Bigr](x;x') \; .
\label{Lorentz}
\end{equation}
If the photon has mass $M_V$ (which can happen due to spontaneous
symmetry breaking) then its propagator obeys the equation,
\begin{equation}
\Bigl[ \square \!-\! (D \!-\! 1) H^2 \!-\! M_V^2 \Bigr] i
\Bigl[\mbox{}_{\mu} \Delta_{\rho}\Bigr](x;x') = \frac{i g_{\mu\rho}
\delta^D( x\!-\! x')}{\sqrt{-g}} + D_{\mu} {D'}_{\rho} 
i\Delta_A(x;x') \; . \label{transeqn}
\end{equation}
The term involving $i\Delta_A$ on the right hand side of 
(\ref{transeqn}) is required for consistency with the gauge 
condition (\ref{Lorentz}) \cite{Tsamis:2006gj} and was missing from 
an earlier analysis \cite{Allen:1985wd}. We can find the transverse
projector $\mathcal{P}_{\alpha\beta}^{~~ \mu}$ by writing the
electromagnetic field strength as $F_{\alpha\beta} \equiv 
\mathcal{P}_{\alpha\beta}^{~~ \mu} A_{\mu}$. The photon
propagator takes the form \cite{Miao:2011fc},
\begin{equation}
i\Bigl[\mbox{}_{\mu} \Delta_{\rho}\Bigr](x;x') =
\mathcal{P}^{\alpha\beta}_{~~\mu} \times
{\mathcal{P}'}^{\kappa\lambda}_{~~\rho} \times 
\frac{D_{\alpha} D'_{\kappa}}{2 H^2} \times \Bigl[ 
\mathcal{S}_1(x;x') \frac{D_{\beta} D'_{\lambda} y}{2 H^2} 
\Bigr] \; . \label{Lorentzphoton}
\end{equation}
The vector structure function can be expressed using a
singly integrated propagator (\ref{Int2}) \cite{Miao:2011fc},
\begin{equation}
\mathcal{S}_1 = + \frac{2 H^2}{M_V^2} i\Delta_{BB} + \frac{2
H^2}{M_V^4} \Bigl[ i\Delta_B - i\Delta_c\Bigr] \quad {\rm where}
\quad c = \sqrt{ \Bigl( \frac{D \!-\! 3}2\Bigr)^2 -
\frac{M_V^2}{H^2} } \; . \label{S1sol}
\end{equation}
It is de Sitter invariant so long as $M_V^2 > -(D-2) H^2$, which
includes the usual massless photon.

Recall from expression (\ref{gravitondef}) that we define the graviton
field $h_{\mu\nu}$ by conformally rescaling. Although this is simplest
for computations, enforcing de Sitter covariant gauge conditions is 
better done with the field $\chi_{\mu\nu} \equiv a^2 h_{\mu\nu}$ whose
indices are raised and lowered with the de Sitter metric. The most 
general de Sitter covariant gauge condition can be parameterized with 
a real number $\beta \neq 2$,
\begin{equation}
D^{\mu} \chi_{\mu\nu} - \frac{\beta}{2} D_{\nu} \chi^{\mu}_{~\mu} =
0 \; . \label{gravitongauge}
\end{equation}
In any gauge of this form the $\chi_{\mu\nu}$ propagator is the sum
of a spin zero part and a spin two part \cite{Miao:2011fc,Mora:2012zi},
\begin{equation}
i \Bigl[\mbox{}_{\alpha\beta} \Delta_{\rho\sigma} \Bigr](x;x') = i
\Bigl[\mbox{}_{\alpha\beta} \Delta^0_{\rho\sigma} \Bigr](x;x') + i
\Bigl[\mbox{}_{\alpha\beta} \Delta^2_{\rho\sigma} \Bigr](x;x') \; .
\label{gravdecomp}
\end{equation}
The spin zero part is diagonal on the primed and unprimed index groups,
\begin{equation}
i\Bigl[\mbox{}_{\mu\nu} \Delta^0_{\rho\sigma}\Bigr](x;x') =
\mathcal{P}_{\mu\nu} \times {\mathcal{P}'}_{\rho\sigma}
\Bigl[\mathcal{S}_0(x;x') \Bigr] \; . \label{Spin0}
\end{equation}
The spin zero projector $\mathcal{P}_{\mu\nu}$ is,
\begin{equation}
\mathcal{P}_{\mu\nu} \equiv D_{\mu} D_{\nu} + g_{\mu\nu} \Biggl[ 
\Bigl(\frac{2 \!-\!\beta}{D \beta \!-\! 2}\Bigr) \square + 
2 \Bigl( \frac{D \!-\! 1}{D \beta \!-\! 2} \Bigr) H^2 \Biggr] 
\; . \label{P0op}
\end{equation}
And the spin zero structure function can be expressed in terms of a
doubly integrated propagator (\ref{Int5b}) 
\cite{Miao:2011fc,Mora:2012zi},
\begin{equation}
\mathcal{S}_0(x;x') = -\frac{2(D\beta-2)^2}{(2-\beta)^2(D-2)(D-1)}
i\Delta_{WNN}(x;x') \; , \label{spin0sol}
\end{equation}
where $b_W \equiv (D+1)/2$ and $b^2_N \equiv (\frac{D-1}2)^2 + 2 
(\frac{D-1}{2-\beta})$. Although $B_W$ corresponds to a tachyonic 
mass of $m^2 = -D H^2$, it turns out that this source of de Sitter 
breaking drops out when acted upon by the projectors 
\cite{Kahya:2011sy}. The index $b_N$ corresponds to a scalar of mass
$m^2 = -2(\frac{D-1}{2-\beta}) H^2$, and it will give rise to 
physical de Sitter breaking for $\beta < 2$.

Constructing the spin two part of the propagator is roughly analogous
to what was done for the photon. I first define the projector by 
expanding the Weyl tensor, $C^{\alpha\beta\gamma\delta} \equiv 
\mathcal{P}_{\mu\nu}^{~~ \alpha\beta\gamma\delta} \kappa \chi^{\mu\nu} 
+ O(\kappa^2 \chi^2)$. This operator can be used to build a manifestly
transverse and traceless quantity,
\begin{eqnarray}
\lefteqn{i\Bigl[\mbox{}_{\mu\nu} \Delta^2_{\rho\sigma}\Bigr](x;x') }
\nonumber \\
& & = \mathcal{P}_{\mu\nu}^{~~ \alpha\beta\gamma\delta} \times
{\mathcal{P}'}_{\rho\sigma}^{~~ \kappa\lambda\theta\phi} \times
\frac{ D_{\alpha} D_{\gamma} {D'}_{\kappa} {D'}_{\theta}}{4 H^4}
\times \Bigl[\mathcal{S}_2(x;x') \frac{D_{\beta} {D'}_{\lambda} y \,
D_{\delta} {D'}_{\phi} y}{4 H^4} \Bigr] \; . \label{Spin2} \qquad
\end{eqnarray}
The (gauge independent) spin two structure function involves 
doubly integrated propagators (\ref{Int5b}) \cite{Miao:2011fc},
\begin{equation}
\mathcal{S}_2(x;z) = \frac{32}{(D \!-\! 3)^2} \Bigl[
i\Delta_{AAA}(x;z) \!-\! 2 i\Delta_{AAB} + i\Delta_{ABB}(x;z) \Bigr]
\; . \label{spin2sol}
\end{equation}
Although the $B$-type propagator is de Sitter invariant, we saw from
expression (\ref{DeltaA}) that the $A$-type propagator is not. An
explicit computation shows that this de Sitter breaking does not 
drop out when all the various derivatives in (\ref{Spin2}) are
acted \cite{Kahya:2011sy}. 

I have already noted that mathematical physicists are {\it very} 
loath to accept de Sitter breaking in the graviton propagator, so 
it is perhaps not surprising that a final attempt was made to avoid 
it \cite{Morrison:2013rqa}. However, the net result was simply to 
clarify the illegitimate analytic continuations which must be 
employed to derive formal de Sitter invariant solutions that are 
not true propagators \cite{Miao:2013isa}. There are still some who 
believe that the de Sitter breaking evident in the all correct 
solutions (\ref{gprop}-\ref{TC}) and 
(\ref{gravdecomp}-\ref{spin2sol}) will drop out when gauge invariant 
operators are studied, as it does from the linearized Weyl-Weyl 
correlator \cite{Mora:2012kr,Mora:2012zh}. I doubt this, and I will 
point out in section \ref{back} that it amounts to re-fighting the 
same controversy over fields versus potentials which was decided for 
electromagnetism by the Aharomov-Bohm effect. However, the important 
thing to note here is that everyone now agrees on the propagators 
which must be used to make such computations.

Just as for their propagators, so it is best to represent
the 1PI 2-point functions of vector and tensor fields in terms of 
differential operators acting on structure functions. When the 
particular loop under study shows no physical de Sitter breaking one
could employ a de Sitter invariant representation. However, long 
experience shows that it is usually superior to a employ simple, de 
Sitter breaking representation, even when there is no physical 
breaking of de Sitter invariance \cite{Leonard:2012si,Leonard:2012ex,
Leonard:2014zua}. For the vacuum polarization that form is,
\begin{equation}
i\Bigl[ \mbox{}^{\mu} \Pi^{\nu}\Bigr] = \Bigl( \eta^{\mu\nu}
\eta^{\rho\sigma} \!-\! \eta^{\mu\sigma} \eta^{\nu\rho}\Bigr) \,
\partial_{\rho} \partial'_{\sigma} F(x;x') + \Bigl(
\overline{\eta}^{\mu\nu} \overline{\eta}^{\rho\sigma} \!-\!
\overline{\eta}^{\mu\sigma} \overline{\eta}^{\nu\sigma} \Bigr) \,
\partial_{\rho} \partial'_{\sigma} G(x;x') \; , \label{vacpol}
\end{equation}
For the graviton self-energy based on the conformally rescaled
field $h_{\mu\nu}$ defined in expression (\ref{gravitondef}) it is,
\begin{eqnarray}
\lefteqn{-i\Bigl[ \mbox{}^{\mu\nu} \Sigma^{\rho\sigma}\Bigr](x;x') =
\mathcal{F}^{\mu\nu} \times {\mathcal{F}'}^{\rho\sigma} \Bigl[
F_0(x;x') \Bigr] } \nonumber \\
& & \hspace{1cm} + \mathcal{G}^{\mu\nu} \times
{\mathcal{G}'}^{\rho\sigma} \Bigl[ G_0(x;x') \Bigr] +
\mathcal{F}^{\mu\nu\rho\sigma} \Bigl[ F_2(x;x') \Bigr] +
\mathcal{G}^{\mu\nu\rho\sigma} \Bigl[ G_2(x;x') \Bigr] \; . \qquad
\label{ourform}
\end{eqnarray}
The scalar differential operators are,
\begin{eqnarray}
\mathcal{F}^{\mu\nu} &\!\!\!\!=\!\!\!\!\!& 
\partial^{\mu} \partial^{\nu} \!+\! 2 f_1 a H
\delta^{(\mu}_{~0} \partial^{\nu)} \!+\! f_2 a^2 H^2 \delta^{\mu}_{~
0} \delta^{\nu}_{~0} - \eta^{\mu\nu} \Bigl[ \partial^2 \!+\! f_3 a H
\partial_0 \!+\! f_4 a^2 H^2\Bigr] \; , \qquad \label{Fansatz} \\
\mathcal{G}^{\mu\nu} & \!\!\!\!\!=\!\!\!\!\! & 
\overline{\partial}^{\mu}
\overline{\partial}^{\nu} \!+\! 2 g_1 a H \delta^{(\mu}_{~0}
\overline{\partial}^{\nu)} \!+\! g_2 a^2 H^2 \delta^{\mu}_{~ 0}
\delta^{\nu}_{~0} - \overline{\eta}^{\mu\nu} \Bigl[ \nabla^2 \!+\!
g_3 a H \partial_0 \!+\! g_4 a^2 H^2\Bigr] \; , \qquad 
\label{Gansatz}
\end{eqnarray}
where ordinary derivative indices are raised using the Minkowski
metric and I remind the reader that a bar denotes the suppression
of temporal indices, $\eta^{\mu\nu} \equiv \eta_{\mu\nu} + 
\delta^{\mu}_0 \delta^{\nu}_0$. The various constants in 
(\ref{Fansatz}-\ref{Gansatz}) are,
\begin{eqnarray}
f_1 = f_3 = f_4 = (D-1) & , &  f_2 = (D-2)(D-1) \; , \\
g_1 = g_3 = g_4 = (D-2) & , & \quad g_2 = (D-2)(D-1) \; .
\end{eqnarray}
The spin 2 operators are constructed using a transverse-traceless
projector obtained by expanding the Weyl tensor, 
$C_{\alpha\beta\gamma\delta} \equiv a^2 
\mathcal{C}_{\alpha\beta\gamma\delta}^{~~~~ \mu\nu} \times \kappa
h_{\mu\nu} + O(\kappa^2 h^2)$,
\begin{eqnarray}
\mathcal{F}^{\mu\nu\rho\sigma} & \equiv &
\mathcal{C}_{\alpha\beta\gamma\delta}^{~~~~~ \mu\nu} \times
{\mathcal{C}'}_{\kappa\lambda\theta\phi}^{~~~~~ \rho\sigma} \times
\eta^{\alpha\kappa} \eta^{\beta\lambda} \eta^{\gamma\theta}
\eta^{\delta\phi} \; , \label{calF} \\
\mathcal{G}^{\mu\nu\rho\sigma} & \equiv &
\mathcal{C}_{\alpha\beta\gamma\delta}^{~~~~~ \mu\nu} \times
{\mathcal{C}'}_{\kappa\lambda\theta\phi}^{~~~~~ \rho\sigma} \times
\overline{\eta}^{\alpha\kappa} \overline{\eta}^{\beta\lambda}
\overline{\eta}^{\gamma\theta} \overline{\eta}^{\delta\phi} \; .
\label{calG}
\end{eqnarray}

\subsection{Results and Open Problems}\label{results} 

As explained in section \ref{why}, inflationary particle production
is significant for gravitons and for massless, minimally coupled
scalars. Any effect due to gravitons is, by definition, quantum 
gravitational, however, scalar effects may or may not be. An 
important example of scalar quantum gravitational effects is 
correlators of the field $\zeta(t,\vec{x})$ which represents the 
gravitational response to a scalar inflaton. On the other hand, it 
is possible to imagine a massless, minimally coupled scalar which is 
a spectator to de Sitter inflation, with gravity considered to be a
nondynamical background. The effects from such scalars are not 
strictly quantum gravitational, but I will mention them as well 
because their physics is so similar, and because they are so much 
simpler to study. For four such spectator scalar models there are 
complete, dimensionally regulated and fully renormalized results:
\begin{itemize}
\item{For a real scalar with a quartic self-interaction both the
the expectation value of the stress tensor \cite{Onemli:2002hr,
Onemli:2004mb} and the self-mass-squared \cite{Brunier:2004sb,
Kahya:2006hc} have been computed at one and two loop orders. These 
calculations show that inflationary particle production pushes the 
scalar up its potential, which increases the vacuum energy and leads 
to a violation of the weak energy condition on cosmological scales
without any instability.}
\item{For a massless fermion which is Yukawa-coupled to a real
scalar, one loop computations have been made of the fermion self-energy
\cite{Prokopec:2003qd,Garbrecht:2006jm}, the scalar self-mass-squared 
\cite{Duffy:2005ue}, and the effective potential \cite{Miao:2006pn}.
There is also a two loop computation of the coincident vertex function 
\cite{Miao:2006pn}. These calculations show that the inflationary 
production of scalars affects super-horizon fermions like a mass, 
while the scalar mass remains small. The fermion mass decreases the 
vacuum energy without bound in such a way that the universe eventually 
undergoes a Big Rip singularity \cite{Miao:2006pn}.}
\item{For scalar quantum electrodynamics one loop computations have
been made of the vacuum polarization from massless scalars 
\cite{Prokopec:2002jn,Prokopec:2002uw} --- and also slightly massive 
scalars \cite{Prokopec:2003tm} --- of the scalar self-mass-squared 
\cite{Kahya:2005kj,Kahya:2006ui,Prokopec:2006ue}, and of the effective
potential \cite{Allen:1983dg,Prokopec:2007ak}. Much more difficult 
two loop computations have been made for the square of the scalar 
field strength and for its kinetic energy \cite{Prokopec:2006ue}, for
two coincident field strength tensors \cite{Prokopec:2008gw}, and 
for the expectation value of the stress tensor \cite{Prokopec:2008gw}. 
These calculations show that the inflationary production of charged 
scalars causes super-horizon photons behave as if they were massive,
while the scalar remains light and the vacuum energy decreases
slightly \cite{Prokopec:2003iu,Prokopec:2007ak}. Co-moving observers,
at an exponentially increasing distance from the sources, perceive 
screening of electric charges and magnetic dipoles, while observers 
at a fixed invariant distance perceive the same sources to be 
enhanced \cite{Degueldre:2013hba}.}
\item{The nonlinear sigma model has been exploited to better
understand the derivative interactions of quantum gravity 
\cite{Tsamis:2005hd}, and explicit two loop results have been 
obtained for the expectation value of the stress tensor 
\cite{Kitamoto:2010et,Kitamoto:2011yx}. These computations show
that while the leading secular corrections to the stress tensor
cancel, there are sub-leading corrections.}
\end{itemize}

Spectator scalar effects are simpler than those of gravitons, and 
generally stronger because they can avoid derivative interactions. 
Although scalar effects avoid the gauge issue, they are less 
universal because they depend upon the existence of light, minimally
coupled scalars at inflationary scales. In five models with gravitons 
there are complete, dimensionally regulated and fully renormalized 
results:
\begin{itemize}
\item{For pure quantum gravity the graviton 1-point function has
been computed at one loop order \cite{Tsamis:2005je}. This shows that
the effect of inflationary gravitons at one loop order is a slight
increase in the cosmological constant.}
\item{For quantum gravity plus a massless fermion the fermion
self-energy has been computed at one loop order \cite{Miao:2005am,
Miao:2006gj}. This shows that spin-spin interactions with 
inflationary gravitons drive the fermion field strength up by an 
amount that increases without bound \cite{Miao:2007az,Miao:2008sp}.}
\item{For quantum gravity plus a slightly massive fermion the fermion
self-energy has been computed to lowest order in the mass 
\cite{Miao:2012bj}. However, the dynamical consequences of this have
not yet been explored.}
\item{For quantum gravity plus a massless, minimally coupled scalar
there are one loop computations of the scalar self-mass-squared
\cite{Kahya:2007bc} and the graviton self-energy \cite{Park:2011ww}.
The effective field equations reveal that the scalar kinetic energy
redshifts too rapidly for it to experience a significant effect from
inflationary gravitons \cite{Kahya:2007cm}, or for a graviton to
experience significant effects from inflationary scalars 
\cite{Park:2011kg,Leonard:2014zua}. The effects of inflationary
scalars on the force of gravity, are still under study.}
\item{For quantum gravity plus electromagnetism there is a one loop
computation of the vacuum polarization \cite{Leonard:2013xsa}. Even
though gravitons are uncharged, they do carry momentum which can be 
added to virtual photons to make co-moving observers perceive an
enhanced force, while static observers experience no secular change
\cite{Glavan:2013jca}. The effect for magnetic dipoles is opposite
\cite{Glavan:2013jca}. The effect on dynamical photons is still
being studied.}
\end{itemize}
It is worth noting that all these computations were made with the 
simple graviton propagator (\ref{gprop}) in the noncovariant gauge 
(\ref{GR}).

Dolgov did a very early computation of the small cosmological 
contribution to the vacuum polarization from fermionic quantum 
electrodynamics \cite{Dolgov:1981nw}. Another early result is Ford's 
approximate evaluation of the one loop graviton 1-point function in 
pure quantum gravity \cite{Ford:1984hs}. A more recent computation 
was made using adiabatic regularization \cite{Finelli:2004bm}. 
Momentum cutoff computations have also been performed of the one 
loop graviton self-energy \cite{Tsamis:1996qm} and the two loop 
graviton 1-point function \cite{Tsamis:1996qk}. For gravity plus 
various sorts of scalars the one loop scalar contribution to the 
noncoincident graviton self-energy has been computed 
\cite{PerezNadal:2009hr}. More recently, Kitamoto and Kitazawa have 
been exploring infrared graviton corrections to various couplings 
in the off-shell effective field equations \cite{Kitamoto:2012ep,
Kitamoto:2012vj,Kitamoto:2012zp,Kitamoto:2013rea,Kitamoto:2014gva}.

Given a Lagrangian, it is quite simple to work out the interaction
vertices in any background. The obstacle to making computations is
finding the propagators. In section \ref{props} I have given all of
them, in all known gauges, so anyone wishing to make computations
should be able to do so. In spite of all the work that has been 
done, many interesting calculations remain which can be done within
the existing formalism:
\begin{itemize}
\item{Use the previously derived one loop graviton contribution to
the vacuum polarization to work out the effects on dynamical
photons \cite{Leonard:2013xsa}.}
\item{Use the previously derived one loop scalar contribution to
the graviton self-energy \cite{Park:2011ww,Leonard:2014zua} to work 
out corrections to the force of gravity.} 
\item{Perform a fully dimensionally regulated and renormalized 
computation of the one loop graviton contribution to the graviton
self-energy in the noncovariant gauge (\ref{GR}) and use it to work
out what happens to dynamical gravitons \cite{Mora:2013ypa} and to 
the force of gravity.}
\item{Re-compute the one loop gravtion contributions to the scalar,
fermion, vector and graviton 1PI 2-point functions in a general de
Sitter invariant gauge (\ref{gravitongauge}). This is especially 
important to check the conjecture \cite{Miao:2012xc} that the leading
infrared logarithms might be gauge independent.}
\item{Compute the two loop expectation value of 
$C^{\alpha\beta\gamma\delta}(x) C_{\alpha\beta\gamma\delta}(x)$ to
see if the secular corrections which are certainly present in 
individual diagrams cancel out all the diagrams are added to produce
a scalar.}
\end{itemize}

There are also two improvements of the basic formalism which need 
to be developed:
\begin{itemize}
\item{Work out corrections to the various initial states; and}
\item{Develop reasonable and consistent approximations for including
the effects of $\epsilon(t) \neq 0$ in realistic models of inflation.}
\end{itemize}
The state corrections would allow us to solve the effective field 
equations at all times, rather than only at asymptotically late times.
This is important to check for corrections to field strengths which 
develop for a while and then approach a constant at late times when 
the particle under study has redshifted too much to interact further 
with the sea of infrared quanta produced by inflation. It has been
suggested that this could produce an observable $k$-dependent tilt to 
the scalar power spectrum \cite{Kahya:2007cm}. The goal of the second 
improvement would be to finally perform complete and fully 
renormalized computations of the one loop corrections to the power
spectra. Note from section \ref{props} that the technology for 
representing vector and tensor propagators and 1PI functions can be 
applied to any geometry, so the real problem is just generalizing the 
scalar propagator to arbitrary $\epsilon(t)$ \cite{Janssen:2008px}.

Finally, there are a number of less focussed but very important 
issues which require study:
\begin{itemize}
\item{Develop gauge independent and physically reasonable ways of 
quantifying changes in particle dynamics due to inflationary 
particle production; and}
\item{Work out observable consequences of the interactions other
particles have with inflationary gravitons and scalars.}
\end{itemize}
The first point is closely related to the important issue I 
discussed in section \ref{nonlinear} of how we correspond the 
observed power spectra to quantities in fundamental theory.
Of course the second issue is crucial if the calculations 
described in this section are ever to be tested. One obvious
point of phenomenological contact is the origin of cosmic
magnetic fields  \cite{Kronberg:1993vk,Grasso:2000wj}. It has 
been suggested that that the vacuum polarization induced during 
inflation by a light, charged scalar --- possibly part of the
Higgs doublet --- might produce the super-horizon correlations
needed to seed cosmic magnetic fields \cite{Davis:2000zp,
Tornkvist:2000js,Dimopoulos:2001wx,Prokopec:2003bx,
Prokopec:2004au}. More needs to be done in connecting the
undoubted inflationary effect to post-inflationary cosmology.
Another point of phenomenological contact might be inflationary 
baryogenesis \cite{Garbrecht:2007gf}. More generally, we need
to understand what happens after inflation to the ubiquitous
factors of ``$H$'' which occur in one loop corrections to scalar
effective potentials from fermions \cite{Miao:2006pn}, from 
gauge particles \cite{Prokopec:2007ak}, and from other scalars
\cite{Janssen:2009pb}.

\subsection{Back-Reaction}\label{back}

I commented in section \ref{single} on the fine-tuning issues of
single-scalar inflation. The worst of these is the problem of the
cosmological constant \cite{Weinberg:1988cp,Carroll:2000fy}. Many
people have suspected that quantum effects associated with a 
positive cosmological constant might lead to the cosmological
constant screening itself. As far as I know, the first person to 
suggest this was Sasha Polyakov back in 1982 \cite{Polyakov:1982ug}.
Variations of the same idea were put forward in the 1980's by 
Myhrvold \cite{Myhrvold:1983hx}, Mottola \cite{Mottola:1984ar},
Ford \cite{Ford:1984hs}, Antoniadis, Ililiopoulos and Tomaras 
\cite{Antoniadis:1985pj}, Mazur and Mottola \cite{Mazur:1986et}, 
and Antoniadis and Mottola 
\cite{Antoniadis:1986sb,Antoniadis:1991fa}.

In 1992 Nick Tsamis and I proposed a model of inflation based on
three contentions \cite{Tsamis:1992sx,Tsamis:1996qq}:
\begin{itemize}
\item{The bare cosmological constant is not unreasonably small;}
\item{This triggered primordial inflation; and}
\item{Inflation was brought to an end by the gradual accumulation
of self-gravitation between infrared gravitons which are ripped 
out of the vacuum by the accelerated expansion and whose 
gravitational attraction is still holding the universe together.}
\end{itemize}
This is the idea that provoked the hostile referee report from 
which I quoted at the beginning of section \ref{howto}. It still
amazes me that our model generates so much opposition because it
really is a natural way to start inflation, and to make it last a 
long time without fine tuning. Our proposal is also wonderfully 
economical, making a virtue out of the absence of any reason for
the bare cosmological constant to be small, and using gravity to
solve a gravitational problem with no new particles. I will 
devote the remainder of this section to discussing our model but
one should take note of significant work on related proposals by 
Polyakov \cite{Polyakov:2007mm,Polyakov:2009nq,Krotov:2010ma,
Polyakov:2012uc}, by Mottola and collaborators 
\cite{Antoniadis:1996dj,Antoniadis:1997fu,Habib:1999cs,
Anderson:2000wx,Anderson:2001th,Anderson:2005hi,Mottola:2006ew,
Antoniadis:2006wq,Anderson:2009ci,Mottola:2010qg,Antoniadis:2011ib,
Anderson:2013ila,Anderson:2013zia}, by Brandenberger and 
collaborators \cite{Zibin:2000uw,Brandenberger:2002sk,
Geshnizjani:2003cn,Brandenberger:2004kx,Martineau:2005aa,
Martineau:2005zu,Martineau:2007dj}, by Boyanovsky, de Vega and
Sanchez \cite{Boyanovsky:2004ph,Boyanovsky:2005sh,
Boyanovsky:2005px}, by Boyanovsky and Holman 
\cite{Boyanovsky:2011xn}, by Marolf and Morrison 
\cite{Marolf:2010zp,Marolf:2010nz,Marolf:2011sh}, and by Akhmedov
and collaborators \cite{Akhmedov:2008pu,Akhmedov:2009be,
Akhmedov:2009ta,Akhmedov:2010ah,Akhmedov:2011pj,Akhmedov:2012hk,
Akhmedov:2012pa,Akhmedov:2012dn,Akhmedov:2013xka,Akhmedov:2013vka}.

Of course scepticism about my proposal with Tsamis is centered on 
the crucial third bullet
point \cite{Garriga:2007zk}, which we have so far been unable to 
prove but is at least not obviously wrong \cite{Tsamis:2008zz}.
Infrared gravitons are certainly produced during inflation and 
it is difficult to understand why they would not attract one 
another, at least a little. It is also difficult to understand 
how that effect, which starts from zero, can avoid growing as 
more and more of the newly created gravitons come into contact 
with one another. Indeed, it is easy to show that as few as 
ten e-foldings of inflation produces enough infrared gravitons
to make the universe collapse if they were all in causal contact 
\cite{Tsamis:2011ep,Woodard:2014wia}.

\begin{figure}[ht]
\vspace{-3cm}
\hspace{2cm} \includegraphics[width=6cm,height=3cm]{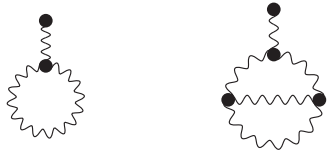} 
\vspace{2cm}
\caption{\label{twoloops} \tiny 
One loop and two loop contributions to the graviton 1-point function
in pure quantum gravity.}
\end{figure}

One reason we have not been able to demonstrate the third point 
is that it requires the finite part of a two loop computation 
in quantum gravity on de Sitter background. This might seem 
confusing after my statement in section \ref{subtree} that the
primordial power spectra are tree order effects but there is not 
any contradiction. The diagram which represents how the tree 
order power spectra of Figure \ref{treegraphs} change the 
background geometry is obtained by attaching the two end points 
to a 3-point vertex and connecting this to an external line, as 
in the leftmost diagram of Figure \ref{twoloops}. That one loop
diagram gives the average gravitational response to the lowest
order stress-energy of inflationary gravitons. It is constant 
and must actually be absorbed into a renormalization of the 
cosmological constant if the universe is really to begin 
inflating at the stated rate \cite{Tsamis:2005je}. (This is 
just like the renormalization condition of flat space quantum
electrodynamics which makes the mass that appears in the 
electron propagator agree with the physical electron mass.)
The effect of self-gravitation between inflationary gravitons 
begins at the next order, for example, in the right hand diagram 
of Figure \ref{twoloops}. This diagram can show secular growth 
in the Schwinger-Keldysh formalism because the internal vertex 
points are integrated over the past light-cone of the external 
point, which grows as one waits to later and later times.  
 
Another reason we have not yet been able to prove the third point
is that it requires an invariant quantification of the expansion
rate. We actually performed a two loop computation of the graviton
1-point function in a fixed gauge (but in $D=4$, with a 
3-momentum cutoff that I don't trust) which does show slowing if we
infer the expansion rate from the expectation value of the metric
the same way one does from a classical metric \cite{Tsamis:1996qm}.
Although the two loop computation consumed a year's time (with a
computer!), using it to get the expansion rate was simple because 
the expectation value of the metric in a homogeneous and isotropic 
state is itself homogeneous and isotropic, so the procedure is 
identical to the passage from expression (\ref{FLRW}) to 
(\ref{cparms}). However, Bill Unruh noted correctly that one cannot 
treat the expectation value of the metric as a metric 
\cite{Unruh:1998ic}. The correct procedure is to instead define an 
invariant {\it operator} which quantifies the expansion rate, and 
then compute its expectation value, just as I did for Green's 
functions in my long-forgotten thesis work \cite{Tsamis:1989yu}. 
So my career has come full circle!

The importance of Unruh's observation was demonstrated by another 
proposal for significant back-reaction in scalar-driven inflation.
Although the lowest order effect in pure gravity is at two loops,
Mukhanov, Abramo and Brandenberger realized in 1996 that mixing 
with dynamical scalars affords scalar-driven inflation the 
opportunity of showing back-reaction at one loop order. Their 
initial work seemed to show such an effect \cite{Mukhanov:1996ak,
Abramo:1997hu}. In 1998 Raul Abramo and I confirmed that their 
result appears as well in different gauges \cite{Abramo:1998hi,
Abramo:1998hj}. However, the effect was absent when we formulated 
an invariant operator to quantify the expansion rate and computed 
its expectation value in 2001 \cite{Abramo:2001db,Abramo:2001dc,
Abramo:2001dd}. A better expansion invariant was proposed and
computed in 2002 by Geshnizjani and Brandenberger, with the same
result \cite{Geshnizjani:2002wp}. Subsequent refinements and
extensions have been made by Morozzi and Vacca 
\cite{Marozzi:2011zb,Marozzi:2012tp,Marozzi:2013uva,
Marozzi:2014xma}.

Pure quantum gravity lacks the scalar upon which the expansion
observable of Geshnizjani and Brandenberger is based. However,
Nick Tsamis and I were eventually able to devise what seems to 
be a reasonable substitute based on a nonlocal functional of the 
metric \cite{Tsamis:2013cka}. With Shun-Pei Miao we are currently
engaged in computing its expectation value at one loop order.
There should be no effect at one loop, but demonstrating that
in a dimensionally regulated and fully renormalized computation 
is an important test of the observable. If it passes then we can 
begin the year-long labor involved in computing its expectation 
value at two loop order. Reaching that order will require a 
determination of the one loop corrections to the initial state.
 
Although computing the expectation value of the expansion
observable is a worthy goal, it will not end the controversy. 
The predicted time dependence takes the form \cite{Tsamis:2005hd,
Prokopec:2007ak},
\begin{equation}
H(t) = H_2 \Biggl\{ 1 + \frac{\hbar G H_2^2}{c^5} \times 0 
+ \Bigl( \frac{\hbar G H_2^2}{c^5}\Bigr)^2 \times K 
\ln\Bigl[ \frac{a(t)}{a(t_2)} \Bigr] + 
O \Bigl( G^3 \ln^2(a)\Bigr) \Biggr\} \; , \label{predict}
\end{equation}
where $H_2$ is the expansion rate of an initially empty 
universe released at time $t = t_2$. Determining the value of 
$K$ is expected to require a year. Assuming it is negative, 
we would have proven that the production of inflationary 
gravitons slows inflation by an amount which eventually 
becomes nonperturbatively strong. However, this only means
that perturbation theory breaks down, not that inflation 
eventually stops. The unknown higher loop contributions might 
push the universe into deceleration, or they might sum up to 
produce only a small fractional decrease in the expansion rate.
Both cases occur in scalar models \cite{Miao:2006pn,
Prokopec:2007ak}.

Working out what happens requires some sort of 
nonperturbative resummation technique. For scalar potential
models (with nondynamical gravity) Starobinsky and Yokoyama
showed how to sum the series of leading logarithms of the 
scale factor $a(t)$ \cite{Starobinsky:1994bd,Woodard:2005cw,
Tsamis:2005hd}. This technique was successfully extended to a
Yukawa-coupled scalar \cite{Miao:2006pn} and to scalar quantum
electrodynamics \cite{Prokopec:2007ak}, but there is still no
generalization to quantum gravity \cite{Miao:2008sp}. One 
important piece of recent progress is the demonstration that 
a classical configuration of gravitational radiation can 
indeed hold the universe together for a least a little while 
\cite{Tsamis:2014kda}. This is crucial because particles
created during inflation decohere so that their physics is 
essentially classical at late times. If Tsamis and I are 
right, it must be a classical configuration of gravitational 
radiation which is holding the universe together today. The
existence of such a configuration also shows that 
inflationary particle production can actually stop inflation,
if only it can produce the necessary configuration.

There is absolutely no doubt that individual two loop 
diagrams make secular contributions to the expansion
observable of the form (\ref{predict}). Sceptics believe that
all such contributions will cancel when the many, many 
diagrams are summed to produce a gauge invariant result. This
certainly does not happen in scalar models which show the 
same sort of secular growth \cite{Miao:2006pn,
Prokopec:2007ak}. Sceptics are convinced that gravity will be 
different because the growth factors derive from long wave 
length gravitons which are nearly constant, and {\it exactly} 
constant gravitons are gauge equivalent to zero. Of course 
the key distinction is between approximately constant and
exactly constant. Sceptics insist that there can be no 
gravitational effects unless there is curvature, and effects
must be small if the curvature is small. I take Lagrangians 
more seriously than pre-conceived opinions, and there is no 
doubt that matter (and the metric itself) couples to the 
metric, not to the curvature. Hence it must be possible, 
under certain circumstances, for these couplings to produce 
effects even in places and at times when the curvature of 
the original source has red-shifted to nearly zero.

The computation will decide who is right, but it is fitting
to end this section by commenting on the similarity of this
controversy to a classic dispute in the history of quantum
electrodynamics. Electromagnetic Lagrangians show 
unambiguously that quantum matter couples to the 
electromagnetic vector potential rather than to the field 
strength, hence it must be possible for these couplings 
to produce effects even at times and in places where the field 
strength is small. However, physicists are as prone to 
prejudice as other humans and generations of theorists ignored
what the theory was telling them and insisted that nothing can
happen unless the field strength is nonzero. In 1949 Ehrenberg 
and Siday predicted \cite{Ehrenberg:1949} what is better known 
today as the Aharonov-Bohm effect \cite{Aharonov:1959fk}. We 
all know how the experiment turned out. I am confident the 
quantum gravitational reprise will have a similar outcome, but 
the important point for this article is that posing such 
classic questions, and assembling the technology to answer 
them, is one more example of how quantum gravity is becoming a 
mature subject.

\section{Conclusions}\label{conclusions}

Another man described another revolutionary era in the following terms: {\it It
was the best of times, it was the worst of times.} Fortunately, no one is likely to
get his head chopped off as quantum gravity makes the difficult passage to maturity!
It is nonetheless a time of great change, and that affects different physicists in 
different ways. Some are delighted by the opportunity to participate in founding a 
new field; others want the changes to stop with the recognition of their own work; 
and some pretend that nothing is changing. The majority of those who call themselves 
quantum gravity researchers are in the last camp. To them the key problem of quantum 
gravity is resolving what happens to black holes in the final stages of evaporation, 
a process for which we have no data and are not likely to acquire data any time soon. 
Many of these people are my friends and I wish them well, but the last three decades 
of unrequited toil in string theory have proven that humans are not good at guessing 
fundamental theory without the guidance of data.

Inflationary cosmology is already changing the way we think about quantum gravity.
For example, neither discretization nor simply refusing to quantize the metric are
any longer viable solutions to the divergence problem of quantum gravity 
\cite{Woodard:2009ns}. Cosmology has required us to abandon in-out quantum field
theory because the universe began with an initial singularity and no one knows how
it will end, or even if it will end. And releasing the Universe in a prepared state 
at a finite time implies we must stop dodging the issue of perturbative corrections 
to the initial state.

Inflationary cosmology is also forcing us to re-think the issue of observables. 
For example, it used to be pronounced, with haughty dogmatism, that the S-matrix 
defines all that is observable about a quantum field theory. That 
statement was always dubious, but we now have counterexamples --- which cannot be 
dismissed because they are being {\it observed} --- in the form of the primordial 
power spectra discussed in section \ref{tree}. The recent controversy over the use 
of adiabatic regularization (see section \ref{adiabatic}) makes it clear that we 
need to think more carefully about how to connect theory to observation even at 
{\it tree order}. The one loop corrections discussed in section \ref{loop} --- 
{\it which may well be observable in a few decades} --- pose the further problem 
of how to define loop corrections to the power spectrum so that they are infrared 
finite, ultraviolet renormalizable, and so that they show the expected pattern of 
$\epsilon$-suppression. And we are just beginning to understand what other quantum 
gravitational effects (see section \ref{other}) might be engendered by 
inflationary gravitons and scalars.

Some people dismiss the impact of inflationary cosmology on quantum gravity
because the detection of primordial gravitons is still tentative \cite{Ade:2014xna,
Flauger:2014qra,Paul} and because fundamental theory has yet to provide a 
compelling model of primordial inflation. I believe these objections are purblind
and transient. Within no more than a few years we will know whether or not BICEP2 
has detected primordial gravitons \cite{IPSIG}. It is not so easy to forecast when 
fundamental theorists will bring forth a compelling model of inflation. The six 
fine tuning problems I mentioned in section \ref{single} indicate that there is 
something very wrong with our current thinking. I suspect only a painful collision 
with data is going to straighten us out. Fortunately, the continuing flow of data 
makes that collision inevitable. When our observational colleagues have finally 
uncovered enough of the truth for us to guess the rest it will be possible to 
replace the rough estimates (\ref{treepower}) of the tree order power spectra 
with precise results, and also to compute loop corrections reliably, both for the 
power spectra and for other quantum gravitational effects. In the fullness of time 
all of these predictions will be tested by the data present in 21 cm radiation. 
When that process has gone to completion perturbative quantum gravity will assume 
its place as a possession for the ages.

\section*{Acknowledgements}

I have profited from conversation and collaboration on this subject with L. R. 
Abramo, B. Allen, R. H. Brandenberger, S. Deser, L. H. Ford, J. Garriga, G.
Geshnizjani,  A. H. Guth, S. Hanany, A. Higuchi, B. L. Hu, E. O. Kahya, H. 
Kitamoto, K. E. Leonard, A. D. Linde, D. Marolf, S. P. Miao, P. J. Mora, I. A. 
Morrison, V. F. Mukhanov, W. T. Ni, V. K. Onemli, S. Park, L. Parker, U. L. Pen, 
T. Prokopec, A. Roura, A. A. Starobinsky, T. Tanaka, N. C. Tsamis, W. G. Unruh, 
Y. Urakawa, A. Vilenkin, C. Wang and S. Weinberg. This work was partially 
supported by NSF grant PHY-1205591, and by the Institute for Fundamental 
Theory at the University of Florida.

\end{document}